\documentclass[a4paper,superscriptaddress,11pt]{article}
\title{ Polymer quantization of the free scalar field and its classical limit}
\author{Alok Laddha\\Raman Research Institute, Bangalore-560 080, India
 \\alok@rri.res.in \and Madhavan Varadarajan\\Raman Research Institute, Bangalore-560 080, India \\madhavan@rri.res.in}
\begin{document}

\maketitle

\begin{abstract}
Building on prior work, a generally covariant reformulation of free scalar field theory on the 
flat Lorentzian cylinder 
is quantized using Loop Quantum Gravity (LQG) type `polymer' representations. 
This quantization of the {\em continuum}
classical theory yields a quantum
theory which lives on a discrete spacetime lattice. We explicitly  
construct a state in the polymer Hilbert space which reproduces
the standard Fock vacuum- two point functions for long wavelength modes of the scalar field. Our construction
indicates that the continuum classical theory emerges under coarse graining. All our considerations are 
free of the ``triangulation'' ambiguities which plague attempts to define quantum dynamics in LQG. 
Our work constitutes the first complete LQG type quantization of a generally covariant field theory together
with a semi-classical analysis of the true degrees of freedom and thus provides a perfect infinite dimensional toy model
to study open issues in LQG, particularly those pertaining to the definition of quantum dynamics.

\end{abstract}

\section{Introduction}
Loop Quantum Gravity (LQG) is a non- perturbative approach to quantum gravity which, in its canonical version, 
attempts to 
construct a Dirac quantization of a
Hamiltonian description of gravity.
LQG techniques have yielded 
many beautiful results such as a satisfactory treatment of spatial diffeomorphisms \cite{RSlooprep,ALM^2T}, 
discrete spatial geometry \cite{RSarea,aajlarea,jlvolume},
a calculation
of black hole entropy \cite{carloentropy,aaentropy} and, in the context of their application to
cosmological minisuperspaces, a resolution of the big bang singularity \cite{martinlqc,aalqc}.
Nevertheless, significant open issues remain, chief among them being (i) a satisfactory definition of the 
quantum dynamics of gravity \cite{thomashamconstraint,ttalqg,spinfoamreview} (ii) issues related to the extraction of
gauge invariant (i.e. spacetime diffeomorphism invariant) physics and (iii) the emergence of general 
relativity in the classical limit and, specifically, that of flat spacetime and its graviton excitations
\cite{megraviton,carlograviton}.  While these issues are involved precisely due to the detailed complicated 
classical dynamics of the gravitational field, one expects that a better understanding of the 
consequences in quantum theory, of  
more general feautures of  gravity such as its general covariance and field theoretic
nature, may be of some use in their elucidation.
It is therefore of interest to analyse generally covariant, field theoretic toy models for which appropriate 
analogs of (i)- (iii) exist and can be resolved. Surprisingly no such toy model exists (at least to our knowledge 
of the literature) for which such an analysis has been carried out to completion.
In this work we continue our investigation \cite{alokme} of free scalar field theory on flat spacetime (cast in
generally covariant disguise) in an LQG type ``polymer'' representation. We show that this system  is indeed a good 
toy model in the context of our remarks above.

The generally covariant reformulation of free scalar field theory considered here 
goes by the name of Parametrised Field Theory (PFT)\cite{karelcm}.
It offers an elegant 
description of free scalar field evolution on {\em arbitrary} (and in general curved) foliations of the 
background spacetime by treating the `embedding variables' which describe the foliation as dynamical variables
to be varied in the action in addition to the scalar field.
Let $X^A= (T,X)$ denote inertial coordinates
on 2 dimensional flat spacetime. In PFT, $X^A$ are parametrized by a new set of arbitrary coordinates
 $x^{\alpha}=(t,x)$
such that for fixed $t$, the embedding variables $X^A(t,x)$ define a spacelike Cauchy slice of flat spacetime.
General covariance of PFT ensues from the arbitrary choice of $x^{\alpha}$ and implies that in its canonical 
description, evolution from one slice of an arbitrary foliation to another is generated by constraints.
Thus, as in the Hamiltonian formulation of General Relativity, evolution and gauge are intertwined and the theory 
must be interpreted through gauge invariant Dirac observables. Since the true dynamical content of PFT is identical
to that of free scalar field theory the standard mode functions
of the latter can be constructed as Dirac observables of
the former \cite{karelcm,alokme}. 

In earlier work \cite{alokme} we used LQG techniques and polymer representations to define and solve the quantum dynamics
of PFT, as well as to construct 
an overcomplete  set of Dirac observables as operators on the space of gauge invariant, physical states. We found that
this set of Dirac observables is so large that at most only a countable subset thereof could possibly display
semiclassical behaviour in any candidate semiclassical state. Since there is no natural choice of a smaller set of
observables, further analysis of semiclassical behaviour would necessarily be based on an ad- hoc choice.
Here we remedy this defect by starting out with a smaller classical algebra at the kinematic level. The set of 
functions chosen are still large enough to seperate points in phase space. They are obtained by restricting 
the embedding and matter charges which multiply the exponents of the holonomy type functions of 
Reference \cite{alokme} to be integer valued rather than real as in
\cite{alokme}. This tighter choice of kinematic algebra has several nice consequences. 
The charge network states of the quantum theory are now also labelled by integer charges, the spectrum of the 
analog of the volume operator for LQG becomes discrete, the physical state space becomes much smaller and
only a countable subset of the  Dirac observables  of Reference \cite{alokme} map the physical state space
into itself. Indeed, the fact that integer charges suffice is also true for treatments of scalar field matter
in LQG and we anticpate that using them instead of the real valued charges of \cite{lqgscalarfield} 
would have similar nice consequences there.

With this improved choice for PFT, we are able to complete the analysis of \cite{alokme} and derive the following
results.
Remarkably, while we started out with a bona- fide quantization of a continuum classical theory,
the quantum theory we end up with can be interpreted as the quantum theory of a free scalar field on a
{\em discrete} spacetime lattice. 
%The final picture is rather similar to the background independence lattice theory obtained by 
%Thiemann et al via loop quantization of Brown-Kuchar model. 
The Poincare group (which for the  $S^1 \times R$ spacetime
topology consists exclusively of (all) spacetime translations) reduces to that of 
discrete spacetime
translations on this lattice. 
Thus,  only discrete spacetime translations are implemented as unitary operators on
the physical Hilbert space. While there are infinitely many discrete- Poincare invariant states, most of them 
are not semiclassical. We explicitly construct a discrete- Poincare invariant state which is  semiclassical 
in the sense that it reproduces the behaviour of the standard Fock vacuum two point functions for slowly varying modes of the 
scalar field. 
%In other words, the Fock vacuum emerges from an appropriate, {\em normalized} 
%state in the physical polymer Hilbert space for coarse grained observables 
thus providing a precise resolution of
the analog of issue (iii) in this model.

The layout of the paper is as follows. After a review of classical PFT in section 2, we construct its  
polymer representation in sections 3- 6 and explore the physical interpretation of this representation in sections 
7- 10.
Section 11 is devoted to a discussion of our results as well as avenues for further research.
As mentioned above, Reference \cite{alokme} contains a detailed construction of the polymer representation 
for PFT. 
In sections 3- 6 the only but {\em key} difference is that we replace the real labels of the PFT 
``spin- networks'' of Reference \cite{alokme} by (modulo some technicalities) integer valued ones. 
While our presentation in sections 3- 6 is self contained, we shall be brief; the interested reader is urged to
consult  Reference \cite{alokme} for details. Indeed, our aim in these sections is 
to introduce our notation to the 
reader unfamiliar with  Reference \cite{alokme} and to 
highlight the consequences of the replacement of the real labels of that work  by integer valued ones.
In section 3, we construct the quantum kinematics and display the unitary action of finite gauge transformations
on the kinematic Hilbert space. In section 4 we construct the Dirac observables of the theory as operators on the 
kinematic Hilbert space. In section 5 we construct the physical Hilbert space by Group Averaging techniques 
\cite{grpavg} and isolate, therein, a physically relevant superselected sector and in section 6 we deal with 
some subtelities related to the zero mode of the scalar field. In section 7 we demonstrate the existence of
infinitely many discrete- Poincare invariant states in the physical Hilbert space. In section 8 we show that 
states in the superselected sector of section 5 can interpreted as quantum excitations of the scalar field on a 
spacetime lattice. Consequently, a detailed analysis of the physical content of such states requires
certain tools from the theory of discrete Fourier transforms which we summarize in section 9.  Section 10 
is devoted to the construction of a (discrete- Poincare invariant) state which yields Fock- vacuum like behaviour 
for long (,with respect to the lattice scale,) wavelength excitations of the scalar field. Various
technicalities and calculational details are collected in the Appendices.

\section{Classical PFT on $S^1\times R$.}
We provide a brief review of classical two dimensional PFT. In section 2.1 we review the Lagrangian formalism for two dimensional PFT and summarise the results related to the hamiltonian
formulation in sections 2.2-2.4. Once again we recommend [1] to those who are interested in the details.
\subsection{The Action for PFT.}

The action for a free scalar field $f$ on
a fixed flat 2 dimensional spacetime ${\cal M}$ in terms of global inertial coordinates $X^A,\;A=0,1$ is 
\begin{equation}
S_0[f] = -\frac{1}{2} \int d^{2}X \eta^{AB}\partial_Af \partial_Bf ,
\label{s0}
\end{equation}
where  the Minkowski metric 
in inertial coordinates, $\eta_{AB}$,  is diagonal with entries $(-1,1)$. If instead, we use coordinates
$x^{\alpha}\;, \alpha= 0,1$ (so that $X^A$ are `parameterized' by $x^{\alpha}$,  $X^A= X^A(x^{\alpha})$), we have
\begin{equation}
S_0[f] = -\frac{1}{2} \int d^{2}x \sqrt{\eta} \eta^{\alpha \beta}\partial_{\alpha}f \partial_{\beta}f ,
\label{s0x}
\end{equation}
where $\eta_{\alpha \beta}= \eta_{AB} \partial_{\alpha}X^A\partial_{\beta}X^B$ and $\eta$ denotes the 
determinant of $\eta_{\alpha \beta}$. The action for PFT is obtained 
by considering the right hand side of (\ref{s0x}) as a functional, not only of $f$, but also of 
$X^A(x)$ i.e. $X^A(x)$ are considered as  2 new scalar fields to be varied in the action ($\eta_{\alpha \beta}$
is a function of $X^A(x)$). Thus 
\begin{equation}
S_{PFT} [f, X^A]= -\frac{1}{2} \int d^{2}x \sqrt{\eta(X)} \eta^{\alpha \beta}(X)\partial_{\alpha}f \partial_{\beta}f .
\label{spft}
\end{equation}
Note that $S_{PFT}$ is a diffeomorphism invariant functional of the scalar fields $f (x), X^A (x)$.
Variation of $f$ yields the equation of motion $\partial_{\alpha}(\sqrt{\eta}\eta^{\alpha \beta}\partial_{\beta}f)= 0$,
which is just the flat spacetime equation $\eta^{AB}\partial_A\partial_B f=0$ written in the coordinates $x^{\alpha}$.
On varying $X^A$, one obtains equations which are satisfied if  $\eta^{AB}\partial_A\partial_B f=0$.
This implies that $X^A(x)$ are   undetermined functions (subject to the condition that determinant of $\partial_{\alpha}X^A$
is non- vanishing). This 2 functions- worth of gauge is a reflection of the 2 dimensional diffeomorphism invariance
of $S_{PFT}$. Clearly the dynamical content of $S_{PFT}$ is the same as that of $S_0$; it is only that 
the diffeomorphism invariance of 
$S_{PFT}$ naturally allows a description of the standard free field dynamics dictated by $S_0$ on {\em arbitrary}
foliations of the fixed flat spacetime.\\
Note that in PFT, $X^{A}(x)$ has a dual interpretation - one as  dynamical variables to be varied in the action, and
the other as inertial coordinates on a flat spacetime. In what follows we shall freely go between these two interpretations.

\subsection{Canonical formulation}
Before we describe the phase-space for the above action, it is useful to keep in mind the following geometric description for $X^{A}(x)$.\\
Consider $S^{1}$ with a ``global" angular co-ordinate $x\in [0,2\pi]$ such that $x=0\sim x=2\pi$. Let $(X^{+}=T+X, X^{-} = T-X)$ be the light-cone co-ordinates on 
${\cal M}= S^{1}\times{\bf R}$. whence X is the angular co-ordinate on T=const. slice which we take to be of length $2\pi L$.\\
Now consider the spacelike embeddings of $(S^{1}, x)$ in ${\cal M}$ via $(X^{+}(x), X^{-}(x))$. Due to the cylinder topology of ${\cal M}$ we require $(X^{+}(x), X^{-}(x))$ to satisfy
the following two conditions.\\
\noindent{\bf (i)} $X^{\pm} (2\pi)- X^{\pm}(0) = \pm 2\pi L$.\\
\noindent{\bf (ii)} Any two sets of embedding data $(X_1^+(x), X_1^-(x))$
and $(X_2^+(x), X_2^-(x))$ are to be identified if there exists an intger $m$ such that
$X_{1}^{+}(x) = X_{2}^{+}(x) + 2m\pi L\ \forall\ x\in\ [0,2\pi]$ and 
$X_{1}^{-}(x) = X_{2}^{-}(x) - 2m\pi L\ \forall\ x\in\ [0,2\pi]$.\\
%Condition {\bf (i)}) infact implies that one could also consider the following extension of the domain of $X^{\pm}$ to the entire real line. 
%\begin{equation}\label{eq:nov16-1}
%X^{\pm}_{ext}(x + 2m\pi)\ =\ X^{\pm}(x) \pm 2m\pi L\  \forall\ x\in [0,2\pi)\\
%\end{equation}
The phase space of PFT splits into two mutually disjoint\footnote{Here we assume that we can consistently ``freeze" the zero mode of the scalar field. This is shown in [1]} sectors
which are co-ordinatized by, $(X^{+}, \Pi_{+}, Y^{+})$, $(X^{-},\Pi_{-}, Y^{-})$.\\
$\Pi_{\pm}$ are the momenta conjugate to the embedding fields $X^{\pm}$ and $Y^{\pm}(x)\ =\ \pi_{f}(x)\pm f'(x)$ describe the right(left) moving modes of the scalar field f, whose
conjugate momentum is $\pi_{f}$. 
$(Y^{+},\ Y^{-})$ do not polarize the phase-space and their Poisson brackets are given by,
\begin{equation}
\begin{array}{lll}
\{Y^{\pm}(x),\ Y^{\pm}(y)\}\ =\ \pm\ (\partial_{x}\delta(x,y)\ -\ \partial_{y}\delta(y,x)\ )\\
\vspace*{0.1in}
\{Y^{\pm}(x),\ Y^{\mp}(y)\}\ =\ 0
\end{array}
\end{equation}
Due to the general co-variance of the Lagrangian, the $(\pm)$ sectors of the theory contain mutually commuting first-class constraints 
\begin{equation}\label{eq:2pm}
{{H^{\pm}}}(x)\ =\ [\ \Pi_{\pm}(x)X^{\pm'}(x)\ \pm\
\frac{1}{4}Y^{\pm}(x)^2\ ].
\end{equation}
and the constraint algebra is
\begin{equation}\label{eq:6}
\begin{array}{lll}
\lbrace {{H}}_{\pm}[N^{\pm}],{{H}}_{\pm}[M^{\pm}]
\rbrace\ =\ {{H}}_{\pm}[\mathcal{L}_{N^{\pm}}M^{\pm}]\\
\vspace*{0.1in}
\lbrace {{H}}_{\pm}[N^{\pm}],{{H}}_{\mp}[M^{\mp}]
\rbrace\ =\ 0
\end{array}
\end{equation}
where $N^{\pm}$ are the Lagrange multipliers. (Geometrically they are vector fields on $(S^{1}, x)$).\\
The action of the constraints $H^{\pm}[N_{\pm}]$ on the canonical variables is given by,
\begin{equation}
\begin{array}{lll}
\{ \Phi^{\pm}(x),\ H_{\pm}[N^{\pm}]\ \}\ =\
\mathcal{L}_{N^{\pm}}\Phi^{\pm}(x)\\
\vspace*{0.1in}
\{ X^{\pm}(x),H_{\pm}[N^{\pm}]\} = N^{\pm}(X^{\pm})^{\prime},\label{c,x1}
\end{array}
\end{equation}
Although the above equations indicate that the constraints generate spatial diffeomorphisms on $S^{1}$, this interpretation is not quite correct due to the quasi-periodic nature of 
$X^{\pm}(x)$. In order to interprete the gauge transformations geometrically, we need to extend domain of the fields (as well as that of $N^{\pm}$) from $S^{1}$ to ${\bf R}$. 
%Such an extension for $X^{\pm}(x)$ is already defined in (\ref{eq:nov16-1}). 
Condition {\bf (i)}) above infact suggests the following extension of the domain of $X^{\pm}$ to the entire real line. 
\begin{equation}\label{eq:nov16-1}
X^{\pm}_{ext}(x + 2m\pi)\ =\ X^{\pm}(x) \pm 2m\pi L\  \forall\ x\in [0,2\pi)\\
\end{equation}
It is straightforward to show that all the remaining fields are periodic in x. Whence the extension of their domain from $[0,2\pi]$ to ${\bf R}$ 
is rather trivial.
\begin{equation}
\begin{array}{lll}
\Phi_{ext}^{\pm}(x + 2m\pi)\ =\ \Phi^{\pm}(x)\\
\vspace*{0.1in}
N_{ext}^{\pm}(x + 2m\pi)\ =\ N^{\pm}(x)
\end{array}
\end{equation}
$\forall x\in [0,2\pi)$. Here $\Phi^{\pm}:= (\Pi_{\pm}, Y^{\pm})$.\\
As we showed in \cite{alokme}, the finite gauge transformations generated by the constraints can be interpreted as spatial diffeomorphisms generated by 
$N_{ext}^{\pm}$ on $(X^{\pm}_{ext}, \Phi_{ext}^{\pm})$.
Let $\Psi^{\pm}(x)\in (X^{\pm}(x), \Pi_{\pm}(x), Y^{\pm}(x))$ 
and let its appropriate quasiperiodic/periodic  extension on $\mathbf{R}$ be $\Psi^{\pm}_{ext}$.
Then  we have that, $\forall x\in[0,2\pi]$,
\begin{equation}\label{eq:13*}
\begin{array}{lll}
(\alpha_{\phi^{\pm}}\Psi^{\pm})(x)\ =\ \Psi^{\pm}_{ext}(\phi^{\pm}(x))\\
\vspace*{0.1in}
(\alpha_{\phi^{\pm}}\Psi^{\mp})(x)\ =\ \Psi^{\mp}(x).
\end{array}
\end{equation}
Where we have labelled every finite gauge transformation by a pair of such 
diffeomorphisms $(\phi^+,\phi^-)$ so that the Hamiltonian flows generated by $H_{\pm}$ are denoted by 
$\alpha_{\phi^{\pm}}$. \\
Equations (\ref{eq:13*}) imply a left representation of the group of periodic diffeomorphisms of $\mathbf{R}$ by the
Hamiltionian flows corresponding to finite gauge transformations:
\begin{eqnarray}
\alpha_{\phi_1^{\pm}}\alpha_{\phi_2^{\pm}}&=& \alpha_{\phi^{\pm}\circ \phi_2^{\pm}} \label{alphaphi1phi2}\\
\alpha_{\phi_1^{\pm}}\alpha_{\phi_2^{\mp}}&=& \alpha_{\phi_2^{\mp}}\alpha_{\phi_1^{\pm}}. 
\end{eqnarray}

\subsubsection{Mode functions as Dirac observables.}
Two dimensional PFT admits a complete set of Dirac observables, which are nothing but the Fourier modes of Klein-Gordon Scalar field.
\begin{equation}
a_{(\pm)n}= \int_{S^1}dx Y^{\pm}(x) e^{inX^{\pm}(x)}, \;n\in {\mathbf Z}, \;n>0
\label{eq:16}
\end{equation}

Together with their complex conjugates, these observables form the (Poisson) algebra,
\begin{equation}\label{eq:17}
\begin{array}{lll}
\lbrace a_{n}, a_{m}* \rbrace\ =\ -4\pi in\delta_{n,m},\\
\lbrace a_{n}, a_{m} \rbrace\ =\ 0,\\
\lbrace a_{n}*, a_{m}* \rbrace\ =\ 0.
\end{array}
\end{equation}

More generally, since finite gauge transformations act as periodic diffeomorphisms 
of $\mathbf{R}$, it follows, directly,
that the integral over $x\in[0,2\pi]$ of any periodic scalar density constructed solely from the phase space
variables, is an observable. In particular 
\begin{equation}
O_{f^{\pm}}:=\int_{S^1}dx Y^{\pm}(x) f^{\pm}(X^{\pm}(x))
\label{defopmf}
\end{equation}
 for all real,
periodic functions $f^{\pm}$ are observables.

\subsubsection{Conformal Isometries.}

Free mass-less scalar field theory in 1+1 dimensions (\ref{s0}) is conformally invariant. As a consequence  
the generators of conformal isometries in PFT are also Dirac observables (for details, see Reference \cite{karelcm}).
Consider the conformal isometry generated by the conformal Killing field ${\vec U}$ on the Minkowskian cylinder.
Let ${\vec U}$ have the components $(U^+(X^+), U^-(X^-))$ in the $(X^+, X^-)$ coordinate system.
$U^{\pm}$ are periodic functions of $X^{\pm}$ by virtue of the fact that ${\vec U}$ is smooth vector field on
the flat spacetime $S^1\times R$.
These components of ${\vec U}$ naturally correspond to the functionals 
 $(U^+(X^+), U^-(X^-))$ on the phase space of PFT. 
The Dirac observable corresponding to the generator of
conformal transformations associated with ${\vec U}$ is given by
\begin{equation}\label{eq:18}
\Pi_{\pm}[U^{\pm}]\ =\ \int_{S^1} \Pi_{\pm}(x)U^{\pm}(X^{\pm}(x))
\end{equation}
%As mentioned above $(U^{+}(X^{+}(x),U^{-}(X^{-})(x))$ are pullbacks
%  of components of  conformal Killing fields on $S^1$ and hence
%are periodic functions of $X^{\pm}(x)$. 
These observables generate a Poisson algebra isomorphic to that of the commutator algebra of conformal Killing fields:
\begin{equation}\label{eq:19}
\begin{array}{lll}
\lbrace \Pi_{\pm}[U^{\pm}], \Pi_{\pm}[V^{\pm}]\rbrace\ =\
\Pi[[U,V]^{\pm}]\\
\vspace*{0.1in}
\lbrace \Pi_{\pm}[U^{\pm}], \Pi_{\mp}[V^{\mp}]\rbrace\ = 0.
\end{array}
\end{equation}
Here $[U,V]^{\pm}$ refer to the $\pm$ components of the commutator of the spacetime vector fields
${\vec U}, {\vec V}$. $[U,V]^{\pm}$ define functions of the embedding variables $X^{\pm}(x)$
in the manner described above.

Note that these observables are weakly equivalent, via the constraints (\ref{eq:2pm}) to
quadratic combinations of the mode functions \cite{karelcm}.
In the 
standard Fock representation of quantum theory (see for e.g. Reference \cite{karelqm}), these
quadratic combinations are nothing but the generators of the Virasoro algebra.

The polymer quantization of PFT  given in \cite{alokme} provides a representation for the finite canonical transformations
generated by $\Pi^{\pm}[U^{\pm}]$. 
\begin{equation}
\alpha_{(\Pi_{\pm}[U^{\pm}],t)} X^{\pm}(x) = (\phi_{(\vec{U},t)} X^{\pm})(x).
\label{alphapiu}
\end{equation}
Here
$\phi_{(\vec{U},t)}$ denotes the one parameter family of conformal isometries generated by the 
conformal Killing field ${\vec U}$ on spacetime. $\phi_{(\vec{U},t)}$ maps the spacetime point $(X^+, X^-)$
to $\phi_{(\vec{U},t)} X^{\pm}$ and hence maps the  spatial slice 
defined by the canonical data $X^{\pm}(x)$ is mapped to the new slice (and hence the new canonical data)
$(\phi_{(\vec{U},t)} X^{\pm})(x)$.
%The action of $\alpha_{(\Pi^{\pm}[U^{\pm}],t)}$ on the embedding momenta is more involved and we do not present it here.
$\phi_{(\vec{U},t)}$ ranges over all conformal isometries connected to identity. Any such conformal isometry
$\phi_c$ is specified by a pair of functions $\phi_c^{\pm}$ so that 
$\phi_c(X^+, X^-) := (\phi_c^+ (X^+), \phi_c^- (X^-))$. 
Invertibilty of $\phi_c$ together with connectedness with identity
 implies that 
\begin{equation}
\frac{d\phi_c^{\pm}}{dX^{\pm}} >0,
\label{phicnondeg}
\end{equation}
and the cylindrical topology of spacetime implies
that
\begin{equation}
\phi_c^{\pm}(X^{\pm}\pm 2\pi L) = \phi_c^{\pm}(X^{\pm})\pm 2\pi L.
\label{phicqp}
\end{equation}
Thus, we may denote the Hamiltonian flows which generate conformal isometries by 
$\alpha_{\phi_c}$ or, without loss of generality, by $\alpha_{\phi_c^{\pm}}$ with
$\alpha_{\phi_c^{\pm}}$ acting trivially on the $\mp$ sector.

To summarise:$\alpha_{\phi_c^{\pm}}$ leave the matter variables untouched, so that
\begin{equation}
\alpha_{\phi_c^{\pm}} Y^{\pm}(x) = Y^{\pm}(x),\;\;\;
\alpha_{\phi_c^{\pm}} Y^{\mp}(x)=Y^{\mp}(x),
\label{phicalpha0}
\end{equation}
 and act on $X^{\pm}(x)$ as
\begin{equation}
\alpha_{\phi_c^{\pm}} X^{\pm}(x) = \phi_c^{\pm} (X^{\pm}(x)),\;\;\;
\alpha_{\phi_c^{\pm}} X^{\mp}(x)=X^{\mp}(x).
\label{phicalpha1}
\end{equation}
Further, since $\Pi_{\pm}[U^{\pm}]$ are observables which commute strongly with the constraints,
the corresponding Hamiltonian flows are gauge invariant. This translates to the condition that for all
\begin{equation}
\begin{array}{lll}
\alpha_{\phi_c^{\pm}}\circ\alpha_{\phi^+}\ =\ \alpha_{\phi^+}\circ\alpha_{\phi_c^{\pm}}\\
\vspace*{0.1in}
\alpha_{\phi_c^{\pm}}\circ\alpha_{\phi^-}\ =\ \alpha_{\phi^-}\circ\alpha_{\phi_c^{\pm}}
\label{phicalpha2}
\end{array}
\end{equation}
where as before $\phi^{\pm}$ label finite gauge transformations.

\section{Polymer quantum kinematics}

In \cite{alokme} we quantized the above theory using techniques of polymer quantization. We quantized the $(\pm)$ sectors separately and within each sector we quantized the embedding and the 
matter phase spaces separately. The elementary operators as well as the basis states in the kinematical Hilbert spaces were labelled by charge-networks which were 
defined as follows.\\
%\subsection{Preliminaries.}
%As in LQG, the polymer quantization is based on suitably defined ``holonomies'' and the polymer Hilbert
%space is spanned by suitably defined ``charge network'' states. In view of the correspondence between 
%finite gauge transformations and periodic diffeomorphisms of $\mathbf{R}$, it is useful to 
%to define periodic and quasiperiodic extensions of charge network labels. Hence we define the following.\\
\noindent
{\bf Definition 1} :  
A charge-network  $s$ is specified by  the  labels
$(\gamma(s),(j_{e_{1}},...,j_{e_{n}}))$ consisting of a graph $\gamma (s)$ (by which we
mean a finite collection of closed, non-overlapping(except in boundary
points) intervals which cover $[0,2\pi]$)  and `charges'  $j_{e}\in {\cal C}$ assigned to each
interval e. (Note that $j_{e}=0$ is allowed.) ${\cal C}$ can be set of reals, or some subset thereof.
Equivalence classes of charge- networks are defined as follows.
The charge- network $s^{\prime}$ is said to be finer than $s$ iff (a) every edge of $\gamma (s)$ is identical to,
or composed of, edges in $\gamma (s^{\prime})$ (b) the charge labels of identical edges in 
$\gamma (s),\gamma (s^{\prime})$
are identical and the charge labels of the edges of $\gamma (s^{\prime})$ which compose to yield an edge of 
$\gamma (s)$ are identical and equal to that of their union in $\gamma (s)$.
Two charge- networks are equivalent  if there exists a charge- network finer than both.
Whenever we refer to charge networks as identical it is understood that we are referring to their
equivalence classes.
We denote, by  $\lambda s$, the charge network $\{\gamma(s), (\lambda j_{e_{1}},...,\lambda j_{e_{n}})\}$
where $\lambda$ is such that $\lambda j_e \in {\cal C}\ \forall\;  e \in \gamma (s) $. 
We denote, by $s+s^{\prime}$, the charge network obtained by dividing 
$\gamma(s)$, $\gamma(s^{\prime})$
into maximal, non-overlapping (upto boundary points) intervals and assigning charge $j_{e}\ +\ j_{e^{\prime}}$ to
$e\cap e^{\prime}$ where $e \in \gamma(s)$, $e^{\prime}\in \gamma(s^{\prime})$.
We also define the `Kronecker delta on charge networks', $\delta_{s,s^{\prime}}$, by $\delta_{s,s^{\prime}}:= 1$ iff $s_1,s_2$ are identical and $:=0$  otherwise.
%\begin{equation}
%\begin{array}{lll}
%\lambda s\ =\ \{\gamma(s), (\lambda j_{e_{1}},...,\lambda j_{e_{n}})\}\\
%\vspace*{0.1in}
%s_{1} + s_{2}\ =\ \{\gamma(s_{1})\cup\gamma(s_{2}),(j_{e_{1}},...,j_{e_{|E(\gamma(s_{1})\cup\gamma(s_{2}))}})\}
%\end{array}
%\end{equation}
%where $\lambda\in\ {\cal C'}$ and where $\gamma(s_{1})\cup\gamma(s_{2})$ is obtained by a decomposing $\gamma(s_{1})$ , %$\gamma(s_{2})$ maximal mutually non-overlapping intervals and assigning charge 
%$j_{e} + j_{e'}$ on the edge $e\in\gamma(s_{1})\cap e'\in\gamma(s_{2})$.\\

As the aim of this paper is to enjoy the fruits of the labour carried out in [1], we provide a very brief 
summary of  the kinematic structures derived therein using the notion of charge networks defined above. 
Section 3.1 summarises the embedding sector, section 3.2 the matter sector, section 3.3 the kinematic Hilbert space
 and section 3.4. the unitary action 
of finite gauge transformations on the kinematic Hilbert space.

%We focus on the right-moving sector. The results for the left-moving sector are exactly analogous.
%\begin{description}
%\item[Embedding sector] : \\ 
%\end{description}
\subsection{Embedding sector}

Charge network : $s^{\pm}\ =\ \{\gamma(s^{\pm}), (k_{e_{1}^{\pm}}^{\pm},...,k_{e_{n}^{\pm}}^{\pm})\}$
%\ k_{e_{I}}^{\pm}\in \frac{2\pi L}{A}{\bf Z}\ \forall I$. 
%A is some fixed integer
\\
\\
Elementary variables : ${\bf (}T_{s^{\pm}}[\Pi_{\pm}],\ X^{\pm}(x){\bf )}$\\
$T_{s^{\pm}}[\Pi_{\pm}]:= \exp[i\sum_{e\in \gamma(s^{+})}k_{e^{\pm}}^{\pm}\int_{e^{\pm}}\Pi_{\pm}]$
\\
\\
Non- Trivial Poisson brackets: 
\begin{eqnarray}
%$\begin{array}{lll}
\lbrace X^{\pm}(x), T_{s^{\pm}}[\Pi_{\pm}]\rbrace & = &
-ik_{e^{\pm}}^{\pm}T_{s^{\pm}}[\Pi_{\pm}]\ if\ x\in \textrm{Interior}(e^{\pm})\nonumber\\
%\vspace*{0.1in}
& = & -\frac{i}{2}(k_{e_{I^{\pm}}^{\pm}}^{\pm}+k_{e_{(I+1)^{\pm}}^{\pm}}^{\pm})T_{s^{\pm}}^{E}[\Pi_{\pm}]\ \textrm{if}\ x\in 
e_{I^{\pm}}^{\pm}\cap e_{(I+1)^{\pm}}^{\pm}\ 1\leq I^{\pm}\leq (n^{\pm}-1)\nonumber\\
%\vspace*{0.1in}
\lbrace X^{\pm}(0),T_{s^{\pm}}[\Pi_{\pm}]\rbrace & = & \lbrace
X^{\pm}(2\pi),T_{s^{\pm}}[\Pi_{\pm}]\rbrace\ =\
-\frac{i}{2}(k_{e_{1}^{\pm}}^{\pm}+k_{e_{n^{\pm}}^{\pm}}^{\pm})T_{s^{\pm}}[\Pi_{\pm}],\nonumber
\end{eqnarray}
\\
\\
Elementary Operators:$\hat{X}^{\pm}(x), \hat{T}_{s^{\pm}}$\\
\\
Charge Network States, Inner Product: 
 $T_{s^{\pm}}$, $\langle T_{s_{1}^{\pm}}, T_{s_{2}^{\pm}} \rangle := \delta_{s_{1}^{\pm}, s_{2}^{\pm}}$\\
\\
%Representation: $\hat{T}_{s_{1}^{+}}\hat{T}_{s_{2}^{+}}\ =\ \hat{T}_{s_{1}^{+}+s_{2}^{+}}$, 
%$[\hat{X}^{+}(x), \hat{T}_{s^{+}}]=-i\hbar\{X^{+}(x), T_{s^{+}}[\Pi_{+}]\}$
%\\
%\\
Representation : $\hat{T}_{s^{\pm}}T_{s_{1}^{\pm}}\ =\ T_{s^{\pm}+s_{1}^{\pm}}$.
\\
\begin{equation}\label{eq:33}
\hat{X}^{\pm}(x)T_{s^{\pm}}\ :=\ \lambda_{x,s^{\pm}}T_{s^{\pm}},
\end{equation}
where, for $\gamma (s^{\pm})$ with $n^{\pm}$ edges,
\begin{equation}\label{eq:33a}
\begin{array}{lll}
\lambda_{x,s^{\pm}}:=\ \hbar k_{e_{I^{\pm}}^{\pm}}^{\pm}\ \textrm{if}\ x\in \textrm{Interior}(e_{I^{\pm}}^{\pm})\ 1\leq I^{\pm}\leq n^{\pm}\\
\vspace*{0.1in}
\hspace*{0.3in}:=\ \frac{\hbar}{2} (k_{e_{I^{\pm}}^{\pm}}^{\pm}\
+\ k_{e_{(I+1)^{\pm}}^{\pm}}^{\pm})\ \textrm{if}\ x\in e_{I^{\pm}}^{\pm}\cap e_{(I^{\pm}+1)}^{\pm}\ 1\leq I^{\pm}\leq (n^{\pm}-1)\\
\end{array}
\end{equation}
\begin{equation}\label{eq:33b}
\begin{array}{lll}
\hspace*{0.3in}:=\ \frac{\hbar}{2} (k_{e_{n^{\pm}}^{\pm}}^{\pm}\ \mp\ \frac{1}{\hbar}2\pi L\ +\ k_{e_{1}^{\pm}}^{\pm})\ \textrm{if}\ x=0\\
\vspace*{0.2in}
\hspace*{0.3in}:=\ \frac{\hbar}{2} (k_{e_{1}^{\pm}}^{\pm}\ \pm\ \frac{1}{\hbar}2\pi L\ +\ k_{e_{n^{\pm}}^{\pm}}^{\pm})\ \textrm{if}\ x=2\pi
\end{array}
\end{equation}
The last two equations, (\ref{eq:33b}), implement the boundary condition $X^{\pm}(2\pi)-X^{\pm}(0)=\pm2\pi L$.\\ 
The embedding Hilbert space ${\cal H}_{E}^{\pm}$ is the Cauchy completion of finite linear combination of charge-network states. \\
\\
{\bf Range of Embedding Charges:} We shall choose 
$\hbar k_{e}^{\pm}\in \frac{2\pi L}{A}{\bf Z}$. 
Here 
A is some fixed positive definite integer. Note that independent of the choice of $A$, the ``holonomy''  variables
$T_{s^{\pm}}[\Pi_{\pm}]$ seperate points in momentum space by virtue of the fact that the graphs underlying the 
charge networks can contain arbitrarily small edges. Note also,
from (\ref{eq:33}), that different values of $A$ obtain different spectra for 
$\hat{X}^{\pm}(x)$. Thus $A$ is the exact analog of the Barbero- Immirizi parameter of LQG \cite{barbero, barb-immir}: while
it does not affect the classical theory, it labels inequivalent representations in quantum theory.
The larger the value of $A$ the smaller the seperation between consecutive eigen values of the embedding coordinates.
Anticipating our analysis of the classical limit of our quantization, we set 
$A$ to be an integer much greater than unity. Finally, we remind the reader again that the only difference between 
our treatment above and the relevant part of \cite{alokme} is that in that work the embedding charges were
in correspondence with the entire real line in contrast to their correspondence with the set of integers here.

%Note that in \cite{am} the ``embedding charges" 
%(eigen-values of the embedding operator $\hat{X}^{+}(x)$) $k_{e}^{+}$ took values in ${\bf R}$. 
%In this paper, we take them to be integral 
%multiples of $\frac{2\pi L}{A}$. Here A is a free dimensionless parameter in the (quantum) theory 
%and we take it to be some large integer. The reason integral charges suffice is because,
%all we require while quantizing a phase-space is that the set of elementary variables separate the 
%points of phase-space. Even with the choice of integer charges, it is easy to show
%that $T_{s^{+}}[\Pi_{+}]$ do form a separating set.\\
%It is straightforward to show that $\hat{T}_{s^{+}}$ are unitary and $\hat{X}^{+}(x)$ are self-adjoint for all x.\\

\subsection{Matter sector}

{\bf Range of matter charges:} Fix a real parameter $\epsilon$ with dimensions $(ML)^{-\frac{1}{2}}$.
Then given any matter charge network, we demand that the {\em difference} between the charge labels of 
any two edges be an integer multiple of $\epsilon$. Thus for any charge network the charge label of any edge is of the
form  $l_{e}+ \lambda$ where 
$l_{e}\in \epsilon{\bf Z}$ and $\lambda \in \epsilon [0,1)$. Note that while  
$\lambda$ is independent of the edge $e$ for any given charge network, the values of $\lambda$ can vary from 
charge network to charge network. Thus we denote the charge network label by $s^{\pm}_{\lambda^{\pm}}$.
With this choice of range for the matter charges, the  holonomies $W_{s^{\pm}_{\lambda^{\pm}}}[Y^{\pm}]$ defined below
seperate points in phase space by virtue of the fact that the edges of the charge networks can be arbitrarily small.
Indeed, as for the embedding charges, we could also have chosen the matter charges to be integer multiples of a 
fixed dimension- ful parameter and still obtained a seperating set of functions on the matter phase space.
Our slightly more involved choice enables the ensuing Hilbert space to carry 
a representation of the transformation generated by
the zero-mode of the scalar field. This in turn helps us to factor out such transformations in section 6, 
thus freezing the zero-mode in quantum theory.
Once again the only difference from the quantization of the matter sector given in \cite{alokme} 
is that here the difference between any two ``matter charges" $l_{e}$ in a given charge network is an
integral multiple of a dimensional-full parameter $\epsilon$ whereas, there, each matter charge could independently 
take values in correspondence with the reals.\\

%Notice that, although with integral charges, the elementary variables would separate the points in matter phase space, we have chosen matter charges to be integral only upto an additive
%charge $\lambda^{+}$ which takes value in $[0,\epsilon)$. We will see that the subsequently constructed Hilbert space carries a representation of the transformation generated by
%the zero-mode of the scalar field. This will in turn help us to factor out such transformations, thus freezing the zero-mode in quantum theory.\\
%Along with A, $\epsilon$ is another free parameter which has gone in our construction of quantum theory. The classial theory is ofcourse independent of these parameters, but they have
%gone into the definition of *-algebras that we quantize. In this sense these two parameters are exactly analogous to the Immirizi parameter in 
%LQG.\\
%We now describe the kinematical Hilbert space obtained by quantizing a 
%certain Weyl algebra of matter fields, as detailed in \cite{am}.\\

\noindent Charge-network :
\begin{equation}
s^{\pm}_{\lambda^{\pm}}\ =\  \{\gamma(s^{\pm}_{\lambda^{\pm}}), 
(l_{e_{1}^{\pm}}^{\pm} + \lambda^{\pm},...,l_{e_{n}^{\pm}}^{\pm}+\lambda^{\pm})\}\ l_{e_{I}}^{\pm}\in \epsilon{\bf Z}\ 
\forall\ I, \;\lambda^{\pm}\in [0,\epsilon).
\label{rangel}
\end{equation}
\\
Elementary variables : $W_{s^{\pm}_{\lambda^{\pm}}}[Y^{\pm}]\ =\exp[i\sum_{e\in \gamma(s^{\pm})}(l_{e}^{\pm}+\lambda^{\pm})
\int_{e}Y^{\pm}]$\\
\\
Charge-network states, Inner product: $W(s^{\pm}_{\lambda^{\pm}})$ ,  $\langle W(s^{\pm}_{\lambda^{\pm}}), W(s'^{\pm}_{\lambda'^{\pm}})\rangle := 
\delta_{s^{\pm}, s'^{\pm}}\delta_{\lambda^{\pm},\lambda'^{\pm}}$\\
\\
Weyl algebra
\footnote{The definition of the Weyl algebra follows in the standard way 
from the Poisson brackets between $Y^{\pm}(x), Y^{\pm}(y)$ 
and an application of the Baker- Campbell- Hausdorff Lemma \cite{bakercampbellhausdorff}}
 of operators:\\
$\hat{W}(s^{\pm}_{\lambda^{\pm}})\hat{W}(s'^{\pm}_{\lambda'^{\pm}})\ =\exp[-i\frac{\hbar}{2}\alpha(s^{\pm}_{\lambda^{\pm}}, s'^{\pm}_{\lambda'^{\pm}})]
\hat{W}(s^{\pm} + s'^{\pm})$.\\
Here the exponent in the phase-factor $\alpha(s^{\pm}_{\lambda^{\pm}},s'^{\pm}_{\lambda'^{\pm}})$ is given by,
\begin{equation}
\alpha(s^{\pm}_{\lambda^{\pm}},s'^{\pm}_{\lambda'^{\pm}})\ :=\ \sum_{e^{\pm}\in \gamma(s^{\pm}_{\lambda^{\pm}})}
\sum_{e^{\prime \pm}\in \gamma(s^{\prime \pm}_{\lambda^{\prime \pm}})}(l_{e}^{\pm}+\lambda^{\pm})(l_{e^{\prime}}^{\pm}+\lambda^{\prime \pm})
\alpha(e^{\pm},e^{\prime \pm}),
\label{defalphass'}
\end{equation}
Here $\alpha(e^{\pm}, e^{\prime \pm})\ =\ (\kappa_{e'^{\pm}}(f(e^{\pm}))-\kappa_{e'^{\pm}}(b(e^{\pm})))-(\kappa_{e^{\pm}}(f(e^{\pm}))-\kappa_{e^{\pm}}(b(e^{\pm})))$.\\ 
Here $f(e)$, $b(e)$ are the final and initial points of the edge $e$ respectively. $\kappa_{e}$ is defined as,
\begin{equation}
\begin{array}{lll}
\kappa_{e}(x)\ =\ 1\ \textrm{if}\ x\ \textrm{is in the interior of e}\\
\hspace*{1.0in} =\ \frac{1}{2}\ \textrm{if}\ x\ \textrm{is a boundary point of e}
\end{array}
\end{equation}
\\
Representation : $\hat{W}(s_{\lambda}^{+\pm})W(s'^{\pm}_{\lambda'^{\pm}})\ =\ \exp(\frac{-i\hbar}{2}\alpha(s_{\lambda}^{\pm}, s'^{\pm}_{\lambda'^{\pm}}))W(s^{\pm}_{\lambda^{\pm}}+s'^{\pm}_{\lambda'^{\pm}})$.\\

The Cauchy completion of finite linear combinations of charge-network states $W(s^{\pm}_{\lambda^{\pm}})$ give ${\cal H}_{M}^{\pm}$.\\

\subsection{The kinematic Hilbert space}
The kinematic Hilbert space ${\cal H}_{kin}$ is the product of the Hilbert spaces
${\cal H}^{\pm}_{kin}$ with 
\begin{equation}\label{hkinpm}
\mathcal{H}^{\pm}_{kin}\ =\ (\mathcal{H}_{E}^{\pm}\ \otimes\ \mathcal{H}_{M}^{\pm})
\end{equation}
so that 
\begin{equation}\label{eq:53*}
\mathcal{H}_{kin}\ =\ 
(\mathcal{H}_{E}^{+}\ \otimes\ \mathcal{H}_{M}^{+})\ \otimes\ (\mathcal{H}_{E}^{-}\ \otimes\ \mathcal{H}_{M}^{-}).
\end{equation}
$\mathcal{H}^{\pm}_{kin}$ is spanned by an orthonormal basis of equivalence classes of charge network states of the 
form $T_{s^{\pm}}\otimes W(s^{\prime\pm}_{\lambda^{\prime\pm}})$
with $s^{\pm}\ =\ \{\gamma(s^{\pm}),(k_{e_{1}^{\pm}}^{\pm},...,k_{e_{n^{\pm}}^{\pm}}^{\pm})\}$,
$s^{\prime\pm}_{\lambda^{\prime\pm}}\ =\ \{\gamma(s^{\prime\pm}),(l_{e_{1}^{\prime\pm}}^{\pm}+\lambda^{\prime\pm},...,l_{e_{m^{\pm}}^{\prime\pm}}^{\pm}+\lambda^{\prime\pm})\}$.\\
The equivalence relation between charge networks was defined above in
Definition 1. Using this equivalence, it is straightforward to see that we can always
choose $s^{\pm}, s^{\prime\pm}_{\lambda^{\prime\pm}}$ such that $\gamma(s^{\pm})=\gamma(s^{\prime\pm}_{\lambda^{\prime\pm}})$. Then each edge $e^{\pm}$ of
$\gamma (s^{\pm})$ is labelled by a pair of charges $(k_e^{\pm},l_e^{\pm} + \lambda^{\pm})$. Note that such a choice
graph and charge pairs is still not unique.
However it is easy to see that a unique choice can be made if we require that 
the  pairs of charges, $(k^{\pm}_{e^{\pm}}, l^{\pm}_{e^{\pm}})$, are such that no two consecutive edges are labelled
by the same pair of charges.
We shall denote this  unique labelling by ${\bf s}^{\pm}_{\lambda^{\pm}}$ so that
\begin{equation}
{\bf s}^{\pm}_{\lambda^{\pm}}:= \{\gamma {(\bf s}^{\pm}_{\lambda^{\pm}}), (k^{\pm}_{e_1^{\pm}},l^{\pm}_{e_1^{\pm}} + \lambda^{\pm}),...,(k^{\pm}_{e_{n^{\pm}}^{\pm}},l^{\pm}_{e_{n^{\pm}}^{\pm}}+\lambda^{\pm})\},
\label{bfs}
\end{equation}
with
\begin{equation}
k_{e_{I^{\pm}}^{\pm}}\neq k_{e_{(I+1)^{\pm}}^{\pm}} \;{\rm or/and }\; l_{e_{I^{\pm}}^{\pm}}\neq l_{e_{(I+1)^{\pm}}^{\pm}}.
\label{unequalpair}
\end{equation}
The corresponding charge network state is denoted by $|{\bf s}^{\pm}_{\lambda^{\pm}}\rangle$ so that
\begin{equation}
|{\bf s}^{\pm}_{\lambda^{\pm}}\rangle =T_{s^{\pm}}\otimes W(s^{\prime\pm}_{\lambda^{\pm}})
\label{s=tw}
\end{equation}
 with ${\bf s}^{\pm}_{\lambda^{\pm}}$ defined from
$s^{\pm}, s^{\prime\pm}_{\lambda^{\prime\pm}}$ in the manner discussed above.

\subsection{Unitary representation of finite gauge transformations.}

The action of finite gauge transformations is most easily specified by introducing the notion of
an extension of a charge network $s$ to the real line. Such an extension is labelled by the graph
$\gamma (s)_{ext}$ which covers the real line and by charge labels on each edge of $\gamma (s)_{ext}$.
Let $T_N(x)\in R$ denote a rigid translation of the point $x\in[0,2\pi]$ by $2N\pi$ so that 
$T_N(\gamma (s) )$ spans $[2N\pi, 2(N+1)\pi]$. Then $\gamma (s)_{ext} = \cup_{N\in\mathbf{Z}}\ T_{N}(\gamma (s))$.
For the  embedding charge network $s^{\pm}$ we define the quasiperiodic extension ${\bar s}^{\pm}_{ext}$
by specifying the embedding charges on $T_N(\gamma (s) )$ by $k^{\pm}_{T_N(e)}:= k_e^{\pm}\pm 2N\pi \frac{L}{\hbar} $
for every edge $e\in \gamma (s)$. Similarly, 
for the  matter charge network $s^{\pm}$ we define the periodic extension $s^{\pm}_{ext}$ by 
setting $l^{\pm}_{T_N(e)}:= l_e$. 

The action of periodic diffeomorphisms, $\phi$, of the real line on ${\bar s}^{\pm}_{ext}$, $s^{\pm}_{ext}$
is defined by mapping $\gamma (s)_{ext}$ to $\phi (\gamma (s)_{ext})$ and setting 
$k^{\pm}_{\phi (e)}= k^{\pm}_e$,$l^{\pm}_{\phi (e)}= l^{\pm}_e$ for every edge $e\in \gamma (s)_{ext}$.

Then  unitary representation of the gauge group is given by,
\begin{equation}\label{eq:56}
\begin{array}{lll}
\hat{U}^{\pm}(\phi^{\pm})T_{s^{\pm}}\ :=\ T_{\phi(\overline{s}^{\pm}_{ext})\vert_{[0,2\pi]}}\\
\vspace*{0.1in}
\hat{U}^{\mp}(\phi^{\mp})T_{s^{\pm}}\ :=\ T_{s^{\pm}}\\
\vspace*{0.1in}
\hat{U}^{\pm}(\phi^{\pm})W(s^{\prime\pm}_{\lambda^{\pm}})\ :=\ W((\phi^{\pm})(s^{\prime\pm}_{\lambda^{\pm}\ ext})\vert_{[0,2\pi]}).\\
\vspace*{0.1in}
\hat{U}^{\mp}(\phi^{\mp})W(s^{\prime\pm}_{\lambda^{\pm}})\ :=\ W(s^{\prime\pm}_{\lambda^{\pm}}),
\end{array}
\end{equation}
where $s^{\prime\pm}_{\lambda^{\pm}\ ext}|_{0.2\pi]}$ is the restriction of $s^{\prime\pm}_{\lambda^{\pm}\ ext}$ to the interval $[0,2\pi]$.\\
Denoting, $T_{s^{\pm}}\otimes W(s^{\prime\pm}_{\lambda^{\pm}})$ by $|{\bf s}^{\pm}_{\lambda^{\pm}}\rangle$ and 
$T_{\phi(\overline{s}^{\pm}_{ext})\vert_{[0,2\pi]}}\otimes W((\phi^{\pm})(s^{\prime\pm}_{\lambda^{\pm}\ ext})\vert_{[0,2\pi]})$ by
$|{\bf s}^{\pm}_{\lambda^{\pm}\ \phi^{\pm}}\rangle$ , the above equations can be written in a compact form as,
\begin{equation}
|{\bf s}^{\pm}_{\lambda^{\pm}\ \phi^{\pm}}\rangle := {\hat U}^{\pm}(\phi^{\pm})|{\bf s}^{\pm}_{\lambda^{\pm}}\rangle .
\label{bfsphi}
\end{equation}
It was shown in \cite{alokme} that the above representation is unitary and its action on elementary operators (via conjugation) precisely mimics the action of finite gauge transformations
on the corresponding (classical) variables.

\section{Unitary representation of Dirac observables}
\subsection{Observables of type $e^{iO_{f^{\pm}}}$}

In \cite{alokme} we showed that the (unitary) action of the operator $\widehat{e^{iO_{{f}^{\pm}}}}$
corresponding to the exponential of the classical observable $O_{f^{\pm}}$ (see equation (\ref{defopmf}))
on the charge network state $T_{s^{\pm}}\otimes W(s^{\prime\pm}_{\lambda^{\pm}})$ is:
\begin{equation}
{\widehat{(\exp {iO_{f^{\pm}}}})}T_{s^{\pm}}\otimes W(s^{\prime\pm}_{\lambda^{\pm}})
:={\hat W}(s^{\pm}_{f^{\pm}})T_{s^{\pm}}\otimes W(s^{\prime\pm}_{\lambda^{\prime\pm}}).
\label{hateof}
\end{equation}
where $s^{\pm}_{f^{\pm}}:= \{\gamma (s^{\pm}), ( f^{\pm}(\hbar k^{\pm}_{e_1^{\pm}}),....,f^{\pm}(\hbar k^{\pm}_{e_{n^{\pm}}^{\pm}}))\}$.
\footnote{A quick way to see this is to expand the exponential in a Taylor series, evaluate the action of 
the embedding part of each term on (its eigen state) $T_{s^{\pm}}$ and resum the series.}

\noindent
{\bf Restriction on $f^{\pm}(X^{\pm})$:} For the real matter charges of Reference \cite{alokme}, the above operator action
is well defined for all real, periodic $f^{\pm}(X^{\pm})$. In that context, we showed \cite{alokme} that no state exists
(whether kinematic or physical) which displays semiclassical behaviour with respect to 
${\widehat{(\exp {iO_{f^{\pm}}}})}$ for all real $f^{\pm}$ and that this lead to the necessity of an ad- hoc choice of
a countable subset of such observables for semiclassical analysis. Here, this problem disappears by virtue of the 
tighter kinematic structure. Specifically, since matter charges are translated by the amount 
$f^{\pm}(\hbar k^{\pm}_{e^{\pm}})$ in equation (\ref{hateof}), the operator action is only well defined for those
$f^{\pm}(X^{\pm})$ such that $f^{\pm}(\hbar k^{\pm}_{e^{\pm}})$ lie in the range sepcified by (\ref{rangel}). Together with the 
integer valuedness of the embedding charges, this implies that 
$f^{\pm}(X^{\pm})$ is  a (smooth periodic real) function of $X^{\pm}$ such that 
\begin{equation}
f^{\pm}(\frac{2\pi L}{A}n)\ \in  {\bf Z} \epsilon\  + \lambda \;\;\;\forall\ n\in {\bf Z}, \;\;\; \lambda \in {\bf R}.
\label{restrictf}
\end{equation}
As we shall see, this vast reduction in the space of observables allows a semiclassical analysis free from any  
ad- hoc choice of the type mentioned above.\\

Note that (\ref{hateof}) is a manifestly regularization/triangulation independent definition.
Moreover, since $s^{\pm}_{f^{\pm}}$ is constructed from the embedding part of the charge network, and
since $f^{\pm}$ are periodic, it is straightforward to check that $\widehat{e^{i O_{f^{\pm}}}}$ commute
with the unitary operators corresponding to finite gauge transformations so that $\widehat{e^{i O_{f^{\pm}}}}$
are Dirac observables in quantum theory.

\subsection{Conformal isometries}
In \cite{alokme} we showed how to represent Hamiltonian flows $\alpha_{\phi_{c}^{\pm}}$ as gauge invariant unitary operators on ${\cal H}_{kin}^{\pm}$. 
The resulting operator $\hat{V}[\phi_{c}^{\pm}]$ had a trivial action on ${\cal H}_{M}^{\pm}$ and its action on $T_{s^{\pm}}\in {\cal H}_{E}^{\pm}$ was given by,
\begin{equation}
\hat{V}^{\pm}[\phi_{c}^{\pm}]T_{s^{\pm}}\ =\ T_{(\phi_{c}^{\pm})^{-1}(s^{\pm})}.
\label{vts}
\end{equation}
Where, $(\phi_{c}^{\pm})^{-1}(s^{\pm})= \{\gamma (s^{\pm}), ((\phi_c^{\pm})^{-1}(k_{e_1^{\pm}}^{\pm}),...,(\phi_c^{\pm})^{-1}(k_{e_n^{\pm}}))\}$.\\
However as $k_{e}^{\pm}$ are now integral multiples of $\frac{2\pi L}{A\hbar}$, generically a monotonically increasing, invertible function $\phi_{c}^{\pm}$ 
will not map   $k_{e}^{\pm}$ to  $k_{e}^{\prime\pm}\in \frac{2\pi L}{A\hbar}\mathbf{Z}$.
In fact it is easy to see that only invertible(1-1) monotonic functions which map integers to integers are constant (integer-valued) translations.\\ 
Whence in the present case only the Hamilton flows corresponding to discrete translations 
(which correspond to a discrete subgroup of the Poincare group) can be unitarily represented on ${\cal H}_{kin}$.\\
\begin{equation}
\hat{V}[\tau^{\pm}]T_{s^{\pm}}\ =\ T_{s_{\tau^{\pm}}^{\pm}}
\end{equation}
Where $\tau^{\pm}\in \frac{2\pi L}{A\hbar}\mathbf{Z}$ and $s_{\tau^{\pm}}^{\pm}= \{ \gamma(s^{\pm}), (k_{e_{1}^{\pm}}^{\pm}+\tau^{\pm},...,k^{\pm}_{e_{n^{\pm}}}+\tau^{\pm})\}$. 
It is straightforward to see that $\hat{V}[\tau^{\pm}]$ are unitary $\forall \tau^{\pm}$ and satisfy,
\begin{equation}
\hat{V}^{\pm}[\tau_{1}^{\pm}]\hat{V}^{\pm}[\tau_{2}^{\pm}]=\hat{V}^{\pm}[\tau_{1}^{\pm}+\tau_{2}^{\pm}],
\label{phicrep}
\end{equation}
so that our definition of  $\hat{V}^{\pm}[\tau^{\pm}]$ implies 
an anomaly free representation of a discrete subgroup of the Poincare group.

Note that in the classical theory the conformal group is homomorphic not only to the group of 
 of conformal isometries but also to that of finite gauge transformations \cite{karelcm}. 
In the quantum theory presented here, we see that while 
the entire group of gauge transformations is faithfully represented on ${\cal H}_{kin}$, 
the entire group of conformal isometries cannot be represented on ${\cal H}_{kin}$; only a discrete Abelian subgroup 
thereof
corresponding to rigid translations is unitarily represented on the Hilbert space. 
We will see later how this discrete subgroup is tied to the emergence of a lattice in the quantum theory

\section{Physical state space by group averaging}
Only gauge invariant states are physical so that physical states $\Psi$  must satisfy the condition 
${\hat U}^{\pm}(\phi^{\pm})\Psi = \Psi, \; \forall \phi^{\pm}$. A formal solution to this condition is to fix some
$|\psi\rangle \in {\cal H}_{kin}$ and set
$\Psi = \sum |\psi^{\prime}\rangle$ where the sum is over all distinct $|\psi^{\prime}\rangle$  which are
gauge related to $\psi$.  A mathematically precise implementation of this idea places the gauge invariant
states in the dual representation (corresponding to a formal sum over bras rather than kets) and goes by the name of
Group Averaging. The reader may consult Reference \cite{alokme} for a quick account of the Group Averaging technique 
as well  its detailed implementation in PFT. The considerations of Reference \cite{alokme} go through unchanged here
with the proviso that both ${\cal H}_{kin}$ and the set of Dirac observables are smaller than those used there;
${\cal H}_{kin}$ is smaller due to the restriction in the range of the charges (see sections 3.1 and 3.2)
and  the set of Dirac observables is smaller since they need to be well defined on ${\cal H}_{kin}$ 
(see sections 4.1 and 4.2).

The Group Average of a  charge network state $\vert {\bf s}^{\pm}_{\lambda^{\pm}}\rangle$ yields 
the physical, gauge invariant distribution
$\eta^{\pm}(\vert {\bf s}^{\pm}_{\lambda^{\pm}}\rangle)$ with
\begin{equation}\label{eq:120}
\begin{array}{lll}
\eta^{\pm} (|{\bf s}^{\pm}_{\lambda^{\pm}}\rangle) & = & \eta^{\pm}_{[{\bf s}^{\pm}_{\lambda^{\pm}}]}\sum_{{\bf s}^{\prime \pm}_{\lambda^{\pm}}\in [{\bf s}^{\pm}_{\lambda^{\pm}}]}
    <\ {\bf s}^{\prime\pm}_{\lambda^{\pm}}|\\
\vspace*{0.1in}
& = & \eta^{\pm}_{[{\bf s}^{\pm}_{\lambda^{\pm}}]}\sum_{\phi^{\pm}\in Diff_{[{\bf s}^{\pm}_{\lambda^{\pm}}]}^{P}\mathbf{R}}<{\bf s}_{\lambda^{\pm}\ \phi^{\pm}}^{\pm}| .
\end{array}
\end{equation}
Here $\eta^{\pm}$ denotes the (antilinear) Group Averaging map from the finite span of charge network states, 
${\cal D}^{\pm}$  into the space of distributions 
\footnote{Distributions are linear maps from ${\cal D}^{\pm}$ to the complex numbers; only finitely many terms in the formal
sum (\ref{eq:120}) contribute to its action on any spin network state thus ensuring that the left hand side is indeed
a distribution.}
${\cal D}^{\pm *}$, $[{\bf s}^{\pm}_{\lambda^{\pm}}]$ is the set  of charge networks defined by
$[{\bf s}^{\pm}_{\lambda^{\pm}}]\: =\ \{ {\bf s}^{\prime\pm}_{\lambda^{\pm}}\vert {\bf s}_{\lambda^{\pm}}^{\prime\pm}\ 
=\  {\bf s}_{\lambda^{\pm} \phi^{\pm}}^{\pm} \; {\rm for\ some\ }\phi^{\pm}\}$, 
$Diff_{[\bf{s}^{\pm}_{\lambda^{\pm}}]}^{P}\mathbf{R}$ is a set of gauge transformations such that
for each  ${\bf s}^{\prime\pm}_{\lambda^{\pm}}\in\ [{\bf s}^{\pm}_{\lambda^{\pm}}]$ there is precisely one gauge transformation  in the set which maps 
${\bf s}^{\pm}_{\lambda^{\pm}}$ to ${\bf s}^{\prime\pm}_{\lambda^{\pm}}$ and $\eta_{[{\bf s}^{\pm}_{\lambda^{\pm}}]}$ is a positive real number depending only on the
gauge orbit $[{\bf s}^{\pm}_{\lambda^{\pm}}]$. The space of such  gauge invariant distributions comes equipped with 
the inner product
\begin{equation}
<\eta^{\pm}(|{\bf s}_{\lambda^{\pm}}^{(1)\pm}\rangle), \eta^{\pm}(|{\bf s}_{\lambda^{\prime\pm}}^{(2)\pm}\rangle)>_{phys}=
\eta^{\pm}(|{\bf s}_{\lambda^{\pm}}^{(1)\pm}\rangle) [|{\bf s}_{\lambda^{\prime\pm}}^{(2)\pm}\rangle ]. 
\label{physip}
\end{equation}
As in Reference \cite{alokme} we shall focus on a superselected sector of $\eta^{\pm} ({\cal D}^{\pm})$
which is obtained by Group Averaging the  superselected sector, ${\cal D}^{\pm}_{ss}$ of ${\cal D}^{\pm}$.
This superselected sector is of physical interest because, as is obvious below, it captures (as well as possible)
 the classical 
non-degeneracy property $\pm X^{\pm \prime}(x) >0$ in the context of quantum theory wherein
${\hat X}^{\pm}(x)$ has the discrete spectrum given by (\ref{eq:33}).

${\cal D}^{\pm}_{ss}$ is defined as follows.
Fix a pair of graphs $\gamma^{\pm}$ with A edges. We will denote the edges by $e_{I}^{\pm}$ with $0\leq\ I\leq (A-1)$.
Place the embedding charges 
$\vec{k}^{\pm}$ such that $k_{e_{I}^{\pm}}^{\pm}-k_{e_{I-1}^{\pm}}^{\pm}=\frac{2\pi L}{A\hbar}\ \forall\ 1\leq I\leq A-1$.
Consider the set of all charge-network states\\ 
$\{|{\bf s}^{\pm}_{\lambda^{\pm}}\rangle\ =\ 
|\gamma^{\pm},\ \vec{k}^{\pm}, (l_{e_{0}^{\pm}}^{\pm}+\lambda^{\pm},...,l_{e_{A-1}^{\pm}}^{\pm}+\lambda^{\pm})\rangle\}$, where
$l_{e_{I}^{\pm}}^{\pm}\in {\bf Z}\epsilon$ and $\lambda^{\pm}\in [0,\epsilon)$ are allowed to take all possible values.
Let ${\cal D}_{ss}^{\pm}$ be finite span of charge network states of the type 
$\{|{\bf s}^{\pm}_{\lambda^{\pm}\ \phi^{\pm}}\rangle\ \forall\ \phi^{\pm}\}$.

It is straightforward to check that  ${\cal D}_{ss}^{\pm}$ 
is invariant under the action of observables as well as gauge transformations. 
Whence ${\cal D}_{ss}^{+}\otimes{\cal D}_{ss}^{-}$ 
constitutes a super-selected (kinematical) sector of the theory.

Next we note that the enormous ``$\eta^{\pm}_{[{\bf s}^{\pm}_{\lambda^{\pm}}]}$'' worth of ambiguity in the Group Averaging map
can be reduced, as in Reference \cite{alokme} by 
requiring that the Group Averaging map commutes with the Dirac observables of the quantum theory.

We now provide a brief sketch  of how this happens in the `$+$' sector; the `$-$' sector can be dealt 
with in a similar manner. 
%We shall occassionally suppress the `$+$' superscript in our discussion. 
From \cite{alokme}, commutatitivity with the Dirac observables of section 4.1 implies that 
$\eta^{+}_{[{\bf s}_{\lambda}^{+}]} = \eta^{+}_{[\widetilde{{\bf s}^{+}}]}$,  where
$\widetilde{{\bf s}^{+}}$ is the embedding charge network given by 
\begin{displaymath}
\widetilde{{\bf s}^{+}}\ =\ \{\gamma({\bf s}^{+}), (k_{e_{0}^{+}}^{+},...,k_{e_{A-1}^{+}}^{+})\}
\end{displaymath}
and the equivalence class $[\widetilde{{\bf s}^{+}}]$ is the the set of all $\widetilde{{\bf s}^{+\prime}}$ 
such that  
$\widetilde{{\bf s}^{+\prime}} = {\hat U}^+(\phi^+ )\widetilde{{\bf s}^{+}}$ for some $\phi^+$.
This is due to the fact (see \cite{alokme} for a proof) that given any two charge networks 
${\bf s}_{\lambda^{+}}^{(1)+}$, ${\bf s}_{\lambda^{\prime +}}^{(2)+}$ such that 
$\widetilde{{\bf s}^{(1)+}}=\widetilde{{\bf s}^{(2)+}}$ ,
 one can always find a function f such that  $(l_{e_{I}^{+}}^{(1) +} + \lambda^{(1) +}) + f(\hbar k_{e_{I}^{+}}^{(1) +})\ =\  (l_{e_{I}^{+}}^{(2) +} + \lambda^{(2) +})$ 
$\forall I\in \{0,...,A-1\}$.
 Here $\vec{l}^{(1) +}+\lambda^{(1) +},\ \vec{l}^{(2) +}+\lambda^{(2) +}$ are the matter charges associated 
to ${\bf s}_{\lambda^{+}}^{(1)+},\ {\bf s}_{\lambda^{\prime +}}^{(2)+}$ 
respectively.
%\footnote{There is a subtlety here which was detailed in \cite{alokme}. We are assuming here that given a $[{\bf s}_{\lambda}^{+}]$ and corresponding 
%$[\widetilde{{\bf s}^{+}}]$, there exists one $\widetilde{{\bf s}^{+}}$ in $[\widetilde{{\bf s}^{+}}]$ such that $k_{e_{A-1}}^{+}-k_{e_{0}}^{+}\ <\ 2\pi L$.} 
Again, from  \cite{alokme},  the commutativity of $\eta$ with the observables of section 4.2  implies that 
\begin{equation}
\eta^+_{[\widetilde{{\bf s}^+}]} = \eta^{+}_{[\gamma^{+}]}
\end{equation}
where $[\gamma^{+}]$ is the equivalence class of all the graphs which are related to $\gamma^{+}$ by a periodic diffeomorphism. 
That is, we say that $\gamma^{+}_{1}\sim\gamma^{+}$
if $\exists\ \phi^{+}\in Diff^{P}{\bf R}$ such that $(\phi^{+}\cdot\gamma^{+}_{ext})|_{[0,2\pi]}\ =\ \gamma^{+}_{1}$, where $(\phi^{+}\cdot\gamma^{+}_{ext})|_{[0,2\pi]}$
 is the restriction of $(\phi^{+}\cdot\gamma^{+}_{ext})$ to $[0,2\pi]$.\\
Once again this is due to the fact that given $\widetilde{{\bf s}^{(1)+}}$, $\widetilde{{\bf s}^{(2)+}}$ such that $\gamma(\widetilde{{\bf s}^{(1)+}})\ =\ 
\gamma(\widetilde{{\bf s}^{(2)+}})$,
one can always find $\tau^{+}\in \frac{2\pi L}{A\hbar}{\bf Z}$, such that $k_{e_{I}^{+}}^{(1) +} + \tau^{+} = k_{e_{I}^{+}}^{(2) +}\ \forall I$.\\ 
(Recall that we are in the super-selected sector, and whence $(\triangle k)_{e_{I}^{+}}^{(1) +}=(\triangle k)_{e_{I}^{+}}^{(2) +}$ = $\frac{2\pi L}{A\hbar}$).\\

Thus we obtain that $\eta^{+}({\cal D}^{+}_{ss})$ is the finite span of states of the form
\begin{equation}\label{eq:jan8-1}
\eta^{+}(\vert {\bf s}^{+}_{\lambda^{+}}\rangle)\ =\ \eta^{+}_{[\gamma^{+}]}\sum_{\phi\in Diff_{[{\bf s}^{+}_{\lambda^{+}}]}^{P}{\bf R}}\langle \phi\cdot {\bf s}^{+}_{\lambda^{+}}\vert
\end{equation}
The physical Hilbert space for right moving sector ${\cal H}_{phy}^{ss+}$ is the Cauchy completion of 
of $\eta^{+}({\cal D}^{+}_{ss})$
An anologous construction yields ${\cal H}_{phy}^{ss-}$. 
\begin{displaymath}
{\cal H}^{ss}_{phy} = {\cal H}_{phy}^{ss+}\otimes {\cal H}_{phy}^{ss-}
\end{displaymath}

\section{Taking care of the Zero-mode constraint}
The zero mode constraint is $p\ =\ 0$. As $p = \int_{S^{1}} Y^{+}\ =\ \int_{S^{1}}Y^{-}$, we can impose $\int_{S^{1}} Y^{\pm} \approx 0$ by Group averaging with respect to the 
one-parameter family of unitaries $\hat{W}(s_{0,\mu^{\pm}})$ where 
$s_{0,\mu^{\pm}}\ =\ \{\gamma = [0,2\pi], l_{S^{1}}^{\pm} = \mu^{\pm}\in {\bf R}\}$ (i.e. $\gamma$ consists of a single edge which covers the circle and 
which is labelled by the real
charge $\mu^{\pm}$).
As explained in \cite{alokme}, one could solve the zero-mode constraint before or after solving the $(H^{+},\ H^{-})$ constraints, as
\begin{equation}\label{eq:sunday-1}
\hat{U}^{\pm}(\phi^{\pm}) \hat{W}(s_{0,\mu^{\pm}})\hat{U}^{\pm}(\phi^{\pm})^{-1}\ =\ \hat{W}(s_{0,\mu^{\pm}})
\end{equation}
$\forall \mu^{\pm}\in {\bf R}$.\\
As we have already averaged over the gauge group, we seek to solve the zero-mode constraint by defining a Group averaging map
\begin{equation}
\overline{\eta}^{\pm} : \eta^{\pm}(\mathcal{D}_{ss}^{\pm})\rightarrow \eta^{\pm}(\mathcal{D}_{ss}^{\pm})^{*}. 
\end{equation}
where ${\cal D}_{ss}^{\pm}$ 
be finite span of charge network states of the type $\{|{\bf s}^{\pm}_{\lambda^{\pm}\ \phi^{\pm}}\rangle\ \forall\ \phi^{\pm}\}$ as defined
in the previous section.\\
Before defining $\overline{\eta}^{\pm}$, note that,
\begin{equation}\label{eq:155}
\hat{W}(s_{0,\mu^{\pm}})\vert {\bf s}^{\pm}_{\lambda^{\pm}}\rangle\ =:\ \vert {\bf s}^{\pm}_{\mu^{\pm} + \lambda^{\pm}}\rangle
\end{equation}
where ${\bf s}^{\pm}_{\mu^{\pm} + \lambda^{\pm}}$ is obtained from ${\bf s}^{\pm}_{\lambda^{\pm}}=\{ \gamma({\bf s})^{\pm},\vec{k}^{\pm},\vec{l}^{\pm}+\lambda^{\pm}\}$ 
by adding $\mu^{\pm}$ to all the matter charges.\\
We now define,
\begin{equation}\label{eq:156}
\begin{array}{lll}
\overline{\eta}^{\pm}(\eta^{\pm}(\vert {\bf s}^{\pm}_{\lambda^{\pm}}\rangle)) = 
\overline{\eta}^{\pm}_{[[{\bf s}^{\pm}]]_{0}}\eta^{\pm}_{[\gamma({\bf s}^{\pm}_{\lambda^{\pm}})]}
\bigoplus _{\mu^{\pm}\in \mathbf{R}}\sum_{\phi^{\pm}\in Diff_{[\gamma({\bf s}^{\pm}_{\lambda^{\pm}})]}^{P}\mathbf{R}} 
\langle ({\bf s}^{\pm}_{\phi^{\pm}})_{\mu^{\pm}+\lambda^{\pm}} |
\end{array}
\end{equation}
The equivalence class $[[{\bf s}^{\pm}]]_{0}$ is defined via following relation.\\
 $[{\bf s}^{\pm}_{\lambda^{\pm}}]\sim [{\bf s}^{(1)\pm}_{\lambda_{1}^{\pm}}]$ 
iff for any $\{\gamma({\bf s}^{\pm}),\vec{k}^{\pm},\vec{l}^{\pm}+\lambda^{\pm}\} 
\in [{\bf s}^{\pm}_{\lambda^{\pm}}]$, $\exists$ $(\{\gamma({\bf s}^{\pm}),\vec{k}^{\pm},\vec{l}^{\pm}+\lambda^{\pm}+\mu^{\pm}\} \in [{\bf s}^{(1)\pm}_{\lambda_{1}^{\pm}}]$
 for some $\mu_{\pm}\in \mathbf{R}$.\\
However in light of (\ref{eq:sunday-1})  the sums in (\ref{eq:156}) can be interchanged as,
\begin{equation}\label{eq:156*}
\begin{array}{lll}
\overline{\eta}^{\pm}(\eta^{\pm}(\vert {\bf s}^{\pm}_{\lambda^{\pm}}\rangle)) = 
\overline{\eta}_{[[{\bf s}^{\pm}]]_{0}}\eta_{[\gamma({\bf s}^{\pm}_{\lambda^{\pm}})]}
\sum_{\phi^{\pm}\in Diff_{[{\bf s}^{\pm}_{\lambda^{\pm}}]}^{P}\mathbf{R}} \bigoplus _{\mu^{\pm}\in \mathbf{R}}
\langle ({\bf s}^{\pm}_{\mu^{\pm}+\lambda^{\pm}})_{\phi^{\pm}} |
\end{array}
\end{equation}
We can define an invariant characterization of the distribution $\bigoplus _{\mu^{\pm}\in \mathbf{R}}
\langle ({\bf s}^{\pm}_{\mu^{\pm}+\lambda^{\pm}}) |$ as follows.\\
Given $\vec{l}^{\pm}+\lambda^{\pm}$ , $\vec{l}_{(1)}^{\pm}+\lambda_{(1)}^{\pm}$ they will both give rise to the same distribution
 $\bigoplus _{\mu^{\pm}\in \mathbf{R}}\langle ({\bf s}^{\pm}_{\mu^{\pm}+\lambda^{\pm}}) |$ 
iff $l_{e_{I}^{\pm}}^{\pm}+\lambda^{\pm}+\mu^{\pm}\ =\ l_{(1),e_{I}^{\pm}}^{\pm} + \lambda_{(1)}^{\pm}$ $\forall\ 0\leq I\leq\ A-1$. 
Which is equivalent to $l_{e_{I+1}^{\pm}}^{\pm}-l_{e_{I}^{\pm}}^{\pm}\ =\ l_{(1),e_{I+1}^{\pm}}^{\pm}-l_{(1),e_{I}^{\pm}}^{\pm}$ $\forall\ 0\leq I\leq\ A-2$.\\
Whence,
\begin{equation}\label{eq:156**}
\begin{array}{lll}
\overline{\eta}^{\pm}(\eta^{\pm}(\vert {\bf s}^{\pm}_{\lambda^{\pm}}\rangle)) = 
\overline{\eta}^{\pm}_{[[{\bf s}^{\pm}]]_{0}}\eta^{\pm}_{[\gamma({\bf s}^{\pm}_{\lambda^{\pm}})]}
\sum_{\phi^{\pm}\in Diff_{[{\bf s}^{\pm}_{\lambda^{\pm}}]}^{P}\mathbf{R}}  ( {\bf (}\gamma(s^{\pm}),\vec{k}^{\pm},\vec{\triangle l}^{\pm})_{\phi^{\pm}} |
\end{array}
\end{equation} 
where $\vec{\triangle l}^{\pm} := \{l_{e_{0}^{\pm}}^{\pm}-l_{e_{A-1}^{\pm}}^{\pm}, l_{e_{1}^{\pm}}^{\pm}-l_{e_{0}^{\pm}}^{\pm},...,
l_{e_{A-1}^{\pm}}^{\pm}-l_{e_{A-2}^{\pm}}^{\pm}\}$, $ {\bf (}\gamma(s^{\pm}),\vec{k}^{\pm},\vec{\triangle l}^{\pm}|$ 
is the distribution $\bigoplus _{\mu^{\pm}\in \mathbf{R}}\langle ({\bf s}^{\pm}_{\mu^{\pm}+\lambda^{\pm}}) |$ on ${\cal D}_{ss}^{\pm}$.\\
Henceforth we will denote the corresponding charge network $\{ \gamma(s^{\pm}), \vec{k}^{\pm}, \vec{\triangle l}^{\pm}\}$ by ${\bf s}_{0}^{\pm}$.\\
The unitary action of gauge group gives rise to a dual action on such distributions. 
\begin{equation}
\hat{U}^{\pm}(\phi^{\pm})|{\bf s}_{0}^{\pm}{\bf )}\ =:\ |({\bf s}_{0})^{\pm}_{\phi^{\pm}}{\bf )}
\end{equation}
In light of (\ref{eq:sunday-1}), this action can be understood by picking any charge-network ${\bf s}^{\pm}$ 
in the equivalence class ${\bf s}_{0}^{\pm}$, applying $\hat{U}^{\pm}(\phi^{\pm})$ to $|{\bf s}^{\pm}\rangle$ and denoting the corresponding equivalence class by 
$({\bf s}_{0}^{\pm})_{\phi^{\pm}}$.\\
It is also easy to see that, $Diff_{[{\bf s}^{\pm}_{\lambda^{\pm}}]}^{P}\mathbf{R}\ =\  Diff_{[{\bf s}^{\pm}_{\lambda^{\pm}=0}]}^{P}\mathbf{R}\ 
\forall\ \lambda^{\pm}$.\\ 
Whence finally the solution to the zero-mode constraints are given by,
\begin{equation}\label{eq:156***}
\overline{\eta}^{\pm}(\eta^{\pm}(\vert {\bf s}^{\pm}_{\lambda^{\pm}}\rangle)) = 
\overline{\eta}^{\pm}_{[[{\bf s}^{\pm}]]_{0}}\eta^{\pm}_{[\gamma({\bf s}^{\pm}_{\lambda^{\pm}})]}
\sum_{\phi^{\pm}\in Diff_{[{\bf s}^{\pm}]}^{P}\mathbf{R}}  \rangle ({\bf s}_{0}^{\pm})_{\phi^{\pm}}|.
\end{equation} 
Since the equivalence classes of charge network under the equivalence defined above are characterized by ${\bf s}_{0}^{\pm}$, we denote 
$\overline{\eta}_{[[{\bf s}^{\pm}]]_{0}}\ \textrm{by}\ \overline{\eta}_{{\bf s}_{0}^{\pm}}$. Once again, the requirement that 
the Group averaging map $\overline{\eta}$ commutes with the action of Dirac observables and (discrete) isometries
reduces the above ambiguity to 
$\overline{\eta}_{\gamma({\bf s}_{0}^{\pm})}$.\\
We will denote the final physical Hilbert space (after solving the zero-mode constraint) by ${\cal H}_{(0) phy}^{ss}\ :=\ {\cal H}_{(0) phy}^{ss +}\otimes{\cal H}_{(0) phy}^{ss -}$. 

\section{Existence of Infinitely many Poincare-invariant states}
In this section we show that ${\cal H}^{ss}_{(0) phy}$ admits infinitely many Poincare invariant states. As the unitaries corresponding to the Poincare group commute with the 
zero mode constraint\footnote{$\hat{W}(s_{0,\mu^{\pm}})\hat{V}[\tau^{\pm}]\hat{W}(s_{0,\mu^{\pm}})^{-1}\ =\ \hat{V}[\tau^{\pm}]$.}, we 
ignore the averaging w.r.t  the zero mode constraint defined in (\ref{eq:jan8-1}) in this section. 
(i.e. we will show the existence of Poincare invariant states in ${\cal H}^{ss}_{phy}$.)  
%Towards the end of the section, we explain how the result can be lifted to 
%${\cal H}_{phy}^{+}\otimes{\cal H}_{phy}^{-}$ (which is the solution space to all the constraints).\\
Once again, we restrict the  analysis to the left-moving sector.\\ 
Consider a charge-network state $|{\bf s}^{+}\rangle$ in ${\cal D}^{+}_{ss}$ with\\ 
${\bf s^{+}}\ =\ \{\gamma^{+}, \left(k_{e_{0}^{+}}^{+}=0,k_{e_{1}^{+}}^{+}=\frac{2\pi L}{A\hbar},...k_{e_{I}^{+}}^{+}=\frac{2\pi LI}{A\hbar},...,
k_{e_{A-1}^{+}}^{+}=\frac{2\pi L(A-1)}{A\hbar}\right),
\vec{l}^{+} = \left(l_{e_{0}^{+}}^{+},...,l_{e_{A-1}^{+}}^{+}\right)\}$\footnote{We set $\lambda^{+}$ to zero.}\\
Also let ${\bf s}^{+}_{cyc}=\{\gamma^{+}, \left(k_{e_{0}^{+}}^{+}=0,k_{e_{1}^{+}}^{+}=\frac{2\pi L}{A\hbar},...k_{e_{I}^{+}}^{+}=\frac{2\pi LI}{A\hbar},...,
k_{e_{A-1}^{+}}^{+}=\frac{2\pi L(A-1)}{A\hbar}\right),
\vec{l}^{+} = \left(l_{e_{A-1}^{+}}^{+},l_{e_{0}^{+}}^{+},...,l_{e_{A-2}^{+}}^{+}\right)\}$.\\
Let us also denote the (dual) representation of Poincare group on ${\cal H}^{ss}_{phy}$ by $\hat{V}[\tau^{+}]'$.\\

\noindent{\bf Lemma 1}: Show that $\hat{V}[\tau^{+}=\frac{2\pi L}{A\hbar}]'\eta^{+}(|{\bf s}^{+}\rangle)\ =\ \eta^{+}(|{\bf s}^{+}_{cyc}\rangle)$\\

\noindent{\bf Proof}:
Note that,
\begin{equation}
\hat{V}[\tau^{+}=\frac{2\pi L}{A\hbar}]'\eta^{+}(|{\bf s}^{+}\rangle) = \eta^{+}(|{\bf s}^{+}_{\tau^{+}}\rangle)
\end{equation}
where ${\bf s}^{+}_{\tau^{+}} = \{\gamma^{+}, \left(k_{e_{0}^{+}}^{+}+\tau^{+},k_{e_{1}^{+}}^{+} + \tau^{+},...,k_{e_{A-1}^{+}}^{+}+\tau^{+}\right), 
\vec{l}^{+} = \left(l_{e_{0}^{+}}^{+},...,l_{e_{A-1}^{+}}^{+}\right)\}$.\\
Notice that $k_{e_{I}^{+}}^{+} + \tau^{+} = k_{e_{I+1}^{+}}^{+}\ \forall\ 0\ \leq\ I\ \leq\ (A-2)$ and $k_{e_{A-1}^{+}}^{+} + \tau^{+} = 2\pi L$.\\
Whence,\\
\begin{equation}
{\bf s}^{+}_{\tau^{+}}\ =\ \{\gamma^{+}, \left(k_{e_{1}^{+}}^{+},k_{e_{2}^{+}}^{+},...,k_{e_{A-1}^{+}}^{+},2\pi L\right), \vec{l}^{+}\}
\end{equation}
Whence we want to show that
\begin{equation}\label{eq:jan15-7} 
\eta^{+}(|{\bf s}^{+}_{\tau^{+}}\rangle)\ =\ \eta^{+}(|{\bf s}^{+}_{cyc}\rangle)
\end{equation}
The definition of the Group averaging map as given in the first equation in (\ref{eq:120}) makes it clear that to prove (\ref{eq:jan15-7}), it suffices to 
show that $\exists\ \phi^{+}$ such that
\begin{equation}
\phi^{+}\cdot{\bf s}_{\tau^{+}}^{+}\ =\ {\bf s}^{+}_{cyc}
\end{equation}
Note that the quasi-periodic extension ${\bf s}_{\tau^{+} ext}^{+}$ of ${\bf s}_{\tau^{+}}^{+}$ as defined in section 3.4, is given by
\begin{equation}
{\bf s}_{\tau^{+} ext}^{+}\ =\ \{\gamma^{+}_{ext}, \left(...,k_{e_{0}^{+}}^{+},(k_{e_{1}^{+}}^{+},...,\frac{2\pi L}{\hbar}),k_{e_{1}^{+}}^{+}+\frac{2\pi L}{\hbar},...\right), 
\left(...,l_{e_{A-1}^{+}}^{+},(l_{e_{0}^{+}}^{+},...,l_{e_{A-1}^{+}}^{+}),l_{e_{0}^{+}}^{+},...\right)\}
\end{equation}
e.g. the embedding charges on $T_{-1}(\gamma^{+})$ are $(k_{e_{1}^{+}}-\frac{2\pi L}{\hbar},...,0=k_{e_{0}^{+}}^{+})$ and the embedding charges on 
$T_{1}(\gamma^{+})$ are $(k_{e_{1}^{+}}^{+}+\frac{2\pi L}{\hbar},...,\frac{4\pi L}{\hbar})$.\\
As $\phi^{+}\cdot{\bf s}_{\tau^{+}}^{+}\ =\ \phi^{+}({\bf s}_{\tau^{+} ext}^{+})\vert_{[0,2\pi]}$. It is easy to envisage a $\phi^{+}$ which is such that
\begin{equation}
\phi^{+}({\bf s}_{\tau^{+} ext}^{+})\vert_{[0,2\pi]}\ =\ {\bf s}^{+}_{cyc}
\end{equation}
This finishes the proof.\\

However the above result implies that any state in ${\cal D}_{ss}$ which is invariant under the cyclic permutations of the matter charges, yields, on Group averaging,  
a Poincare invariant Physical state.\\
Consider any state in (the closure of) ${\cal D}_{ss}$ of the form,
\begin{equation}
|\Phi^{+}\rangle\ =\ \sum_{\vec{l}^{+}}c(\vec{l}^{+})|\gamma^{+},\vec{k}^{+},\vec{l}^{+}\rangle
\end{equation}
where each charge-network state in the sum has the same underlying graph $\gamma^{+}$ and same embedding charges $\vec{k}^{+}$.  $|\Phi^{+}\rangle$ is clearly invariant under
cyclic permutations of matter charges, if the function $c(\vec{l}^{+})$ is symmetric. As there are infinitely many functions of $\{l_{e_{0}^{+}}^{+},...,l_{e_{A-1}^{+}}^{+}\}$ 
which are invariant under cyclic permutations, each corresponding $|\Phi^{+}\rangle$ on Group averaging will yield a Poincare invariant Physical state. Thus there are
infinite number of Poincare invariant states in the theory.\\

\section{Emergence of spacetime lattice}
In section 8.1 we show that the Group average of a (suitably defined) charge-network state is associated with a discrete spacetime
\footnote{We note here that similar arguments in \cite{alokme}, while incomplete, yielded the same picture}. In section 8.2 we show in a precise sense that the Dirac
observables of the theory cannot resolve the spacetime at scales finer then the minimum embedding charge separation $a\ =\ \frac{2\pi L}{A}$. Sections 8.1 and 8.2 together
imply that the quantum theory of the true degrees of freedom is a lattice field theory. 
%At this point it is good to summarise what we have achieved so far. In the kinematical Hilbert space for Parametrized field theory, 
%there exists a super-selected sector defined by a regular distribution of embedding charges. 
%Group averaging with respect to the gauge group and the abelian group generated by zero-mode constraint gave us the Physical Hilbert space of the theory.
%This physical Hilbert space is the Cauchy completion of the inner-product space spanned by charge network states 
%$\sum_{\phi^{\pm}}{\bf (}(\gamma(s^{\pm}), \vec{k}^{\pm}, \vec{\triangle l}^{\pm})_{\phi^{\pm}}\vert$.\\ 
%There exists a  canonical choice of elementary Dirac observables $exp(iO_{F_{I}^{\pm}})$ which are quantum counter-parts of classical fields that are
%localized at discrete points($k^{\pm}_{I}=\pm\frac{2\pi I L}{A})$ in the embedding (continuum) spacetime.
%Whence naively we expect the quantum theory to be a lattice field theory defined on a regular spacetime lattice (with lattice length being a).
%In fact, as 
%reviewed in the section 8.1 below, the embedding data corresponding to a physical charge network state can be interpreted as defining a
%lattice in underlying spacetime.
%However in order to show that the quantum theory really is a lattice field theory, 
%we need to show that any Dirac observable can not ``resolve" the spacetime at scales finer then the the embedding scale a. This is what we show in section 8.2 (wherein  we restrict ourselves to the right-moving sector).

\subsection{Discrete Cauchy slices  and discrete  spacetime}

The polymer quantization of the embedding variables replaces the classical (flat) spacetime continuum with 
a discrete structure consisting of a countable set of points. This can be seen as follows.

Consider the classical canonical data $(X^+(x), X^-(x), Y^+(x), Y^-(x))$.
The  data $(X^+(x), X^-(x))$ is a map from $S^1$ into the flat spacetime $(S^1\times R, \eta )$ and embeds the former
into the latter as a spatial Cauchy slice with coordinate $x$ on which $(Y^+(x), Y^-(x))$ serve as initial data for 
scalar field evolution.
 Any gauge transformation generated by the constraints maps this 
data to new data which, in turn, defines matter data on a new Cauchy slice in the flat spacetime. 
In particular, the action of
the one parameter family of gauge transformations generated by smearing the constraints with some choice
of ``lapse-shift'' type functions $N^A$ (see section 2) generates a foliation of $(S^1\times R, \eta )$. The matter 
data on each slice of this foliaton together define a single solution $f$ to the flat spacetime wave equation, and are 
related to this solution through the equation 
$Y^{\pm}(x)= \pm 2 X^{\pm\prime}(x) \frac{\partial f}{\partial X^{\pm}}$ \cite{karelcm}. 
We shall refer to this property  (that the same flat spacetime and the same solution arise independent of the choice of $N^A$) as spacetime covariance. Spacetime covariance is
guaranteed
by the detailed nature of the constraints and their algebra \cite{karelcm}. 

%Next , consider 
%the image set in $(S^1\times R, \eta )$ of the set of all embeddings which are gauge related to a given one.
%From the above discussion it follows that this image set is exactly the flat spacetime $( S^1\times R, \eta )$ itself.

Next, consider the corresponding quantum structures. 
%Any charge network state is an eigen state of ${\hat X}^{\pm}(x)$. 
Consider a charge network state with n edges,
$|{\bf s}^+_{\lambda^{+}}\rangle\otimes |{\bf s}^-_{\lambda^{-}}\rangle$ with $|{\bf s}^{\pm}_{\lambda^{\pm}}\rangle= T_{s^{\pm}}\otimes W_{s^{\prime\pm}_{\lambda^{\pm}}}$,
where ${\bf s^{\pm}_{\lambda^{\pm}}}$ 
is such that \\
{\bf (a)} the embedding labels of successive edges are unequal and montonically increasing (or decreasing) for the left (or right) moving sectors, and\\
{\bf (b)}
$|\hbar k^{\pm}_{e^{\pm}_n}-\hbar k^{\pm}_{e^{\pm}_{0}}|<2\pi L$. \\
From \cite{alokme} it follows that the Group Average of
such charge networks span a superselected subspace of physical states
(Clearly, 
the sector ${\cal H}^{ss \pm}_{phys}$ (as well as ${\cal H}^{ss \pm}_{(0) phy}$) obtained by averaging states in 
${\cal D}_{ss}^{\pm}$ is of this type).
The action of the Dirac Observables of section 4.1 contains the physical information in the charge network state.
This action depends on the pairs $(k^{\pm}_{e^{\pm}},l^{\pm}_{e^{\pm}}+\lambda^{\pm})$ of embedding and matter charges for each 
edge $e^{\pm}\in \gamma ({\bf s}^{\pm}_{\lambda^{\pm}})$. Consider the coarsest graph, $\gamma ({\bf s}^{+}_{\lambda^{+}},{\bf s}^{-}_{\lambda^{-}})$, 
which is finer than $\gamma ({\bf s}^{+}_{\lambda^{+}}),\gamma ({\bf s}^{-}_{\lambda^{-}})$. Then the pairs $(k^+_e,k^-_e)$ for each edge of this
graph define a point $(X^+, X^-)=(\hbar k^+_e, \hbar k^-_e)$ in the flat spacetime by virtue of the embedding 
charges being 
eigen values of the embedding coordinate operators. Hence, we may place on each such spacetime point, the matter charges
$(l^+_e + \lambda^{+},l^-_e+\lambda^{-})$.  
%From (\ref{eq:33}), 
%the set of eigen values,
%$\lambda_{x,s^{\pm}}$, of ${\hat X}^{\pm}(x)$ for all $x\in [0,2\pi]$ describes a finite set of points on a spacelike Cauchy surface in 
%$(S^1\times R, \eta )$. These points have light cone coordinates $(X^+, X^-) = (\lambda_{x,s^{+}}, \lambda_{x,s^{-}})$.
The action of any gauge transformation on such a charge network state yields another charge network state
for which a similar association may be made. Note that the set of spacetime points associated with each
such charge network are either spacelike or null related and constitute the quantum version of a classical Cauchy
slice.
%
%
%whose eigen values lie, once again, on a Cauchy slice in $(S^1\times R, \eta )$. From equation (\ref{eq:56})
%it follows that the set of  eigen values for all possible gauge related charge network states is countable
%and defines a corresponding set of points in $(S^1\times R, \eta )$. 
The gauge invariant state obtained by
group averaging a charge network state is a sum over all distinct gauge equivalent states. It is straightforward to 
see that the set of labels $(k^+_e,k^-_e, l^+_e + \lambda^{+}, l^-_e + \lambda^{-})$ for all the edges of all the charge networks in the sum yield
a {\em consistent} spacetime association. In other words, the different sets of labels  coming from different
gauge related states all fit into a {\em single} labelling of a countable set of points in flat spacetime
each point being labelled by a single pair of matter charges. This property is the analog of classical spacetime
covariance.

It is easy to see that  for states in ${\cal H}_{phy}^{ss}$ (as well as in ${\cal H}_{(0) phy}^{ss}$) the countable set of points correspond to those of 
 of a regular
spacetime lattice with lattice spacing $\frac{2\pi L}{A}$ and that no finer discrete structure is available
for states which satisfy the ``nondegeneracy conditions'' {\bf (a),(b)} above. The existence of this 
{\em finest lattice} is tied to the integer- valuedness of the embedding charges used in this work; no such 
finest lattice exists for the real charges of Reference \cite{alokme}.

\subsection{Dirac observables and the emergence of the spacetime lattice}

 Consider the Weyl algebra generated by observables of the type $e^{iO_{f^{+}}}\ =\ \exp(i\int_{x}f^{+}(X^{+}(x))Y^{+}(x))$ where 
$f^{+}$ is a continuous periodic function (in the $X^{+}$ variable) such that 
$f^{+}(Ia)\ \in \mathbf{Z}\epsilon$ $\forall I$. The abstract Weyl algebra generated by such observables is
\begin{equation}
\widehat{e^{iO_{f^{+}}}}\widehat{e^{iO_{g^{+}}}}\ =\ e^{-\frac{i\hbar}{2}\alpha(\hat{f^{+}}, \hat{g^{+}})}\widehat{e^{iO_{f^{+}+g^{+}}}}
\end{equation}
where $\alpha(\hat{f}^{+}, \hat{g}^{+})\ =\ \int_{x}\left(f^{+}(\hat{X}^{+})(x)\partial_{x}g^{+}(\hat{X}^{+})(x)\ -\ g^{+}(\hat{X}^{+})(x)\partial_{x}f^{+}(\hat{X}^{+})(x)\right)$.\\ 
%\hspace*{0.2in} Note that classically the phase factor 
%can also be written as $\alpha(f,g)\ =\ \int_{X^{+}}(\ f^{+}(X^{+})\partial_{X^{+}}g^{+}(X^{+})\ -\ \ g^{+}(X^{+})\partial_{X^{+}}f^{+}(X^{+})\ )$.\\
\\
Note that action of the phase factor $e^{-\frac{i\hbar}{2}\alpha(\hat{f^{+}}, \hat{g^{+}})}$ 
on a charge-network state $|{\bf s}^{+}_{\lambda^{+}}\rangle\ =\ |\gamma({\bf s}^{+}_{\lambda^{+}}), \vec{k}^{+}, \vec{l}^{+}+\lambda^{+}\rangle$ is given by,
\begin{equation}\label{eq:jan10-3}
\begin{array}{lll}
 e^{-\frac{i\hbar}{2}\sum_{e^{+}\in \gamma({\bf s}^{+}_{\lambda^{+}})}\left(f^{+}(\hbar k_{e^{+}}^{+})\left(g^{+}(\frac{\hbar k_{e^{+}}^{+}+\hbar k_{(e+1)^{+}}^{+}}{2})- 
g^{+}(\frac{\hbar k_{e^{+}}^{+}+\hbar k_{(e-1)^{+}}^{+}}{2})\right) - 
f^{+}\leftrightarrow g^{+} \right)}|{\bf s}^{+}_{\lambda^{+}}\rangle
\end{array}
\end{equation}
 
Let $|{\bf s}^{+}_{\lambda^{+}}\rangle\ \in\ {\cal D}^{+}_{ss}$.\\
A simple computation shows that action of $\widehat{e^{iO_{f^{+}}}}\widehat{e^{iO_{g^{+}}}}$ on $|{\bf s}^{+}_{\lambda^{+}}\rangle$ is given by,
\begin{equation}\label{eq:jan18-1}
e^{-\frac{i\hbar}{2}\sum_{I=0}^{A-1}\left(f^{+}(\hbar k_{e_{I}^{+}}^{+})\left(\frac{g^{+}(\hbar k_{e_{I}^{+}}^{+})+g^{+}(\hbar k_{e_{I+1}^{+}}^{+})}{2}-
\frac{g^{+}(\hbar k_{e_{I}^{+}}^{+})+g^{+}(\hbar k_{e_{I-1}^{+}}^{+})}{2}\right)- f^{+}\leftrightarrow g^{+}\right)}\widehat{e^{iO_{f^{+}+g^{+}}}}|{\bf s}^{+}_{\lambda^{+}}\rangle
\end{equation}
where we have identified $e_{-1}^{+}$ with $e_{A-1}^{+}$ and $e_{A}^{+}$ with $e_{0}^{+}$.\\
Using (\ref{eq:jan10-3}) and (\ref{eq:jan18-1}) it is straightforward to see that 
the Weyl-algebra of observables is faithfully represented on ${\cal D}_{ss}^{+}$ iff the test functions satisfy a set of discrete consistency conditions,
\begin{equation}\label{eq:oct7}
\begin{array}{lll}
\frac{f^{+}(\hbar k_{e_{I}^{+}}^{+})+f^{+}(\hbar k_{e_{I+1}^{+}}^{+})}{2}\ =\ f^{+}(\frac{\hbar k_{e_{I}^{+}}^{+}+\hbar k_{e_{I+1}^{+}}^{+}}{2})\ 
\forall\ 0\leq I\leq (A-2)\\
\vspace*{0.1in}
\frac{f^{+}(\hbar k_{e_{0}^{+}}^{+})+f^{+}(\hbar k_{e_{A-1}^{+}}^{+})}{2}\ =\ f^{+}(\frac{\hbar k_{e_{0}^{+}}^{+}+\hbar k_{e_{A-1}^{+}}^{+}}{2})
\end{array}
\end{equation}
Thus the class of functions for which the Weyl algebra generated by $\widehat{e^{iO_{f^{+}}}}$ can be represented on ${\cal D}_{ss}^{+}$ 
is tied to the structure of the finest lattice.\\
As the Weyl algebra strongly commutes with the Group averaging map, the above consistency conditions continue to hold at the Physical state space level in the following sense.
 A basis state in ${\cal H}_{phy}^{+ ss}$ is given by,
\footnote{We ignore the zero-mode constraint in this section as the corresponding Group averaging map does not effect the graph or the embedding charges. The results 
described here can be easily lifted to ${\cal H}_{(0) phy}^{ss}$.}
\begin{equation}
\eta^{+}(|{\bf s}^{+}_{\lambda^{+}}\rangle )\ =\ \eta_{[\gamma({\bf s}^{+}_{\lambda^{+}})]}^{+}\sum_{\phi^{+}}\langle{\bf s}^{+}_{\lambda^{+} \phi^{+}}| 
\end{equation}
Due to periodicity of $f^{+}$, conditions given in (\ref{eq:oct7}) are satisfied for all ${\bf s}^{+}_{\lambda^{+} \phi^{+}}$ if they are satisfied for 
${\bf s}^{+}_{\lambda^{+}}$.
%The Weyl algebra of Dirac observables is faithfully represented on the above state iff the consistency conditions \ref{eq:oct7} are satisfied for embedding data associated with $|{\bf s}^{+}_{\phi}\rangle$ for some $\phi$. \\
\\
Now let us look at the commutator of two Dirac observables.
\begin{equation}
[\widehat{e^{iO_{f^{+}}}}, \widehat{e^{iO_{g^{+}}}}]\ =\ \sin(\frac{\hbar}{2}\alpha(f^{+},g^{+}))\widehat{e^{iO_{f^{+}+g^{+}}}}
\end{equation}
Let us assume that there exist a semi-classical state in which expectation value of $\widehat{e^{iO_{f^{+}}}}$ 
equals its classical value (at certain point in phase-space) plus corrections.\\ 
To the leading order in $\hbar$  the above commutator will equal $i\hbar$ times the corresponding Poisson bracket iff the quantum phase factor
$\alpha(f^{+},g^{+})$ equals the classical phase factor $\int_{X^{+}}(\ f(X^{+})\partial_{X^{+}}g(X^{+})\ -\ \ g(X^{+})\partial_{X^{+}}f(X^{+})\ )$.\\
 From (\ref{eq:jan10-3}) and (\ref{eq:oct7}), 
\begin{displaymath}
\alpha(f^{+},g^{+})\ =\ \sum_{I=0}^{A-1}\left[ f(\hbar k_{e_{I}^{+}}^{+})\left(g(\hbar k_{e_{I+1}^{+}}^{+})-g(\hbar k_{e_{I-1}^{+}}^{+})\right)- f\leftrightarrow g\right].
\end{displaymath}
(Where $e_{-1}^{+} = e_{A-1}^{+}$ and $e_{A}^{+} = e_{0}^{+}$.)\\
It is straightforward to show that $\alpha(f^{+},g^{+})$ equals the classical phase factor iff $f^{+}$ and $g^{+}$ are piecewise constant functions i.e.\\
$f^{+}(X^{+}) = f^{+}(\hbar k_{e_{I}^{+}}^{+})\ \forall\ X^{+}$ such that $\frac{\hbar k_{e_{I}^{+}}^{+}+\hbar k_{e_{I-1}^{+}}^{+}}{2}\ <\ X^{+}\ <\ 
\frac{\hbar k_{e_{I}^{+}}^{+}+\hbar k_{e_{I+1}^{+}}^{+}}{2}$.
(Here once again $e_{-1}^{+} = e_{A-1}^{+}$ and $e_{A}^{+} = e_{0}^{+}$.)\\
The upshot of the above arguements is that the classical symplectic structure (on the reduced phase space) can only emerge for (periodic) 
functions satisfy following two conditions.\\
\noindent {\bf a} Given the set of embedding data $\{k_{e_{0}^{+}}^{+}=0,...,k_{e_{A-1}^{+}}^{+}=\frac{2\pi(A-1) L}{A\hbar}\}$ 
$f^{+}$ has to satisfy $\frac{f^{+}(\hbar k_{e_{I}^{+}}^{+})+f^{+}(\hbar k_{e_{I+1}^{+}}^{+})}{2}\ =\ f^{+}(\frac{\hbar k_{e_{I}^{+}}^{+}+\hbar k_{e_{I+1}^{+}}^{+}}{2})\ \forall\ I$.\\
\noindent {\bf b} $f^{+}(X^{+})$ has to be piecewise constant. In other words it cannot probe the underlying continuum spacetime at scales finer then 
$\frac{2\pi L}{A}$.\\
The above two conditions describe the precise sense in which the quantum theory is a lattice field theory.\footnote{Also notice that 
$\hat{V}[\tau^{+}=\frac{2\pi L}{A\hbar}]\widehat{e^{iO_{F_{I}^{+}}}}\hat{V}[\tau^{+}=\frac{2\pi L}{A\hbar}]^{-1}\ =\ \widehat{e^{iO_{F_{I+1}^{+}}}}\ \forall\ I$ 
where I=(A+1) is identified with I=0. Whence
$\widehat{e^{iO_{F_{I}}}}$ also satisfy discrete evolution equations under time translations. This is yet another indication that we are dealing with a lattice field theory.}

%We now consider the dynamics of the elementary observables, and argue how the quantum evolution equations (\ref{eq:oct15}) are really difference equations.  
%So let us go back to the ``evolution" equations in (\ref{eq:oct5})
%\begin{equation}
%\hat{V}_{m}[\triangle X^{+}=\frac{2\pi}{A}]\widehat{e^{iO_{F_{e_{I}}^{+}}}}\hat{V}_{m}[\triangle X^{+}=\frac{2\pi}{A}]^{-1}\ =\ \widehat{e^{iO_{F_{e_{I+1}}^{+}}}}
%\end{equation}

%Classically, the above elemenraty observables are 
%\begin{equation}
%\begin{array}{lll}
%e^{iO_{F_{I}}}\ =\ exp[i\int_S^{1} Y^{+}(x)F_{I}(X^{+})(x)]\\
%\vspace*{0.1in}
%=\ exp[i\int_{\frac{k_{I}^{+}+k_{I-1}}{2}}^{\frac{k_{I}^{+}+k_{I+1}}{2}}\frac{\partial \Phi}{\partial X^{+}}
%\end{array}
%\end{equation}
%where $\Phi(X^{+},X^{-})$ is the Klein-Gordon scalar field.
%To leading order in the parameter a, the integrand of the above equation is given by $\Psi(k_{I}^{+}):=\frac{\partial\Phi}{\partial X^{+}}(\frac{2\pi I}{A})$.\\
\section{Tool-kit for the lattice theory}
As we saw in the last section, the polymer quantization of parametrized field theory leads to a lattice field theory. 
Detailed analysis of this theory requires certain tools from Discrete Fourier transforms which we summarise in this section.\\
Let $F(X^{\pm})$ be a real-valued periodic function with periodicity $L$. 
\begin{equation}\label{eq:oct10*}
\tilde{F}(n)\ =\ \frac{1}{2\pi L}\int_{0}^{2\pi L} F(X^{+})e^{\frac{in X^{+}}{L}}\ \forall n\in\ {\bf Z}
\end{equation}
Discrete Fourier transform is an approximation of the above integral by Riemann sum.
\begin{equation}\label{eq:oct10**}
\tilde{F}_{D}(n)\ :=\ \frac{1}{2\pi L}\sum_{I=0}^{A-1}F(k_{I}^{+})e^{\frac{2\pi inI}{A}}a
\end{equation}
where $k_{I}^{+} = \frac{2\pi I L}{A}$ and $a = \frac{2\pi L}{A}$ is the lattice spacing. The subscript D in $\tilde{F}_{D}(n)$ stands for discrete. \\
Notice that whereas in the case of continuous Fourier transform , $n\in {\bf Z}$, in the discrete case we have $-(A-1)\leq n\leq (A-1)$. 
Finally note that, using
\begin{equation}\label{eq:oct10***}
\sum_{m=0}^{A-1}e^{\frac{2\pi in(I-J)}{A}}\ =\ A\delta_{I,J}
\end{equation}
We can write inverse (discrete) Fourier transform as,
\begin{equation}\label{eq:oct10****}
F(k_{J}^{+})\ =\ \sum_{n=0}^{A-1}\tilde{F}_{D}(n) e^{-\frac{2\pi inJ}{A}}
\end{equation} 
Finally if the function F doesnot have a zero-mode, i.e.
\begin{equation}
\int_{X^{+}}F(X^{+})\ =\ 0
\end{equation}
then $\tilde{F}(0)\ =\ \tilde{F}_{D}(0)\ =\ 0$.

We will use these equations extensively in the subsequent sections.

\section{Fock vacuum as a polymer state}

This section is devoted to the construction of a state in the polymer Hilbert space which approximates
the Fock vacuum in the behaviour of its (1 and)  2 point functions. 
The nature of the approximation is precisely defined as follows. 
Recall that the two Barbero- Immirizi
parameters, $a=\frac{2\pi L}{A},\epsilon$ dictate the spacing of the embedding and matter charges, the former also being the lattice
spacing of the  spacetime lattice associated with states in ${\cal H}_{(0) phy}^{ss}$ (see section 8).
We define the {\em continuum limit} to be the limiting behaviour as $a,\epsilon\rightarrow 0$. Thus, the 
continuum limit is a limiting property of a 2 parameter family of quantum theories.
Accordingly, our aim is to construct a 2 parameter set of states (each state in an inequivalent quantum theory) 
and hence a 2 parameter family of 2 point functions which approach the Fock vacuum 2 point function in the 
continuum limit.

We are interested here in the Fourier transform of the two point function. 
This quantity is determined by the expectation values of all operators quadratic in the creation-annihilation modes.
An added complication in the polymer quantization is that these modes themselves have to be approximated by 
suitable polymer operators (recall \cite{alokme} that only exponentials of the modes are well defined operators not the modes
themselves). As we shall see, in the spirit of lattice field theory,  the approximants depend on $\epsilon, a$.

Specifically we aim to construct a 2 parameter family of states $|\Psi^{\pm}\rangle$ 
and a corresponding 2 parameter family of 
suitable approximants $\hat{a}_{n}^{poly}$ to $\hat{a}_{n}$ such that 
\begin{equation}\label{eq:jan11-3}
\begin{array}{lll}
\lim_{continuum}||\hat{a}_{(\pm)n}^{poly}\frac{|\Psi^{\pm}\rangle}{||\Psi^{+}||}|| = 0\ \forall\ 0\ n <\infty \\
\vspace*{0.1in}
\lim_{continuum}\frac{1}{||\Psi^{\pm}||^{2}}{\langle\Psi^{\pm}|[ \hat{a}_{(\pm)n}^{poly},\hat{a}_{(\pm)m}^{\dagger poly}]|\Psi^{\pm}\rangle}\ =\ \hbar n\delta_{n,m}
\end{array}
\end{equation}
That it suffices to satisfy the above equations to obtain the correct Fock  (1 and) 2 point function 
follows straightforwardly from a judicious use of the Cauchy Schwarz inequality. 
More in detail, the vanishing of the one point function follows from the first equation above and various 
quadratic combinations of ${\hat a}_{(\pm)n}, {\hat a}^{\dagger}_{(\pm)m}$ obtain the correct expectation values in the 
continuum limit using both the above equations and the Cauchy- Schwarz inequality.

Note that in the Fock representation the analog of the first equation constitutes the definition of the Fock vacuum
and the second equation is a trivial consequence of the canonical commutation relations. In contrast, the 
polymer representation (as we shall see explicitly in section 10.3)  supports the canonical commutation
relations in expectation value only in the continuum limit. Note also that in the Fock representation 
the analog of the first equation together with the canonical commutation relations, generates all the $n$- point 
functions. This gives us hope that the computations here will help in establishing whether or not the 
behaviour of all $n$- point functions can be adequately approximated in the polymer representation.
Finally, note that obtaining the 2 point function is more ambitious than obtaining semiclassical behaviour 
peaked at trival classical data.

In section 10.1 we summarise our results. In section 10.2 we discuss the polymer approximant to the 
mode function.
In section 10.3 we outline a heuristic set of calculations which allow
us to make an educated guess for the solution of section 10.1. We show in section 10.4 that the candidate solution 
guessed in section 10.3 and displayed 
in  section 10.1 actually solves the annihilation operator condition (i.e the first equation of 
(\ref{eq:jan11-3})), and in 10.5 that it solves the commutation relation condition (i.e. the second
equation of (\ref{eq:jan11-3})).

With out loss of generality, we work with the solution to the zero-mode constraint $|\gamma^{\pm}, \vec{k}^{\pm},\vec{\triangle l}^{\pm}{\bf )}$ 
instead of the working with states in ${\cal H}_{(0)phy}^{ss}$. This is only in the interest of pedagogy. Towards the end of the 
section we describe how to generalise these constructions to physical states (which are solutions to all the constraints.)

\subsection{Summary of main results}
Fix the graph $\gamma^{+}$ and the embedding data $\vec{k}^{+}\ =\ \left(k_{e_{0}^{+}}^{+}=0,...,k_{e_{A-1}^{+}}^{+}=\frac{2\pi (A-1)L}{\hbar}\right)$ once and for all.\\
Consider a state in the left moving sector
\begin{equation}
|\Psi^{+}\rangle\ =\ \sum_{\vec{\triangle l}^{+}}c(\vec{\triangle l}^{+})|\gamma^{+}, \vec{k}^{+}, \vec{\triangle l}^{+}{\bf )}
\label{polyfockstate}
\end{equation}
where 
\begin{equation}\label{eq:bored1}
c(\vec{\triangle l}^{+})\ =\ \exp[-\frac{\hbar 2\pi L^{2}\epsilon^{2}}{a^{2}}\sum_{n>0}\frac{|\tilde{\triangle l}^{+}(n)|^{2}}{f(n)}]
\end{equation}
where $f(n)$ is given by 
\begin{equation}
\begin{array}{lll}
f(n)\ =\ \frac{\tan(\frac{n\pi}{A})}{\frac{n\pi}{A}}\ \forall n \leq N_{0}\\
\vspace*{0.1in}
f(n)\ =\ n\ \forall n\geq\ N_{0}, 
\end{array}
\end{equation}
and $N_0$ is a fixed integer independent of $\epsilon ,a$. 
This state(as shown in appendix A) is normalizable.\\

%\begin{equation}
%\begin{array}{lll}
%c(\vec{\triangle l})\ =\ exp[-\frac{\hbar L^{2}\epsilon^{2}}{2\pi a^{2}}\sum_{n=0}^{A-1} \frac{|\tilde{\triangle l}(n)|^{2}}{\frac{\tan(\frac{n\pi a}{L})}{\frac{\pi a}{L}}}] 
%\forall n\ <\ N_{0},\\
%\vspace*{0.1in}
%c(\vec{\triangle l})\ =\ exp[-\frac{\hbar L^{2}\epsilon^{2}}{2\pi a^{2}}\sum_{n=0}^{A-1} \frac{|\tilde{\triangle l}(n)|^{2}}{n}] \forall n\ \geq\ N_{0},
%\end{array}
%\end{equation}
%and $N_0$ is a fixed integer independent of $\epsilon ,a$. 
%This state(as shown in appendix A) is normalizable.\\
In the next section we show that a suitable choice of approximants to the mode functions which can be quantized on ${\cal H}_{(0) phy}^{ss}$ are given by,
\begin{equation}
\hat{a}_{n}^{poly}\ =\ \frac{1}{A\epsilon}\sum_{I=0}^{A-1}\ a(\widehat{e^{i O_{F_{I}^{+}}}} - \widehat{e^{-i O_{F_{I}^{+}}}})e^{in \hbar k_{e_{I}^{+}}^{+}}
\end{equation}
with the functions $F_{I}^{+}$ defined as follows.
\begin{equation}
\begin{array}{lll} 
F_{I}^{+}(X^{+}) = \epsilon\ $if$\ \frac{\hbar k_{e_{I}^{+}}^{+} +\hbar k_{e_{I-1}^{+}}^{+}}{2} <\ X^{+}\ <\ \frac{\hbar k_{e_{I}^{+}}^{+}+\hbar k_{e_{I+1}^{+}}^{+}}{2}\\
\vspace*{0.1in}
\hspace*{0.5in} = \frac{\epsilon}{2}\ $if$\ X^{+}(x)= \frac{\hbar k_{e_{I}^{+}}^{+}+\hbar k_{e_{I-1}^{+}}^{+}}{2}\\
\vspace*{0.1in}
\hspace*{0.5in} = 0\ $otherwise$ 
\end{array}
\end{equation}
$\forall\ 0\leq I\leq A-1$.
Here $\hbar k_{e_{-1}^{+}}^{+}\ :=\ -\frac{-2\pi L}{A}$ and $\hbar k_{e_{A+1}^{+}}^{+}=\frac{2\pi L}{A}$.\\

The precise statement about the continuum limit and emergence of Fock vacuum (in the left moving sector) is the following.\\

\begin{equation}\label{eq:jan11-4}
\begin{array}{lll}
\hat{a}_{n}^{poly} \frac{|\Psi^{+}\rangle}{||\Psi^{+}||}\ =\  0\ +\ |\delta\Psi^{+}\rangle\ \forall\ 0\ <\ n\ < N_{0}\\
\vspace*{0.1in}
\frac{1}{||\Psi^{+}||^{2}}\langle\Psi^{+}|\hat{a}_{n}^{poly}\hat{a}_{m}^{poly\ \dagger}|\Psi^{+}\rangle\ =\ \frac{\hbar}{\pi L^{2}} n\delta_{n,m} + \textrm{error-terms}
\end{array}
\end{equation}
where $n,\ m\ >\ 0$,\\
and $||\delta\Psi^{+}||$ as well as the error-terms in (\ref{eq:jan11-4}) vanish in the limit
\begin{equation}\label{eq:cont}
\begin{array}{lll}
a\rightarrow\ 0\\
\vspace*{0.1in}
\epsilon =  C_{0}a^{\triangle}\ \textrm{with}\ \triangle\ >\ 4. 
\end{array}
\end{equation}
Henceforth we shall assume that the continuum limit (i.e. $\epsilon,\ a\rightarrow\ 0$) is governed by (\ref{eq:cont}).

We note that even though the form of $|\Psi^{+}\rangle$ is reminiscent  of 
a complexifier coherent state \cite{ttcoherent}, we have not been 
able to obtain it  via the complexifier techniques, the reason being that we do not know of any function whose 
operator correspondent possesses the matter charges  $\triangle l_{I}^{+}$ as its eigen values.\\
Another note-worthy feature of $|\Psi^{+}\rangle$ is its invariance under cyclic permutations of $(l_{e_{0}^{+}}^{+},...,l_{e_{A-1}^{+}}^{+})$, which implies that it is 
(discrete) Poincare invariant.

\subsection{Polymer approximants to the mode functions.}
Consider the classical Fourier modes of K-G scalar field $\Phi(X^{+},X^{-})$\footnote{In section 2 we denoted the spacetime scalar field by $f$. 
Henceforth we shall denote it by $\Phi$.}
\begin{equation}
a_{(\pm) n}\ =\ \pm\frac{1}{2\pi L}\int_{0}^{2\pi L}\frac{\partial \Phi}{\partial X^{\pm}}e^{\frac{inX^{\pm}}{L}}
\end{equation} 
where $n\ >\ 0$.\\
Let us restrict our attention to the (+)-sector.\\
Whence $a_{n}$ is just Fourier transform of $\frac{\partial \Phi}{\partial X^{+}}$. The corresponding discrete Fourier transform is (\ref{eq:oct10**}),
\begin{equation}
a_{n,D}\ =\ \frac{1}{A}\sum_{I=0}^{A-1}(\frac{\partial \Phi}{\partial X^{+}})(k_{e_{I}^{+}}^{+})e^{\frac{in \hbar k_{e_{I}^{+}}^{+}}{L}}
\end{equation}
Where $\hbar k_{e_{I}^{+}}^{+}=\frac{2\pi IL}{A}$. As the computations performed subsequently do not change the graph or the embedding charges, we 
will supress the edge label on the embedding data. Whence we will denote $\hbar k_{e_{I}^{+}}^{+}$ by $k_{I}^{+}$ from now on.\\
Further more, as the only combination in which $k_{I}^{+}$ will come up in all the subsequent calculations, is 
$\frac{k_{I}^{+}}{L}=\frac{2\pi I}{A}$, 
\emph{we will abuse the notation slightly  more and denote $\frac{2\pi I}{A}$ by $k_{I}^{+}$.}\\
Whence the suitable approximant to $a_{n}$ which can be quantized on ${\cal H}_{phy}^{+}$ is given by,
\begin{equation}
a_{n}^{poly}\ :=\ \frac{1}{A}\sum_{I=0}^{A-1}(\frac{e^{iO_{F_{I}^{+}}}-e^{-iO_{F_{I}^{+}}}}{2ia\epsilon})e^{ink_{I}^{+}}
\end{equation}
where we have approximated 
$\frac{\partial \Phi}{\partial X^{+}}$ by its lattice approximant
\begin{equation}
(\frac{\partial \Phi}{\partial X^{+}})^{poly}:= 
\frac{e^{iO_{F_{I}^{+}}}-e^{-iO_{F_{I}^{+}}}}{2ia\epsilon}
\label{latticescalarfield}
\end{equation}
The corresponding quantum operator is
\begin{equation}\label{eq:jan9-1}
\hat{a}_{n}^{poly}\ :=\ \frac{1}{A}\sum_{I=0}^{A-1}(\frac{\widehat{e^{iO_{F_{I}^{+}}}}-\widehat{e^{-iO_{F_{I}^{+}}}}}{2ia\epsilon})e^{ink_{I}^{+}}
\end{equation}

\subsection{Heuristic Guesswork}

This section could be ommitted without affecting the logic and completeness of the paper. We include it
in the hope that similar heuristics might prove useful for constructions in LQG. 
In this section our aim is to provide an educated guess for a solution to equations (\ref{eq:jan11-4}).
Since the argumentation is heuristic and intuitive, we do not attempt to summarise it precisely and the reader is 
advised to temporarily relax standards of rigor and simply ``go with the flow''. 
The argumentation uses an intuitive idea of a continuum limit wherein discrete structures are summarily 
replaced by intuitive continuum analogs. This transforms  the first equation of (\ref{eq:jan11-4}) into a 
functional differential 
equation for the continuum analog of the state; this equation can easily be solved and its discrete correspondent
provides an educated guess for the polymer state.\\
{\bf As we are working in $\overline{\eta}^{+}({\cal D}^{+}_{ss})$ with a fixed graph $\gamma^{+}$ and the fixed embedding data 
$\left(k_{0}^{+}=0,...,k_{A-1}^{+} = \frac{2\pi(A-1)}{A}\right)$, we will supress the (+)-label on the $\vec{k}^{+}$, $\vec{\triangle l}^{+}$ 
charges in all our subsequent calculations.}\\

It is easy to show that the action of $\hat{a}_{n}^{poly}$ on an arbitrary state 
$|\Psi^{+}\rangle\ =\ \sum_{\vec{\triangle l}}c(\vec{\triangle l})|\gamma,\vec{k}, \vec{\triangle l}{\bf )}$ in $\overline{\eta}({\cal D}_{ss}^{+})$ is given by,
\begin{equation}\label{eq:oct11-1}
\begin{array}{lll}
\hat{a}_{n}^{poly}|\Psi^{+}\rangle\ =\\ 
\vspace*{0.1in}
\hspace*{0.6in} \frac{1}{2i\epsilon a A}\sum_{I=0}^{A-1}e^{ink_{I}}\sum_{\vec{\triangle l}}\left[e^{-\frac{i\hbar\epsilon^{2}}{2}(\triangle l_{I} + \triangle l_{I+1})}c(\vec{\triangle l}-\vec{\triangle \epsilon}^{(I)})-\right.\\
\vspace*{0.1in}
\hspace*{1.7in} \left. e^{\frac{i\hbar\epsilon^{2}}{2}(\triangle l_{I} + \triangle l_{I+1})}c(\vec{\triangle l}+\vec{\triangle \epsilon}^{(I)})\right]\\
\vspace*{0.1in}
\hspace*{2.3in} |\gamma,\vec{k},\vec{\triangle l}{\bf )}
\end{array}
\end{equation}
where $(\triangle \epsilon)^{(I)}_{J} :=\ \epsilon(\delta^{I}_{J}-\delta^{I}_{J-1})$.\\

Now as, $a\rightarrow\ 0$, the embedding charges behave as,
\begin{equation}\label{eq:oct12-0}
\begin{array}{lll}
k_{I}\ L\ =\ \frac{2\pi I L}{A}\ &=&\ Ia\\
\vspace*{0.2in}
&\rightarrow&\  x_{I}
\end{array}
\end{equation} 
where $x_{I}$ is a continuous variable whose range is $[0,2\pi L]$.\\
\\  
The exponents in the phase factors in (\ref{eq:oct11-1}) are $\pm\frac{\hbar\epsilon^{2}}{2}(\triangle l_{I} + \triangle l_{I+1})$ which can be written as,
$\pm\frac{\hbar\epsilon a }{2}(\frac{\epsilon\triangle l_{I}}{a} + \frac{\epsilon\triangle l_{I+1}}{a})$.\\

Hence the phase factor can be written as a power-series in $\epsilon,\ a$ only if 
$\frac{\epsilon\triangle l_{I}}{a}\ =\ \frac{\epsilon(l_{I}-l_{I-1})}{a}$ remains finite as $\epsilon, a\rightarrow 0$.\\ 
Whence let us assume that,
\begin{equation}\label{eq:oct11-2}
\lim_{\epsilon,a\rightarrow 0}(\frac{\epsilon\triangle l_{I}}{a}) = O(1)
\end{equation}
More precisely the above assumption hints at $\lim_{\epsilon,a\rightarrow 0}(\frac{\epsilon\triangle l_{I}}{a}) = g(x_{I})$ 
where g is a continuous function of the (continuous)  variable $x_{I}$. 
As we want to try the simplest possible scenarios under which (\ref{eq:oct11-1}) admits a solution, we also assume that $g(x)$ is a smooth function. 
Whence in the limit $\epsilon,\ a\rightarrow 0$,
\begin{equation}
\frac{\epsilon\triangle l_{I+1}}{a}\ =\ \frac{\epsilon\triangle l_{I}}{a} + \mathrm{error-terms}
\end{equation} Where the error-terms vanish in the limit $a\rightarrow\ 0$.
%This is an additional restriction on the set of matter charges we are summing over in the state $|\Psi\rangle$.\\
% Note that \ref{eq:oct11-2} is clearly satisfied if we look at the one-parameter family of states defined in \ref{eq:oct11-3} above. That is, in the continuum limit we only sum over ``matter charge" configurations bounded from below.\\
%Thus (\ref{eq:oct11-2}) is satisfied if $\epsilon$ tends to zero faster then $a$ and as $\frac{\epsilon}{a}$ tends to zero, only those charge configurations
%$\vec{\triangle l}$ contribute which satisfy 
%$\triangle l_{I}$ tends to infinity at the rate of $\frac{a}{\epsilon}$ so that the quantity $\frac{\epsilon\triangle l_{I}}{a}$ remains finite $\forall\ I$. 
%\emph{This is our notion of continuum limit.}\\
%From here it immediately follows that $\frac{(\triangle \epsilon)^{(I)}_{J}}{a}\rightarrow\ 0$ as $\frac{\epsilon}{a}\rightarrow\ 0$.
Another quantity, whose continuum limit will be crucial to subsequent calculations is the following.
\begin{equation}\label{eq:oct12-1}
\begin{array}{lll}
\lim_{a\rightarrow 0}\epsilon^{(I)}_{J}\ =\ \lim_{a\rightarrow 0}\epsilon(\delta_{I,J}-\delta_{I,J-1})\\
\vspace*{0.1in}
\hspace*{0.6in} = \epsilon\frac{d}{d x_{I}}\delta(x_{I},x_{J})
\end{array}
\end{equation}

where we have used $\lim_{a\rightarrow 0}\frac{\delta_{I,J}}{a}\ =\ \delta(x_{I}, x_{J})$ and $\lim_{a\rightarrow 0} \frac{\delta(x_{I}, x_{J})\ -\ \delta(x_{I}, x_{J}-a)}{a}\ =\ \frac{d}{d x_{J}}\delta(x_{I},x_{J})$.

%Thus in light of \ref{eq:oct11-3},  \ref{eq:oct11-2} can also be written as,
%\begin{equation}
%lim_{\lambda\rightarrow \infty}(\frac{\epsilon_{\lambda}(\triangle l_{I})_{\lambda}}{a_{\lambda}}) = O(1)
%\end{equation}

In order to ensure that in the continuum limit we can also taylor expand $c(\vec{\triangle l})$ in terms of (a certain combination of) $\epsilon$ and a, 
we take, 
\begin{equation}\label{eq:oct11-4}
c(\vec{\triangle l})\ =\  c(\frac{\epsilon\vec{\triangle l}}{a})
\end{equation}
Whence in the continuum limit,
\begin{equation}\label{eq:oct11-5}
\begin{array}{lll}
c(\vec{\triangle l}\pm \vec{\triangle \epsilon}^{(I)})\ =\ c(\frac{\epsilon\vec{\triangle l}}{a}\pm \frac{\vec{\triangle \epsilon}^{(I)}}{a})\\
\vspace*{0.1in}
\hspace*{1.1in} \rightarrow c(g(x)\pm a\epsilon\frac{d}{d x_{I}}\delta(x,x_{I}))\\
\vspace*{0.1in} 
\hspace*{1.0in} =\ c(g(x))\pm\ a\epsilon\frac{d}{d x_{I}}\delta(x,x_{I})\int dy\frac{\delta c(g(x))}{\delta c(g(y))}  
\end{array}
\end{equation}

The continuum limit of $\hat{a}_{n}^{poly}|\Psi^{+}\rangle$ (i.e. the continuum limit of (\ref{eq:oct11-1})) is given by,
\begin{equation}
\begin{array}{lll}
\lim_{\epsilon, a\rightarrow 0}\hat{a}_{n}^{poly}|\Psi^{+}\rangle\ = \\ 
\vspace*{0.1in}
=\lim_{C}\frac{1}{2i\epsilon (2\pi L)}\sum_{I=0}^{A-1}\sum_{\triangle l}e^{ink_{I}}\left[e^{\frac{-i\hbar\epsilon a}{2}
(\frac{\epsilon\triangle l_{I}}{a}+\frac{\epsilon\triangle l_{I+1}}{a})}c(\vec{\triangle l}- \vec{\triangle \epsilon}^{(I)})\ -\right.\\
\vspace*{0.1in} 
\hspace*{1.6in} \left. e^{\frac{i\hbar\epsilon a}{2}(\frac{\epsilon\triangle l_{I}}{a}+\frac{\epsilon\triangle l_{I+1}}{a})}
c(\vec{\triangle l}+ \vec{\triangle \epsilon}^{(I)})\right]|\vec{\triangle l}\rangle\\
\vspace*{0.1in}
= \lim_{C}\frac{1}{2i\epsilon (2\pi L)}\sum_{I=0}^{A-1}\sum_{\triangle l} e^{ink_{I}}\left[\left( c(\vec{\triangle l}-\vec{\triangle \epsilon}^{(I)})\ -\ 
c(\vec{\triangle l}+ \vec{\triangle \epsilon}^{(I)})\right)\right.\\
\vspace*{0.1in}
\hspace*{1.6in}\left. -\ i\hbar\epsilon a\frac{\epsilon{\triangle l_{I}}}{a}\left(c(\vec{\triangle l}- \vec{\triangle \epsilon}^{(I)})\ +\ 
c(\vec{\triangle l}+ \vec{\triangle \epsilon}^{(I)})\right)\right]|\vec{\triangle l}\rangle
\end{array}
\end{equation}
where $\lim_{C}$ is an abbreviation for the limit $\epsilon,\ a\rightarrow 0$.\\ 
%that are to be summed over, which in light of (\ref{eq:oct11-2}) is dependent on $\epsilon$ and a. 
We have expanded the phase factor in $\epsilon a$ and kept only the leading order terms. $|\vec{\triangle l}\rangle$ is 
an abbreviation for the charge network state $|\gamma, \vec{k}, \vec{\triangle l}{\bf )}$.
Whence the equation $\lim_{C}\hat{a}_{n}^{poly}|\Psi^{+}\rangle\ =\ 0$ implies
\begin{equation}
\begin{array}{lll}
\lim_{C}\frac{1}{2i\epsilon (2\pi L)}\sum_{I=0}^{A-1}\sum_{\triangle l} e^{ink_{I}}\left[\left( c(\vec{\triangle l}-\vec{\triangle \epsilon}^{(I)})\ -\ 
c(\vec{\triangle l}+ \vec{\triangle \epsilon}^{(I)})\right)\right.\\
\vspace*{0.1in}
\hspace*{1.6in}\left. -\ i\hbar\epsilon a\frac{\epsilon{\triangle l_{I}}}{a}\left(c(\vec{\triangle l}- \vec{\triangle \epsilon}^{(I)})\ +\ 
c(\vec{\triangle l}+ \vec{\triangle \epsilon}^{(I)})\right)\right]|\vec{\triangle l}\rangle\\
\vspace*{0.1in}
\hspace*{2.7in} = 0
\end{array}
\end{equation}
We can now use (\ref{eq:oct11-2}), (\ref{eq:oct11-5}) ,(\ref{eq:oct12-0}) and, (\ref{eq:oct12-1}) to turn the above discrete equation into an integral equation.\\
\begin{equation}
\begin{array}{lll}
\lim_{C}\frac{1}{2i\epsilon a 2\pi (2\pi L)}\int dx_{I} e^{\frac{in x_{I}}{L}}\left[c(g(x)-\epsilon a
\frac{d}{dx_{I}}\delta(x,x_{I}))-c(g(x)+\epsilon a\frac{d}{dx_{I}}\delta(x,x_{I}))\right.\\
\vspace*{0.1in}
\hspace*{1.3in} \left. -i\hbar\epsilon a g(x_{I}) (c(g(x)-\epsilon a\frac{d}{dx_{I}}\delta(x,x_{I}))+c(g(x)+\epsilon a\frac{d}{dx_{I}}\delta(x,x_{I})))\right]\ =\ 0
\end{array}
\end{equation}
Using (\ref{eq:oct11-5}) above equation further simplifies to,
\begin{equation}
\begin{array}{lll}
\frac{1}{2\pi iL}\int dx_{I} [\ e^{in\frac{ x_{I}}{L}}\left(\frac{d}{dx_{I}}\frac{\delta c}{\delta g(x_{I})}\ -\ i\hbar g(x_{I}) c(g(x))\right)\ ]\ =\ 0
\end{array}
\end{equation} 

Using integration by parts for the first term and $\frac{1}{2\pi L}\int dx_{I} g(x_{I})e^{\frac{in x_{I}}{L}} =: \tilde{g}(n)$, we get,
\begin{equation}\label{eq:oct14-1}
-\int dx_{I}(in\frac{1}{L})e^{in\frac{ x_{I}}{L}}(\frac{\delta c}{\delta g(x_{I})})\ -\ i\hbar (2\pi L) \tilde{g}(n)c(g)\ =\ 0
\end{equation}
Now notice the following,
\begin{equation}
\begin{array}{lll}
\frac{\delta c}{\delta g(x_{I})}\ =\ \sum_{m\in {\bf Z}}\frac{\delta c}{\delta \tilde{g}(m)}\frac{\delta \tilde{g}(m)}{\delta g(x_{I})}\\
\vspace*{0.1in}
\hspace*{1.2in} =\sum_{m}\frac{\delta c}{\delta \tilde{g}(m)}\frac{1}{2\pi L}e^{im\frac{x_{I}}{L}}
\end{array}
\end{equation}
Above equation is satisfied by,
\begin{equation}
c(g)\ =\ {\cal A}\exp[-\hbar2\pi L^{2}\sum_{n>0}\frac{|\tilde{g}(n)|^{2}}{n}]
\end{equation}
Where $|\tilde{g}(n)|^{2}\ =\ \tilde{g}(n)\tilde{g}^{*}(n)$ and ${\cal A}$ is an arbitrary constant.
As $\tilde{g}(n)\ =\ \frac{1}{2\pi L}\int dx g(x)e^{in\frac{x}{L}}$,
The corresponding discrete Fourier transform is given by,
\begin{equation}
\tilde{g}_{D}(n)\ =\ \frac{1}{2\pi L}\sum_{I=0}^{A-1}a(\frac{\epsilon\triangle l_{I}}{a})e^{ink_{I}}
\end{equation}
But as the (discrete) Fourier transform of $\vec{\triangle l}$ is given by,
\begin{equation}
\tilde{\triangle l}(n)\ =\ \frac{a}{2\pi L}\sum_{I}\triangle l_{I}e^{ink_{I}}
\end{equation}
This implies that $\tilde{g}_{D}(n)\ =\ \frac{\epsilon}{a}\tilde{\triangle l}(n)$.
This suggests that we consider $c(\vec{\triangle l})$ of the form,
\begin{equation}\label{eq:oct14-2}
c(\vec{\triangle l})\ =\ \exp[-\frac{\hbar 2\pi L^{2}\epsilon^{2}}{a^{2}}\sum_{n>0}\frac{|\tilde{\triangle l}(n)|^{2}}{n}]
\end{equation}
In fact there is more freedom in the choice of these weights. We can consider $c(\vec{\triangle l})$ of the form,

\begin{equation}\label{eq:bored}
c(\vec{\triangle l})\ =\ \exp[-\frac{\hbar 2\pi L^{2}\epsilon^{2}}{a^{2}}\sum_{n>0}\frac{|\tilde{\triangle l}(n)|^{2}}{f(n)}]
\end{equation}
where $f(n)$ should be such that $\lim_{a\rightarrow 0} f(n) = n\ \forall n$. One such choice of $f(n)$ is given by,
\begin{equation}
\begin{array}{lll}
f(n)\ =\ \frac{\tan(\frac{n\pi}{A})}{\frac{n\pi}{A}}\ \forall n \leq N_{0}\\
\vspace*{0.1in}
f(n)\ =\ n\ \forall n\ >\ N_{0}
\end{array}
\end{equation}
Where $N_{0} << A $ but is otherwise arbitrary. Thus choice of f(n) will be justified aposteriori in the subsequent section.
The corresponding approximant to the Fock-vacuum is given by,
\begin{equation}
|\Psi^{+}\rangle\ =\ \sum_{\vec{\triangle l}}c(\vec{\triangle l})|\vec{\triangle l}\rangle
\end{equation}
The naive, intuitive continuum calculation suggests that 
the only non-trivial contributions to $\hat{a}_{n}^{poly}|\Psi^{+}\rangle$ 
come from the matter charges which are such that \\
$\frac{\epsilon\triangle l_{I}}{a} \sim O(1)\ \forall\ I\ \textrm{as}\ \epsilon, a\rightarrow\ 0$.
As we will see in the next section, certain highly delicate cancellations 
between such contributions will contrive to ensure that $\hat{a}_{n}^{poly}\frac{|\Psi^{+}\rangle}{||\Psi^{+}||}\ =\ 0\ +\ |\delta\Psi^{+}\rangle$ 
in the polymer Hilbert space.

\subsection{The annihilation operator condition}
In this section we compute the action of 
$\hat{a}_{n}^{poly}$ 
on 
$|\Psi^{+}\rangle\ =\ \sum_{\vec{\triangle l}}c(\vec{\triangle l})|\gamma,\vec{k},\vec{\triangle l}{\bf )}$
with $c(\vec{\triangle l})$ 
given in (\ref{eq:bored1}). 
Our aim is to show that
\begin{equation}\label{eq:jan20-1}
\hat{a}_{n}^{poly}\frac{|\Psi^{+}\rangle}{||\Psi^{+}||}\ =\ 0 + |\delta\Psi^{+}\rangle\ \forall n\leq N_{0}
\end{equation}
and the $||\delta\Psi^{+}||$ vanishes in the continuum limit.\\
Using (\ref{eq:jan9-1}) we have, 
\begin{equation}\label{eq:oct14-4}
\begin{array}{lll}
\hat{a}_{n}^{poly}|\Psi^{+}\rangle\ =\ \frac{1}{2i\epsilon (2\pi L)}\sum_{I=0}^{A-1}e^{ink_{I}}\sum_{\vec{\triangle l}}\\
\vspace*{0.1in}
\hspace*{1.1in}\left[\ e^{-\frac{i\hbar\epsilon a}{2}(\frac{\epsilon\triangle l_{I}}{a}\ +\ \frac{\epsilon\triangle l_{I+1}}{a})}
\exp[-\frac{\hbar 2\pi L^{2}\epsilon^{2}}{a^{2}}\sum_{m>0}\frac{|\tilde{\triangle l}(m)-\ \tilde{\triangle \epsilon}^{(I)}(m)|^{2}}{f(m)}]\right.\\
\vspace*{0.1in}
\hspace*{1.2in}-\left.\ e^{\frac{i\hbar\epsilon a}{2}(\frac{\epsilon\triangle l_{I}}{a}\ +\ 
\frac{\epsilon\triangle l_{I+1}}{a})}\exp[-\frac{\hbar 2\pi L^{2}\epsilon^{2}}{a^{2}}
\sum_{m>0}\frac{|\tilde{\triangle l}(m)+\tilde{\triangle \epsilon}^{(I)}(m)|^{2}}{f(m)}]\right]\\
\vspace*{0.1in}
\hspace*{4.7in} |\vec{\triangle l}\rangle
\end{array}
\end{equation}
Note that $\triangle\epsilon^{(I)}_{J}\ :=\ \epsilon(\delta_{I,J}-\delta_{I,J-1})$. However in this and all the subsequent sections, we will slightly abuse the
notation and denote $(\delta_{I,J}-\delta_{I,J-1})$ by $\triangle\epsilon^{(I)}_{J}$. 

%Using,
%\begin{equation}
%\begin{array}{lll}
%\frac{|\tilde{\triangle l}(n)\mp\frac{\tilde{\triangle \epsilon}^{(I)}(n)}{\epsilon}|^{2}}{f(n)}\ =\\
%\vspace*{0.1in}  
%\hspace*{1.0in}\frac{|\tilde{\triangle l}(n)|^{2} + |\tilde{\triangle\epsilon}^{(I)}(n)|^{2}\mp\ (\tilde{\triangle l}(n)\tilde{\triangle\epsilon}^{(I)}(-n)\ +\ c.c.\ )}{f(n)}
%\end{array}
%\end{equation} 
%where, using $(\triangle \epsilon)^{(I)}_{J} :=\ \epsilon(\delta^{I}_{J}-\delta^{I}_{J-1})$ and (\ref{eq:oct10****})
%\begin{equation}
%\begin{array}{lll}
%\frac{\tilde{\triangle\epsilon}^{(I)}(n)}{\epsilon}\ =\ \frac{1}{A}\sum_{J=0}^{A-1} e^{ink_{J}^{+}}(\delta_{I,J}\ -\ \delta_{I,J-1})\\
%\vspace*{0.1in}
%\hspace*{0.5in} =\ \frac{1}{A}e^{ink_{I}^{+}}(1-e^{in\frac{2\pi}{A}})
%\end{array}
%\end{equation}
(\ref{eq:oct14-4}) can be easily re-written as,
\begin{equation}\label{eq:nov-14}
\begin{array}{lll}
\hat{a}_{n}^{poly}|\Psi^{+}\rangle\ =\ \frac{1}{2i\epsilon (2\pi L)}\sum_{I=0}^{A-1}e^{ink_{I}}\sum_{\vec{\triangle l}}\ 
c(\vec{\triangle l})\exp[-\frac{\hbar 2\pi L^{2}\epsilon^{2}}{a^{2}}\sum_{m>0}\frac{|\tilde{\triangle \epsilon}^{(I)}(m)|^{2}}{f(m)}|]\\
\vspace*{0.1in}
\hspace*{1.7in}\left[e^{-\frac{i\hbar\epsilon a}{2}(\frac{\epsilon\triangle l_{I}}{a}\ +\ \frac{\epsilon\triangle l_{I+1}}{a})}
\exp[+\frac{\hbar 2\pi L^{2}\epsilon^{2}}{a^{2}}\sum_{m>0}[\frac{\tilde{\triangle l}(m)\triangle\epsilon^{(I)}(-m) + c.c.}{f(m)}]\right.\\
\vspace*{0.1in}
\hspace*{1.7in}-\left. e^{\frac{i\hbar\epsilon a}{2}(\frac{\epsilon\triangle l_{I}}{a}\ +\ \frac{\epsilon\triangle l_{I+1}}{a})} 
\exp[-\frac{\hbar 2\pi L^{2}\epsilon^{2}}{a^{2}}\sum_{m>0}[\frac{\tilde{\triangle l}(m)\tilde{\triangle\epsilon^{(I)}}(-m) + c.c.}{f(m)}]\right]\\
\vspace*{0.1in}
\hspace*{4.7in} |\vec{\triangle l}\rangle
\end{array}
\end{equation}
where, using $(\triangle \epsilon)^{(I)}_{J} :=\ (\delta^{I}_{J}-\delta^{I}_{J-1})$ and (\ref{eq:oct10****})
\begin{equation}
\begin{array}{lll}
\tilde{\triangle\epsilon}^{(I)}(n)\ =\ \frac{1}{A}\sum_{J=0}^{A-1} e^{ink_{J}}(\delta_{I,J}\ -\ \delta_{I,J-1})\\
\vspace*{0.1in}
\hspace*{0.5in} =\ \frac{1}{A}e^{ink_{I}}(1-e^{in\frac{2\pi}{A}})
\end{array}
\end{equation}
Whence, 
\begin{equation}\label{eq:jan15-1}
\begin{array}{lll}
\exp[-\frac{\hbar 2\pi L^{2}\epsilon^{2}}{a^{2}}\sum_{m>0}\frac{|\tilde{\triangle \epsilon}^{(I)}(m)|^{2}}{f(m)}|]=\\
\vspace*{0.1in}
\hspace*{0.6in} \exp[-\frac{\hbar \pi L^{2}\epsilon^{2}}{a^{2}}\sum_{m>0}|\frac{1}{A}\frac{e^{imk_{I}}(1-e^{im\frac{2\pi}{A}})|^{2}}{m}]\ =\ 
\exp[-\frac{\hbar \pi L^{2}\epsilon^{2}}{a^{2}}\sum_{m>0}\frac{1}{A^{2}}4\frac{\sin^{2}(\frac{m\pi}{A})}{m}]
\end{array}
\end{equation}

%Note that the sum
%\begin{equation}
%\begin{array}{lll}
%\sum_{m=1}^{A-1}\frac{1}{A^{2}}4\frac{sin^{2}(\frac{m\pi}{A})}{m}\ =\ \frac{\pi}{A}\frac{4}{A^{2}}\sum_{m=1}^{A-1} \frac{sin^{2}(\frac{m\pi}{A})}{\frac{m\pi}{A}}\\
%\vspace*{0.1in} 
%\hspace*{1.2in} \leq\  \frac{\pi}{A}\frac{4}{A^{2}}\sum_{m=1}^{A-1} (\frac{m\pi}{A})\\
%\vspace*{0.1in}
%\hspace*{1.2in} = \frac{\pi^{2}}{A^{2}}\frac{4}{A^{2}}\frac{A(A-1)}{2}\ <\ \frac{\pi^{2}}{A^{2}}\frac{4}{A^{2}}\frac{A^{2}}{2}
%\hspace*{1.8in} = \frac{2\pi^{2}}{A^{2}}
%\end{array}
%\end{equation}
Before proceeding, we state a lemma whose proof is provided in Appendix-C. This lemma makes the computations a lot easier.\\
Consider a disjoint union of the set of all matter charges into two mutually exclusive subsets, 
$S_{\mathbf{1}}(\epsilon, a) := \{\vec{\triangle l}|\frac{\epsilon^{2}}{a^{2}}\sum_{I}a(\triangle l_{I})^{2}\ <\ \frac{C}{a^{\delta}}\}$, 
$S_{\mathbf{2}}(\epsilon, a)\ =\ S_{\mathbf{1}}(\epsilon, a)^{c}$,\footnote{$S_{\mathbf{2}}(\epsilon, a)$ is the compliment of $S_{\mathbf{1}}(\epsilon, a)$ in the set of all 
$\vec{\triangle l}$.} where C is a  (dimensionful) constant which is positive and is independent of $\epsilon$ and $a$. 
We also choose $2 > \delta > 1$. 
%\footnote{As there is no mass-scale in
%the theory, there is no canonical choice for C. However we will see that all our results are independent of C.}
\footnote{As shown in appendix-A, choosing $\delta > 1$ ensures that contribution of 
$S_{\mathbf{2}}(\epsilon, a)$ to the norm of the state is finite and tends to zero as $\epsilon$ and a tend to zero.}\\
(\ref{eq:nov-14}) can be written in a schematic form as,
\begin{equation}\label{eq:nov-17*}
\begin{array}{lll}
\hat{a}_{n}^{poly}|\Psi^{+}\rangle\ =\ \frac{1}{2i\epsilon (2\pi L)}\sum_{\vec{\triangle l}}\sum_{I}e^{ink_{I}}
[e^{-\frac{i\hbar}{2}\alpha(\vec{\triangle\epsilon},\vec{\triangle l})}c(\vec{\triangle l}-\vec{\triangle \epsilon}^{(I)})\\
\vspace*{0.1in}
\hspace*{1.2in}-\ e^{+\frac{i\hbar}{2}\alpha(\vec{\triangle\epsilon},\vec{\triangle l})}c(\vec{\triangle l}+\vec{\triangle \epsilon}^{(I)})]|\vec{\triangle l}\rangle\\
\vspace*{0.1in}
=\ [\sum_{\vec{\triangle l}\in S_{\mathbf{1}}} + \sum_{\vec{\triangle l}\in S_{\mathbf{2}}}]
(\sum_{I}e^{ink_{I}}[e^{-\frac{i\hbar}{2}\alpha(\vec{\triangle\epsilon},\vec{\triangle l})}c(\vec{\triangle l}-\vec{\triangle \epsilon}^{(I)})\\
\vspace*{0.1in}
\hspace*{1.2in}-\ e^{+\frac{i\hbar}{2}\alpha(\vec{\triangle\epsilon},\vec{\triangle l})}c(\vec{\triangle l}+\vec{\triangle \epsilon}^{(I)})]|\vec{\triangle l}\rangle\\
\vspace*{0.1in}
=\ |\Psi_{1}\rangle + |\Psi_{2}\rangle
\end{array}
\end{equation}

\pagebreak 
\noindent{\bf Lemma}:Show that,
$||\Psi_{2}||\rightarrow\ 0$ rapidly as $a\rightarrow 0$

\noindent{\bf Proof}:\\
Given in Appendix C.
\\
\\
Whence it suffices to restrict the attention to $|\Psi_{1}\rangle$. Whence from now on, 
\emph{we will assume that we are summing over $\vec{\triangle l}\in\ S_{\mathbf{1}}(\epsilon, a)$.}\\
We will indicate this explicitly only where we think its necessary. 
%Now recall that we are summing over only those matter-charge configurations $\vec{\triangle l}$ which are such that in the limit $\epsilon$ and $a$ going to zero,\\
%({\bf a}) $\frac{\epsilon\triangle l_{I}}{a}$ remain finite $\forall\ I$.\\
%({\bf b}) $\triangle l_{I+1}\ -\ \triangle l_{I}\ =\ O(a)\ \forall\ I$.\\
Because $\frac{\epsilon\triangle l_{I}}{a}$ are bounded from above,  
to leading order in $\epsilon a$ we can expand the phase factors
$e^{\mp\frac{i\hbar\epsilon a}{2}(\frac{\epsilon\triangle l_{I}}{a}+\frac{\epsilon\triangle l_{I+1}}{a})}$ as
\begin{equation}\label{eq:dec25-1}
e^{\mp\frac{i\hbar\epsilon a}{2}(\frac{\epsilon\triangle l_{I}}{a}\ +\ \frac{\epsilon\triangle l_{I+1}}{a})}\ =\ 1\ \mp\ \frac{i\hbar\epsilon a}{2}[\ (\frac{\epsilon\triangle l_{I}}{a} + \frac{\epsilon\triangle l_{I+1}}{a}) + O(\epsilon^{2}a^{2})\ ]
\end{equation}
%Hence our definition of continuum limit requires that $\epsilon\rightarrow\ 0$ faster then a.\footnote{More precisely 
%while estimating sub-leading terms (contained in $|\delta\Psi\rangle$) in the appendix, we will see that $\epsilon\sim\ a^{\triangle}$ with $\triangle\ >\ 4$}\\

%\footnote{One way to ensure this is to take one parameter family of quantum theories via, 
%$\epsilon_{\lambda}=\frac{\epsilon_{0}}{\lambda^{M+1}}$, $a_{\lambda}=\frac{a_{0}}{\lambda}$, where $\epsilon_{0}$ and $a_{0}$ are fixed once and for all, $\lambda$ 
%is dimensionless and M is some large integer. continuum limit then corresponds to $\lambda\rightarrow\ \infty$.}\\ 
Substituting (\ref{eq:dec25-1}) in (\ref{eq:nov-14}), we get
\begin{equation}\label{eq:oct26-1}
\begin{array}{lll}
\hat{a}_{n}^{poly}|\Psi^{+}\rangle\ =\\
\vspace*{0.1in}
\frac{1}{2i\epsilon (2\pi L)}\sum_{I=0}^{A-1}e^{ink_{I}}\sum_{\vec{\triangle l}}\\ 
\vspace*{0.1in}
\hspace*{0.4in} c(\vec{\triangle l})\exp[-\frac{\hbar 2\pi L^{2}\epsilon^{2}}{a^{2}}\sum_{m>0}\frac{|\tilde{\triangle \epsilon}^{(I)}(m)|^{2}}{f(m)}|]\\
\vspace*{0.1in}
\hspace*{1.4in}\left[\left(1-i\frac{\hbar\epsilon a}{2}(\frac{\epsilon(\triangle l_{I}+\triangle l_{I+1})}{a})+ O(\epsilon^{2}a^{2})\right)
e^{\frac{\hbar\epsilon^{2}2\pi L^{2}}{a^{2}}\sum_{m}\frac{1}{f(m)}[\tilde{\triangle l}(m)\tilde{\triangle\epsilon^{(I)}}(-m)+c.c.]}\right.\\
\vspace*{0.1in}
\hspace*{1.4in}\left.- \left(1+i\frac{\hbar\epsilon a}{2}(\frac{\epsilon(\triangle l_{I}+\triangle l_{I+1})}{a})+ O(\epsilon^{2}a^{2})\right)
e^{-\frac{\hbar\epsilon^{2} 2\pi L^{2}}{a^{2}}\sum_{m}\frac{1}{f(m)}[\tilde{\triangle l}(m)\tilde{\triangle\epsilon^{(I)}}(-m)+c.c.]}\right]\\
\vspace*{0.1in}
\hspace*{4.7in}|\vec{\triangle l}\rangle
\end{array}
\end{equation}

We now show that the exponent $\frac{\hbar\epsilon^{2}2\pi L^{2}}{a^{2}}\sum_{m}\frac{1}{f(m)}[\tilde{\triangle l}(m)\tilde{\triangle\epsilon^{(I)}}(-m)+c.c.]$ is bounded.\\
Using Cauchy-Schwarz,

\begin{equation}
\begin{array}{lll}
(\frac{\hbar\epsilon^{2}2\pi L^{2}}{a^{2}})^{2}|\sum_{n}\frac{1}{f(n)}[\tilde{\triangle l}(n)\tilde{\triangle\epsilon^{(I)}}(-n)+c.c.]|^{2}\ \leq
[\frac{\hbar\epsilon^{2}2\pi L^{2}}{a^{2}}\sum_{n}\frac{1}{f(n)}|\tilde{\triangle l}(n)|^{2}]\\
\vspace*{0.1in}
\hspace*{3.0in}[\frac{\hbar\epsilon^{2}2\pi L^{2}}{a^{2}}\sum_{n}\frac{1}{f(n)}|\tilde{\triangle\epsilon}^{(I)}(n)|^{2}]\\
\vspace*{0.1in}
\hspace*{2.5in}= [\frac{\hbar\epsilon^{4}2\pi L^{2}}{a^{4}}\sum_{n}\frac{1}{f(n)}|\tilde{\triangle l}(n)|^{2}]\\
\vspace*{0.1in}
\hspace*{3.0in}[\hbar2\pi L^{2}\sum_{n}\frac{1}{f(n)}|\tilde{\triangle\epsilon}^{(I)}(n)|^{2}]
\end{array}
\end{equation}
Using, $f(n)\geq\ \frac{n}{2}\geq \frac{1}{2}\ \forall\ n$, $\sum_{n}|\tilde{\triangle l}(n)|^{2} = \frac{1}{A}\sum_{I}\triangle l_{I}^{2}$ and 
$\frac{\epsilon^{2}}{a^{2}}\triangle l_{I}^{2}\ <\ \frac{C}{a^{1+\delta}}$ we have,
\begin{equation}\label{eq:jan10-1}
\begin{array}{lll}
\frac{\hbar\epsilon^{4}2\pi L^{2}}{a^{4}}\sum_{n}\frac{1}{f(n)}|\tilde{\triangle l}(n)|^{2}<\\
\vspace*{0.1in}
\hspace*{1.5in}\frac{\hbar\epsilon^{4}2\pi L^{2}}{a^{4}}\left(\frac{2}{A}\frac{C}{a^{1+\delta}}\right)=\\
\vspace*{0.1in}
\hspace*{1.5in}\ (C\hbar L)\frac{\epsilon^{4}}{a^{4+\delta}}
\end{array}
\end{equation}
And the sum
\begin{equation}\label{eq:oct26-0}
\begin{array}{lll}
\sum_{m}\frac{1}{f(m)}|\tilde{\triangle\epsilon}^{(I)}(m)|^{2}\ =\\
\vspace*{0.1in} 
\hspace*{1.2in}\sum_{m=1}^{A-1}\frac{1}{A^{2}}4\frac{\sin^{2}(\frac{m\pi}{A})}{f(m)}\ =\ \sum_{m=1}^{A-1} \frac{\sin^{2}(\frac{m\pi}{A})}{\frac{m\pi}{A}}\\
\vspace*{0.1in} 
\hspace*{2.5in} \leq\  \sum_{m}\frac{2}{A\pi}(\frac{2m\pi}{A})\\
\vspace*{0.1in}
\hspace*{2.5in} = \frac{4}{A^{2}}\sum m\\
\vspace*{0.1in}
\hspace*{2.5in} \sim O(1)
\end{array}
\end{equation}
Together (\ref{eq:jan10-1}) and (\ref{eq:oct26-0}) imply that,
\begin{equation}
\begin{array}{lll}
(\frac{\hbar\epsilon^{2}2\pi L^{2}}{a^{2}})^{2}|\sum_{m}\frac{1}{f(m)}[\tilde{\triangle l}(m)\tilde{\triangle\epsilon^{(I)}}(-m)+c.c.]|^{2}\ \leq 
[(C\hbar L)\frac{\epsilon^{4}}{a^{4+\delta}}]\\
\vspace*{0.1in}
\hspace*{3.2in}[\hbar\pi L^{2}O(1)]
\end{array}
\end{equation}
As $1\ <\ \delta\ <\ 2$, $\epsilon\sim\ a^{\triangle}$ with $\triangle\ >\ 4$ means that
$e^{\pm\frac{\hbar\epsilon^{2}2\pi L^{2}}{a^{2}}\sum_{m}\frac{1}{f(m)}[\tilde{\triangle l}(m)\tilde{\triangle\epsilon^{(I)}}(-m)+c.c.]}$
can be expanded as a power-series in $\frac{\epsilon}{a}$. This observation simplifies (\ref{eq:oct26-1}) to,

\begin{equation}\label{eq:oct26-2}
\begin{array}{lll}
\hat{a}_{n}^{poly}|\Psi^{+}\rangle\ =\\
\vspace*{0.1in}
\frac{1}{2i\epsilon (2\pi L)}\sum_{I=0}^{A-1}e^{ink_{I}}\sum_{\vec{\triangle l}}\ 
c(\vec{\triangle l})\exp[-\frac{\hbar 2\pi L^{2}\epsilon^{2}}{a^{2}}\sum_{m>0}\frac{|\tilde{\triangle \epsilon}^{(I)}(m)|^{2}}{f(m)}|]\\
\vspace*{0.1in}
\hspace*{1.0in}\left[\left(1-i\frac{\hbar\epsilon a}{2}(\frac{\epsilon(\triangle l_{I}+ \triangle l_{I+1})}{a} + O(\epsilon^{2}a^{2})\right)\right.\\
\vspace*{0.1in}
\hspace*{1.6in}\left.\left(1 + \frac{\hbar\epsilon^{2} 2\pi L^{2}}{a^{2}}\sum_{m}\frac{1}{f(m)}[\tilde{\triangle l}(m)\tilde{\triangle\epsilon^{(I)}}(-m)+c.c.]+ 
O(\frac{\epsilon^{4}}{a^{4}})\right)\right]\\
\vspace*{0.1in}
\hspace*{1.0in}-\left[\left(1+i\frac{\hbar\epsilon a}{2}(\frac{\epsilon(\triangle l_{I}+\triangle l_{I+1})}{a})+ O(\epsilon^{2}a^{2})\right)\right.\\
\vspace*{0.1in}
\hspace*{1.6in}\left.\left(1 - \frac{\hbar\epsilon^{2} 2\pi L^{2}}{a^{2}}\sum_{m}\frac{1}{f(m)}[\tilde{\triangle l}(m)\tilde{\triangle\epsilon^{(I)}}(-m)+c.c.]
+ O(\frac{\epsilon}{a}^{4})\right)\right]\\
\vspace*{0.1in}
\hspace*{4.7in}|\vec{\triangle l}\rangle
\end{array}
\end{equation}
If we assume for a moment that all the sub-leading terms (that is terms which on using $\epsilon\sim\ a^{\triangle}$ vanish in the limit $a\rightarrow 0$) are 
bounded\footnote{Notice that this is a highly non-trivial assumption as although each term inside the sum over 
$\vec{\triangle l}$ is finite and vanishes in $a\rightarrow 0$ limit (with $\epsilon\sim\ a^{\triangle}$), it is not at all clear if after 
summing over all possible matter charges, such terms remain bounded. 
That this is the case is shown in Appendix-B.}, then the leading order terms in the above equation are given by,
\begin{equation}\label{eq:oct26-3}
\begin{array}{lll}
\hat{a}_{n}^{poly}|\Psi^{+}\rangle\ \approx\\
\vspace*{0.1in}
\frac{1}{2i\epsilon (2\pi L)}\sum_{I=0}^{A-1}e^{ink_{I}}\sum_{\vec{\triangle l}\in S_{\mathbf 1}(\epsilon, a)}\ 
c(\vec{\triangle l})\exp[-\frac{\hbar 2\pi L^{2}\epsilon^{2}}{a^{2}}\sum_{m>0}\frac{|\tilde{\triangle \epsilon}^{(I)}(m)|^{2}}{f(m)}|]\\
\vspace*{0.1in}
\hspace*{1.5in} \left(-i\hbar\epsilon a(\frac{\epsilon(\triangle l_{I}+\triangle l_{I+1})}{a})\ +\ 
\frac{2\hbar\epsilon^{2} 2\pi L^{2}}{a^{2}}\sum_{m}\frac{1}{f(m)}[\tilde{\triangle l}(m)\tilde{\triangle\epsilon^{(I)}}(-m)+c.c.]\right)\\
\vspace*{0.1in}
\hspace*{4.7in} |\vec{\triangle l}\rangle
\end{array}
\end{equation}
where $\tilde{\triangle\epsilon}^{(I)}(\pm m)\ =\ \mp\frac{2i}{A}e^{\pm imk_{I}^{+}}e^{\frac{\pm im\pi}{A}}\sin(\frac{m\pi}{A})$.\\
We use the following formulae in (\ref{eq:oct26-3}) 
\begin{equation}\label{eq:oct26-4}
\begin{array}{lll}
\frac{1}{2\pi L}\sum_{I}e^{in k_{I}}(\triangle l_{I} + \triangle l_{I+1})\ =\ \frac{1}{a}e^{-\frac{i\pi n}{A}}2\cos(\frac{2\pi n}{A})\tilde{\triangle l}(n)\\
\vspace*{0.1in}
\sum_{I}(a\ e^{ink_{I}})\tilde{\triangle\epsilon}^{(I)}(\pm m)\ =\ \sum_{I}\frac{\mp2i}{A}a\ e^{i(n\pm m)k_{I}}e^{\frac{\pm im\pi}{A}}\sin(\frac{m\pi}{A})\\
\hspace*{1.0in}\ =\ (\mp 2i)(a)\delta_{n,\mp m}e^{\frac{\pm im\pi}{A}}\sin(\frac{m\pi}{A})\\
\vspace*{0.1in}
\textrm{and}\\
\vspace*{0.1in}
\sum_{I}e^{ink_{I}}\sum_{m>0}\tilde{\triangle l}(-m)\tilde{\triangle\epsilon}^{(I)}(m)\ \alpha\ \sum_{m>0}\delta_{n,-m}\ =\ 0
\end{array}
\end{equation}
and obtain,
\begin{equation}\label{eq:oct26-5}
\begin{array}{lll}
\hat{a}_{n}^{poly}|\Psi^{+}\rangle\ \approx\\
\vspace*{0.1in}
\frac{1}{2i\epsilon a L}\sum_{\vec{\triangle l}\in S_{\mathbf 1}(\epsilon, a)}\ 
c(\vec{\triangle l})\exp[-\frac{\hbar 2\pi L^{2}\epsilon^{2}}{a^{2}}\sum_{m>0}\frac{|\tilde{\triangle \epsilon}^{(I)}(m)|^{2}}{f(m)}|]\\
\vspace*{0.1in}
\hspace*{1.2in}\left((-i\hbar\epsilon a)(A)e^{-\frac{i\pi n}{A}}(2\cos(\frac{2\pi n}{A})(\frac{\epsilon\tilde{\triangle l}(n)}{a})\right.\\
\vspace*{0.1in}
\hspace*{2.1in}\left.+ (4\pi iL)(\frac{\hbar\epsilon^{2}}{a})\tilde{\triangle l}(n)e^{-\frac{i\pi n}{A}}\cos(\frac{2\pi n}{A})\right)|\vec{\triangle l}\rangle
\end{array}
\end{equation}
Using $A = \frac{2\pi L}{a}$, it is easy to see that L.H.S. of (\ref{eq:oct26-5}) is exactly zero!\footnote{ From appendix-A $\Psi^{+}\rangle$ is normalizable. 
It is easy to see that $||\Psi^{+}||\ >\ 1$. Hence the vanishing of R.H.S. of (\ref{eq:oct26-5}) establishes vanishing of (\ref{eq:jan20-1}) to leading order.} 
Notice that the form of $f(m)\ =\ \frac{\tan(\frac{m\pi}{A})}{\frac{\pi}{A}}$ is crucial for the above result. 
Whence for long-wavelength modes (our assumption on $N_{0}$ is that it is much smaller then A) of the scalar field,
$|\Psi^{+}\rangle$  is a suitable approximant to Fock vacuum (for $\hat{a}_{n}^{poly}$ with $n\ >\ 0$) 
provided that all the terms that we have neglected above are really error-terms. That is, they are bounded and tend to 
zero in the continuum limit.\\
One can do exactly analogous analysis for the right-moving modes ($\hat{a}_{(-)n}^{poly}$) and obtain a corresponding approximant to the Fock vacuum
$|\Psi^{-}\rangle$. The error-terms which arise in the (+)-sector computation, are exactly the same error-terms that arise in the (-)-sector computation.
This implies that $|\Psi^{+}\rangle\ \otimes\ |\Psi^{-}\rangle$
is the discrete approximant to the Fock vacuum for all the long wavelength modes of the scalar field.

\subsection{The commutation relation condition}
In this section we show that:
\begin{equation}
\begin{array}{lll}
\frac{\langle\Psi^{+}|\ [\hat{a}_{n}^{\dagger\ poly}, \hat{a}_{m}^{poly}]\ |\Psi^{+}\rangle}{||\Psi^{+}||^{2}}\ = -i\hbar\{a_{n}^{*}, a_{m}\}\ +\ O(a)\\
\vspace*{0.1in}
\hspace*{1.2in}\ =\ \frac{\hbar}{\pi L^{2}}n\delta_{n,m} + O(a)
\end{array}
\end{equation}
where we indicate the error-terms by O(a).\\
Recall that,\\
\begin{equation}
e^{iO_{F_{I}^{+}}}|{\bf s}_{0}\rangle\ =\ e^{-i\frac{\hbar\epsilon^{2}}{2}(\triangle l_{I}+\triangle l_{I+1})}
|\vec{\triangle l} + \vec{\triangle\epsilon}^{(I)}\rangle
\end{equation}
where $|{\bf s}_{0}\rangle = |\gamma,\vec{k},\vec{\triangle l}{\bf )}\ =:\ |\vec{\triangle l}\rangle$ 
(Recall that we are supressing all the (+)-labels on the charge-networks), 
and $(\triangle\epsilon^{(I)})_{J}\ =\ (\delta_{I,J}\ -\ \delta_{I,J-1})$.\\
A straightforward computation reveals,
\begin{equation}\label{eq:jan9-2}
[e^{iO_{F_{I}^{+}}}, e^{iO_{F_{J}^{+}}}]|\vec{\triangle l}\rangle\ =\ (-2i)\sin(\frac{\hbar\epsilon^{2}}{2}[\delta_{J,I+1}-\delta_{J,I-1}])
e^{iO_{F_{I}^{+}+F_{J}^{+}}}|\vec{\triangle l}\rangle
\end{equation}

Using (\ref{eq:jan9-1}), (\ref{eq:jan9-2}) it is easy to see that,
\begin{equation}\label{eq:jan9-3}
\begin{array}{lll}
[\hat{a}_{n}^{\dagger\ poly}, \hat{a}_{m}^{poly}]|{\bf s}_{0}{\bf )}\ =\\
\vspace*{0.1in}
-\frac{1}{\epsilon^{2}4\pi L^{2}}\sum_{I,J}e^{-ink_{I}+imk_{J}}
\left([\ e^{iO_{F_{I}^{+}}}-e^{-iO_{F_{I}^{+}}},\ e^{iO_{F_{J}^{+}}}-e^{-iO_{F_{J}^{+}}}\ ]\right)|{\bf s}_{0}{\bf )}\\
\vspace*{0.1in}
\hspace*{0.8in} =\ -\frac{1}{\epsilon^{2}4\pi L^{2}}\sum_{I,J}e^{-ink_{I}+imk_{J}}\left((-2i)\sin(\frac{\hbar\epsilon^{2}}{2}[\delta_{J,I+1}-\delta_{J,I-1}])\right.\\
\vspace*{0.1in}
\hspace*{1.5in}\left.\left[e^{iO_{F_{I}^{+}+F_{J}^{+}}}\ +\  e^{iO_{F_{I}^{+}+F_{J}^{+}}}\ +\ e^{iO_{F_{I}^{+}+F_{J}^{+}}}\ +\  e^{iO_{F_{I}^{+}+F_{J}^{+}}}
\right]\right)|{\bf s}_{0}{\bf )}
\end{array}
\end{equation}
Let $\left[e^{iO_{F_{I}^{+}+F_{J}^{+}}}\ +\  e^{iO_{F_{I}^{+}+F_{J}^{+}}}\ +\ e^{iO_{F_{I}^{+}+F_{J}^{+}}}\ +\  e^{iO_{F_{I}^{+}+F_{J}^{+}}}\right]=:\ \hat{{\bf F}}(I,J)$.\\
Then (\ref{eq:jan9-3}) can be written as,
\begin{equation}\label{eq:jan9-4}
\begin{array}{lll}
[\hat{a}_{n}^{\dagger\ poly}, \hat{a}_{m}^{poly}]|{\bf s}_{0}{\bf )}\ =\\
\vspace*{0.1in}
\frac{1}{\epsilon^{2}4\pi L^{2}}\sum_{I,J}e^{-ink_{I}+imk_{J}}\left((-2i)\sin(\frac{\hbar\epsilon^{2}}{2}[\delta_{J,I+1}-\delta_{J,I-1}]){\bf 1}\right)|{\bf s}_{0}{\bf )}\\
\vspace*{0.1in}
\hspace*{0.2in}+\ \frac{1}{\epsilon^{2}4\pi L^{2}}\sum_{I,J}e^{-ink_{I}+imk_{J}}\left((-2i)\sin(\frac{\hbar\epsilon^{2}}{2}[\delta_{J,I+1}-\delta_{J,I-1}])
( \hat{{\bf F}}(I,J)-1)\right)|{\bf s}_{0}{\bf )}\\
\vspace*{0.1in}
= \left((-2i)\frac{1}{\epsilon^{2}4\pi L^{2}}\frac{\hbar\epsilon^{2}}{2}\sum_{I}e^{i(n-m)k_{I}}\left[\exp(\frac{2\pi im}{A})-\exp(-\frac{2\pi im}{A})\right]{\bf 1}
+ O(\epsilon^{2}){\bf 1}\right)|{\bf s}_{0}{\bf )}\\
\hspace*{0.2in}+\ \frac{1}{\epsilon^{2}4\pi L^{2}}\sum_{I,J}e^{-ink_{I}+imk_{J}}\left((-2i)\sin(\frac{\hbar\epsilon^{2}}{2}[\delta_{J,I+1}-\delta_{J,I-1}])
( \hat{{\bf F}}(I,J)-1)\right)|{\bf s}_{0}{\bf )}\\
\vspace*{0.1in}
= \left(4\frac{\hbar}{8\pi L^{2}}\sin(\frac{2\pi m}{A})A\delta_{n,m}{\bf 1} + O(\epsilon^{2}){\bf 1}\right)|{\bf s}_{0}{\bf )}\\
+\frac{1}{\epsilon^{2}4\pi L^{2}}\sum_{I,J}e^{-ink_{I}+imk_{J}}
\left((-2i)\sin(\frac{\hbar\epsilon^{2}}{2}[\delta_{J,I+1}-\delta_{J,I-1}])(\hat{{\bf F}}(I,J)-1)\right)|{\bf s}_{0}{\bf )}\\
\end{array}
\end{equation}
where ${\bf 1}$ is the identity operator.\\
As $m<< A$, 
\begin{equation}
\begin{array}{lll}
[\hat{a}_{n}^{\dagger\ poly}, \hat{a}_{m}^{poly}]|{\bf s}_{0}{\bf )}\ =\\
\vspace*{0.1in}
\frac{\hbar}{\pi L^{2}}n\delta_{n,m}{\bf 1}|{\bf s}_{0}{\bf )} + O(a){\bf 1}|{\bf s}_{0}{\bf )}\\
+\ \frac{1}{\epsilon^{2}4\pi L^{2}}\sum_{I,J}e^{-ink_{I}+imk_{J}}\left((-2i)\sin(\frac{\hbar\epsilon^{2}}{2}[\delta_{J,I+1}-\delta_{J,I-1}])
( \hat{{\bf F}}(I,J)-1)\right)|{\bf s}_{0}{\bf )}\\
\end{array}
\end{equation}
Whence,

\begin{equation}
\begin{array}{lll}
\frac{\langle\Psi^{+}\ [\hat{a}_{n}^{\dagger\ poly}, \hat{a}_{m}^{poly}]\ \Psi^{+}\rangle}{||\Psi^{+}||^{2}}=\\
\vspace*{0.1in}
\frac{\hbar}{\pi L^{2}}n\delta_{n,m} + O(a)\\
\hspace*{0.5in}+\ \frac{1}{\epsilon^{2}4\pi L^{2}}\sum_{I,J}e^{-ink_{I}+imk_{J}}(-2i)\left(\sin(\frac{\hbar\epsilon^{2}}{2}[\delta_{J,I+1}-\delta_{J,I-1}])
\frac{\langle\Psi^{+}( \hat{{\bf F}}(I,J)-1)\Psi^{+}\rangle}{||\Psi^{+}||^{2}}\right)\\
\end{array}
\end{equation}
Thus we need to show that\\
 $\frac{1}{\epsilon^{2}4\pi L^{2}}\sum_{I,J}e^{-ink_{I}+imk_{J}}\left((-2i)\sin(\frac{\hbar\epsilon^{2}}{2}[\delta_{J,I+1}-\delta_{J,I-1}])
\frac{\langle\Psi^{+}( \hat{{\bf F}}(I,J)-1)\Psi^{+}\rangle}{||\Psi^{+}||^{2}}\right)$ are finite and vanish in $a\rightarrow 0$ limit. 
Once again the various terms in the above expression can be
extimated using exactly the same techniques that are used in appendix-B. The upshot is that, 
two point functions of low frequency (left-moving) modes in the state $|\Psi^{+}\rangle$ approximate the corresponding Fock-space functions upto arbitrary accuracy.\\
Exactly analogous computation shows that two point functions of low frequency (right-moving) modes in the state $|\Psi^{-}\rangle$ 
are equal to the Fock-space functions.\\

%Once again, we approximate $a_{m}^{\dagger}$ by,
%\begin{equation}
%\hat{a}_{m}^{P,\dagger}\ =\ -\frac{1}{2i\epsilon L}\sum_{J}(e^{iO_{\epsilon_{J}}}\ -\ e^{-iO_{\epsilon, J}})e^{-imk_{J}}
%\end{equation}

The existence of state $|\Psi^{+}\rangle\otimes\Psi^{-}\rangle$ 
and its properties under the action of long-wavelengh mode operators shows us there is a continuum limit of the polymer quantized Parametrized field
theory, and this is the classical theory plus fluctuations given by the Fock vacuum.\\ 
It is also easy to lift the above constructions to the physical Hilbert space.\\
Define
\begin{equation}
|\Psi^{+}\rangle_{phy}\ :=\ \sum_{\vec{\triangle l}}c(\vec{\triangle l})\eta^{+}(|\vec{\triangle l}\rangle)
\end{equation}
Note that $|\Psi^{+}\rangle_{phy}\in\ {\cal H}^{+ ss}_{(0)phy}$.\\
As $\hat{a}_{n}^{poly}$ commute with the group averaging map we have,
\begin{equation}\label{eq:jan20-2}
\hat{a}_{n}^{poly '}|\Psi^{+}\rangle_{phy}\ =\ \sum_{\vec{\triangle l}}c(\vec{\triangle l})\eta^{+}(\hat{a}_{n}^{poly}|\vec{\triangle l}\rangle)
\end{equation}
Using (\ref{eq:jan20-2}) it is straight forward to show that all our results continue to hold in the physical Hilbert space ${\cal H}^{+ ss}_{(0)phy}$.

\section{Discussion and Open Issues}
A detailed summary of our results is available in Section 1; the interested
reader is invited to peruse that section again for an informed global 
perspective of our work in this paper. We proceed directly
to a discussion of our results in section 11.1 and of avenues for further
research in section 11.2\\

\subsection{Discussion of Results.}
\noindent
{\bf (i)} {\bf Role of the quantum constraints}: In the  classical theory the 
constraints ensure {\em foliation independence} of flat spacetime free scalar field dynamics \cite{karelcm}.
This, together with the 
interpretation of the embedding variables as spacetime coordinates, is responsible
for the emergence of a classical {\em spacetime} picture from the Hamiltonian 
theory in which explicit spacetime covariance is absent. 
Our treatment of the quantum 
constraints via Group Averaging together with the fact that the embedding coordinate operators
are well defined and admit a complete set of eigen states, leads us to ascribe exactly the same role to these
structures in quantum theory. From section 8.1 it follows that 
any charge network state ${\bf s}_0$ defines a `quantum slice' (consisting of the set of points 
$(X^+,X^-)= (k^+_e, k^-_e)$, one point for every edge $e$ of the charge network), that the 
1 parameter set of gauge related charge networks generated by the action of 
a 1 parameter set of gauge transformations on  ${\bf s}_0$
defines a 1 parameter set of discrete slices 
(a `quantum foliation') and 
that  the Group Averaging of the charge network defines a `quantum spacetime' 
(consisting of the set of embedding charges for
all the gauge related charge networks). 

The fact that the set of gauge related discrete slices
correspond to a single discrete {\em spacetime} is extremely non- trivial.
Note that  our treatment of the quantum constraints ensures that the group of finite gauge tranformations
is represented correctly. This is how the classical constraint algebra is encoded in quantum theory.
We believe that this faithful encoding of the group of finite gauge transformations is a key component in 
the emergence of a spacetime picture at the quantum level.\\

\noindent 
{\bf (ii) Scalar field dynamics on discrete spacetime}: 
%Let us restrict attention to the finest weave
%(see section 8.1) i.e. to states in ${\cal H}_{\phys}^{ss\pm}$ (see section 5).
While spacetime discreteness can be seen at the level of physical states in {\bf (i)} above,
the arguments of sections 4.2, 8.2 show that the polymer scalar field evolves on this discrete spacetime.

This may be seen as follows.
From section 4.2 it follows that only discrete Poincare translations are represented as well defined unitary
operators on the superselected physical states of interest. From section 8.2, it follows that classical
symplectic structure which emerges from the polymer quantization of the true degrees of freedom of PFT (which are
coordinatized by the Dirac Observables of section 4.1) corresponds to that of a lattice scalar field theory
\footnote{Note that while the original continuum symplectic structure is  faithfully represented by the basic 
{\em kinematic} operators of the polymer representation on ${\cal H}_{kin}$, this need not (and is not) true of
the Dirac observables of section 4.1 which are non-trivially constructed from the quantum kinematics.}.
Using the action of a  1 (integer- valued) parameter family  
discrete Poincare transformations of section 4.1 on any Dirac observable of section 8.2
yields a 1 (integer- valued) parameter family of ``evolving Dirac observables. It is then straightforward to see
that these evolving Dirac observables (which encode the dynamics of the true degrees of freedom)  evolve
on the underlying discrete spacetime. A particularly simple set of Dirac observables for which this can be checked, 
are those corresponding  to equation (\ref{latticescalarfield}).\\

\noindent
{\bf (iii) Fock like behaviour and the Continuum Limit}: As discussed in section  10.1, the polymer state 
(\ref{polyfockstate}) approximates the Fock vacuum at the level of coarse grained 2 point functions. The nature of the
approximation is endowed with precision due to the availability of the notion of a continuum limit. We think that  
two properties of the continuum limit are noteworthy in the context of semiclassical issues in LQG:\\
\noindent
(a)The limit has the crucial property of being {\em independent of} $\hbar$ thus allowing a seperation of the 
notions of {\em continuum} and {\em quantum}.\\
\noindent
(b) The limit is {\em not} a property of a 1 parameter set of ad- hoc triangulations in a {\em single} 
quantum theory; rather, it is a property of a 1 parameter family of unitarily inequivalent quantizations where
the parameter is anologous to the  Barbero- Immirizi parameter in LQG.\\

%\noindent{\bf (iv)}
%Semiclassicality:``fluxes'' {\em vs}  ``holonomies'': Our notion of semiclassicality for PFT is developed in terms of
%lattice approximants to continuum scalar field operators. Instead we could   base semiclassical 
%considerations directly on the operator correspondents of  Dirac observables which are
%exponentials of (coarsely smeared) scalar fields. The strategy would be to approximate the standard free scalar field
%Fock space Weyl algebra vacuum expectation values (for suitably coarsely smeared scalar field operators).
%We did try this and failed  to identify a suitable (1 parameter set of) polymer states. Indeed, the quantum 
%state (\ref{polyfockstate}) does not reproduce such ``Weyl algebra''  expectation values!
%The LQG analog of Weyl algebra elements are holonomies and those of the lattice scalar field approximants
%are magnetic flux approximants through plaquettes (see, for example, \cite{mejose,ttcoherent}). Thus it may be
%appropriate to code semiclassicality in terms of fluxes rather than holonomies in LQG (see \cite{mejose} for
%a discussion of this issue).\\

\noindent
{\bf (iv) The absence of ad- hoc triangulations}: Our construction of the 
physical state space and the action of Dirac observables thereon is free
of the ``triangulation'' type ambiguities which plague LQG. The only ambiguity
in our work is in the detailed choice of lattice approximants (\ref{latticescalarfield}) to the
true (scalar field) degrees of freedom. Since this arises only at the level of semiclassical
analysis, it is different from potential ambiguities in the very definition of the polymer 
quantum dynamics.
Indeed, the freedom of choice in lattice approximants is akin that encountered in 
elementary point particle quantum mechanics wherein semiclassical considerations are based
on the ``${\hat q}, {\hat p}$ '' operators rather than more complicated  functions of
${\hat q}, {\hat p}$.

\subsection{Directions for Further Research}
\noindent
{\bf (i) Thiemann Quantization and Spacetime Covariance}: This, to us, is the most interesting issue in that 
we believe the issue in polymer PFT closely mirrors that in LQG. The issue is as follows. As indicated in {\bf (i)}
of section 11.1 above, our treatment of the quantum constraints via Group Averaging implemented the correct
constraint algebra  and obtained spacetime covariance at the quantum level. The constraints (see section 2)
are density weight two constraints. An equivalent set of constraints exist at the classical level which 
are closer to those of classical Hamiltonian gravity. Specifically, the density weight 2 constraint, $C_{diff}$, 
obtained as the linear combination $H^{+}\ +\ H^{-}$ is given by,
\begin{equation}
C_{diff}(x)\ =\ \left[\Pi_{+}(x)X^{+'}(x)\ +\ \Pi_{-}(x)X^{-'}(x)\ +\ \pi_{f}(x)f^{'}(x)\right].
\end{equation}
It generates spatial diffeomorphisms of the Cauchy slice. The density weight one constraint, $C_{ham}$ obtained by rescaling a linear
combination,$H^{+}\ -\ H^{-}$, of the density weight 2 ones by the square root of the determinant of the induced spatial metric is given by
\begin{equation}
C_{ham}(x)\ =\ \frac{1}{\sqrt{X^{+'}(x)X^{-'}(x)}}\left[\Pi_{+}(x)X^{+'}(x)\ -\ \Pi_{-}(x)X^{-'}(x)\ +\ \frac{1}{2}(\pi_{f}^{2}+f^{' 2})\right].
\end{equation}
It generates evolution normal to the Cauchy slice. Further, the Poisson algebra generated by 
 $C_{diff}$ and $C_{ham}$ is the  Dirac algebra,
\begin{equation}\label{eq:structure}
\begin{array}{lll}
\{C_{diff}[\vec{N}],\ C_{diff}[\vec{M}]\}\ =\ C_{diff}[\vec{N},\vec{M}]\\
\vspace*{0.1in}
\{C_{diff}[\vec{N}],\ C_{ham}[M]\}\ =\ C_{ham}[{\it L}_{\vec{N}}M]\\
\vspace*{0.1in}
\{C_{ham}[N],\ C_{ham}[M]\}\ =\ C_{diff}[\vec{\beta(N,M)}]
\end{array}
\end{equation}
wherein the structure function $\beta^{a}(N,M):=\ q^{ab}(N\nabla_{b}M\ -\ M\nabla_{b}N)$ in (\ref{eq:structure}) is defined by the induced spatial metric $q_{ab}$. 
This is the exact analog 
of the constraint algebra of gravity. It is easy to see that spatial diffeomorphisms are unitarily 
implemented in the polymer kinematics developed here, that they can be solved by Group Averaging to yield 
spatial difffeomorphism invariant distributions  and that the physical states constructed by 
Group Averaging of the density weight 2 constraints are invariant under the (dual) action
of finite spatial diffeomorphisms. A key question  is: {\em Can we implement ${\hat C}_{ham}$ as an operator
on the space of spatially diffeomorphism invariant distributions in a manner similar to that developed
by Thiemmann \cite{thomashamconstraint} in LQG ?}. Preliminary work seems to indicate 
that the answer is in the affirmative, 
that the definition of ${\hat C}_{ham}$ suffers from regularization  ambiguities similar to the LQG ones, that 
the most straightforward regularizations do not admit the physical states constructed here as solutions 
and  that there is almost certainly a {\em non- trivial} regularization  which does. Clearly there are extremely
interesting open issues related to the relation between regularization ambiguities in the definition of
${\hat C}_{ham}$, the implementation of the correct constraint algebra and spacetime covariance at the quantum level.
We hope to confirm our preliminary findings and analyse such issues in a future publication.\\

\noindent
{\bf (ii) Semiclassical States}: Within the context of our work in section 10, we would like to see if the 
polymer state (\ref{polyfockstate}) yields coarse grained $n$- point functions for $n>2$. 
Distinct from this, our quantization offers a test bed for proposals to define semiclassical states (see for example
\cite{aashadow}) and, thus, avenues for more work.\\

\noindent
{\bf (iii) Non- compact spatial topology}: Generalization of our work here to the case of planar spacetime 
topology is an open problem of particular interest because of its relation to the Callen-Giddings-Harvey-Strominger
model of black holes \cite{cghs, mejoekarel, mecghs, meaavictor}. Preliminary work shows that the non-trivial asymptotic boundary conditions
can be implemented on a suitably generalised space of states; it would be of interest to compare our 
preliminary results with the Infinite Tensor Product arena proposed by Thiemann in \cite{ITP1}.
Our (preliminary) implementation of the asymptotic boundary conditions ensures that we admit 
asymptotically boosted families of Cauchy slices; this is distinct from the issue of unitary implementation of 
Global Lorentz transformations because the former are implemented in the sense of {\em gauge} transformations
of PFT, whereas the latter are symmetries of the classical theory of true degrees of freedom.\\

\noindent
{\bf (iv) Local Lorentz Invariance (LLI)}: This is an open issue and our comments here are 
of a slightly imprecise nature. Recall that, as mentioned in Section 1, 
the true degrees of freedom of classical PFT are exactly those of flat spacetime free scalar field theory.
Note that the compact spatial topology ensures, already at the classical level, 
that there is no global Lorentz symmetry;
there are only Poincare translations. However the classical theory of a free scalar field on the Minkowskian cylinder 
has local Lorentz invariance.
In contrast our quantization yields a field theory on (the light cone) lattice so that LLI is broken.
Note however that our choice of quantization splits the degrees of freedom in left and right movers,
and, as result there is no anomalous dispersion in the polymer dynamics. Our intuition is that 
`long time but low energy' phenomena (such as time of delay) do not magnify the effects of this broken LLI
by virtue of the unitary implementation of discrete Poincare translations. 
Thus we believe that only phenomena localised in spacetime at the lattice scale 
(i.e. effects of very high energy processes
as opposed to accumulated effects of low energy ones over large spacetime regions) would capture this 
propery of broken LLI in a noticeable way.
However an exhaustive effective
field theory analysis of our polymer quantization still needs to be performed and constitutes an open problem.
Note also that for the case of compact spatial toplogy studied here, the absence of Lorentz boosts implies that 
there is a unique global inertial rest frame
(Indeed, $X^{\pm}$ are coordinates in this rest frame). However for the non- compact case (see {\bf (iii)} above),
our preliminary work indicates that global Lorentz invariance would be broken; again, we do not yet know 
in which physical processes such a broken invariance would manifest.

\pagebreak

\section*{Appendix}
\section*{A. Normalizability of the candidate Fock vacuum}
In this appendix, we show that our candidate Fock vacuum state has a finite norm. Note that this statement is not at all obvious as one is summing over 
all possible matter charge configurations and each term in the sum is positive definite. 
As always, we only look at $|\Psi^{+}\rangle$, as $||\Psi^{+}||^{2} = ||\Psi^{-}||^{2}$. 
\begin{equation}
\vert \Psi^{+}\rangle\ =\ [\sum_{\vec{\triangle l}\in{\bf S}_{1}(\epsilon, a)} + \sum_{\vec{\triangle l}\in{\bf S}_{2}(\epsilon, a)}]c(\vec{\triangle l})|\vec{\triangle l}\rangle
\end{equation}
Where we have denoted the charge-network state $|\gamma, \vec{k},\vec{\triangle l}\rangle$ by $|\vec{\triangle l}\rangle$ 
\begin{equation}
\begin{array}{lll}
||\Psi^{+}||^{2}\ =\ \sum_{\vec{\triangle l}\in{\bf S}_{1}(\epsilon, a)}|c(\vec{\triangle l})|^{2}\ +\ \sum_{\vec{\triangle l}\in{\bf S}_{2}(\epsilon, a)}|
c(\vec{\triangle l})|^{2}\\
\vspace*{0.1in}
\hspace*{0.4in}=: ||\Psi_{1}||^{2} + ||\Psi_{2}||^{2}
\end{array}
\end{equation}
Recall that ${\bf S_{2}}(\epsilon, a)\ =\ \{\vec{\triangle l}|\frac{\epsilon^{2}}{a^{2}}\sum_{I}|\triangle l_{I}|^{2}a\ >\ \frac{C}{a^{\delta}}\}$ and 
${\bf S}_{1}=S\ -\ {\bf S}_{2}$ where S is the set of all possible matter-charge configurations. We choose $\delta > 1$ and C is a positive constant which is independent 
of $\epsilon$ and $a$.\\
(None of the results, here or elsewhere in the appendix depend on the choice of C).
\\
So to show that $|\Psi^{+}\rangle$ has a finite norm, it suffices to show that $||\Psi_{2}||^{2}\ <\ \infty$. We will first show that 
\begin{equation}\label{eq:dec26-1}
c(\vec{\triangle l})\leq \exp[-\frac{1}{A}\hbar\frac{\epsilon^{2}}{a^{2}}(2\pi L^{2})\sum_{n}|\tilde{\triangle l}(n)|^{2}]
\end{equation}
and then prove that 
\begin{equation}\label{eq:dec26-2}
\sum_{\vec{\triangle l}\in {\bf S}_{2}(\epsilon, a)}\exp[-\frac{1}{A}\hbar\frac{\epsilon^{2}}{a^{2}}(2\pi L^{2})\sum_{n}|\tilde{\triangle l}(n)|^{2}]\ <\ \infty
\end{equation}
To prove (\ref{eq:dec26-1}), notice that (for sufficiently large A)
\begin{equation}
f(n)=\frac{\tan(\frac{n\pi}{A})}{\frac{n\pi}{A}}\ \leq n_{max}=A
\end{equation}
Whence (\ref{eq:dec26-1}) follows.\\
%Before proving (\ref{eq:dec26-2}), let us ensure that $exp[-\frac{1}{A}\hbar\frac{\epsilon^{2}}{a^{2}}(2\pi L^{2})\sum_{n}|\tilde{\triangle l}(n)|^{2}]$ remains bounded as
%$\epsilon$, a tend to zero when $\vec{\triangle l}\in {\bf S}_{2}(\epsilon, a)$.\\
%Using (inverse) discrete Fourier transform, it is easy to see that
%\begin{equation}
%\sum_{n}|\tilde{\triangle l}(n)|^{2}\ =\ \frac{2\pi a}{L}\sum_{I}(\triangle l_{I} )^{2}
%\end{equation}
%Whence for $\vec{\triangle l}\in {\bf S}_{2}(\epsilon,a)$, this means that
%\begin{equation}
%\frac{\epsilon^{2}}{a^{2}}(2\pi L^{2})\sum_{n}|\tilde{\triangle l}(n)|^{2}\ >\ (4\pi^{2} L)\frac{1}{a^{\delta}}
%\end{equation}
%This in-turn implies that
%\begin{equation}
%c(\vec{\triangle l})\ <\ exp[-\hbar(2\pi)a^{1-\delta}
%\end{equation}
%Thus by choosing $\delta\ <\ 1$ we ensure that $c(\vec{\triangle l})\ \forall \vec{\triangle l}\in {\bf S}_{2}(\epsilon, a)$ remains bounded in the limit 
%$a \righarrow\ 0$.\\
Using, $\sqrt{\sum_{n}a_{n}^{2}}\leq\ \sum_{n}a_{n}$ if all $a_{n}$'s are positive,
\begin{equation}
\begin{array}{lll}
||\Psi_{2}||\ \leq\ \sum_{\vec{\triangle l}}c(\vec{\triangle l})\\
\vspace*{0.1in}
<\ \sum_{\vec{\triangle l}\in {\bf S}_{2}}\exp[-\frac{\hbar}{A}\frac{\epsilon^{2}}{a^{2}}(4\pi^{2} L)\sum_{I}a(\triangle l_{I})^{2}]
\end{array}
\end{equation}
 Now set $\sum_{I}(\triangle l_{I})^{2} = r^{2}$ and $\frac{\epsilon^{2}}{a^{2}}r^{2}=R^{2}$.\\
As $\frac{\epsilon^{2}}{a^{2}}\sum_{I}(\triangle l_{I})^{2} > \frac{C}{a^{1+\delta}}$, $R^{2}$ varies (almost) continuously as 
$\vec{\triangle l}\rightarrow\ \vec{\triangle l} + \vec{\triangle l}_{min}$. Whence the number of $vec{\triangle l}$ which lie between R and R + dR
is,
\begin{displaymath}
D^{A}\frac{R^{A-1}dR}{(\frac{\epsilon}{a})^{A}}
\end{displaymath}
Here D is a constant of O(1).\\
Thus
\begin{equation}
\begin{array}{lll}
\sum_{\vec{\triangle l}}c(\vec{\triangle l})\ <\ \sum_{\vec{\triangle l}\in {\bf S}_{2}}\exp[-\hbar 2\pi a^{2}\sum_{I}\frac{\epsilon^{2}}{a^{2}}\triangle l_{I}^{2}]\\
\vspace*{0.1in}
\approx\ \int_{R_{>}}^{\infty}D^{A}\frac{R^{A-1}dR}{(\frac{\epsilon}{a})^{A}}\exp[-\hbar(2\pi a^{2})R^{2}]\\
\end{array}
\end{equation}
where $R_{>}\ =\ \frac{C}{a^{1+\delta}}$.\\
Let $aR\ =\ X$. It is easy to see that,
\begin{equation}
\begin{array}{lll}
\sum_{\vec{\triangle l}\in {\bf S}_{2}}c(\vec{\triangle l})\ <\ (\frac{a}{\epsilon})^{A}\frac{D^{A}}{a^{A}}\int_{X_{>}}^{\infty}X^{A-1}dX \exp[-\hbar (2\pi X^{2})]\\
\vspace*{0.1in}
\hspace*{0.5in} < (\frac{D}{\epsilon})^{A}\ \int_{X_{>}}^{\infty}\exp[-\hbar (2\pi X^{2})]\ <\ (\frac{D}{\epsilon})^{A}\int_{X_{>}}^{\infty}2\ X\ dX\ \exp[-\hbar(2\pi X^{2})]\\
\vspace*{0.1in}
\hspace*{0.5in} =\ (\frac{D}{\epsilon})^{A}(\frac{1}{2\pi\hbar})\exp[-2\pi\hbar X_{>}^{2}]
\end{array}
\end{equation}
where $X_{>}^{2}\ =\ a^{2}R_{>}^{2}\ =\ \frac{1}{a^{\delta-1}}$. So finally we get the following bound on $||\Psi_{II}||$.
\begin{equation}
\begin{array}{lll}
||\Psi_{II}||\leq \sum_{\vec{\triangle l}\in{\bf S}_{2}}c(\vec{\triangle l})\ <\ (\frac{1}{2\pi\hbar})\exp[-\frac{2\pi\hbar}{a^{\delta-1}}\ +\ A\ln(\frac{D}{\epsilon})]
\end{array}
\end{equation}
As $\delta > 1$ and $\epsilon\sim\ a^{\triangle}$ with $\triangle\ >\ 4$ implies that the R.H.S of above equation is finite  and bounded from above 
$\forall\ \epsilon$ and a $\neq\ 0$. This proves normalizability. Infact it is easy to see that R.H.S tends to zero rapidly as $a\rightarrow\ 0$.
\pagebreak

\section*{B. Estimation of sub-leading terms}
In this section we derive bounds on the sub-leading terms in (\ref{eq:nov-14}), and show that these terms vanish in the limit $\epsilon$, a $\rightarrow$ 0.\\
Once again recall that (as shown in appendix-C) it suffices to restrict attention to $|\Psi_{1}\rangle\ =\ \sum_{\vec{\triangle l}\in {\bf S}_{1}(\epsilon, a)}
c(\vec{\triangle l})\hat{a}_{n}^{poly}|\vec{\triangle l}\rangle$.\\
\vspace*{0.1in}
Our strategy to show that the sub-leading terms vanish in the continuum limit will be as follows.\\
Notice that, $|\Psi_{1}\rangle$ can be succintly written in the following form. 
\begin{equation}\label{eq:nov17-2}
|\Psi_{1}\rangle\ =\ \exp[-\frac{\hbar\epsilon^{2}}{a^{2}}(2\pi L^{2})\sum_{m}\frac{|\tilde{\triangle\epsilon}^{(I)}(m)|^{2}}{f(m)}]
\sum_{j=1}^{N}\sum_{\vec{\triangle l}\in S_{1}(\epsilon a)}c(\triangle l)F_{j}(\triangle l)|\vec{\triangle l}\rangle
\end{equation}
where $\frac{|\tilde{\triangle\epsilon}^{(I)}(m)|^{2}}{f(m)} = \frac{\sin^{2}(\frac{m\pi}{A})}{\frac{m\pi}{A}}$  and is independent of I.\\
In the above equation N is the number of sub-leading O(a) terms  in (\ref{eq:nov-14}). (we showed in the main text that the leading order term in $\Psi_{1}$ is zero).\\ 
Whence
\begin{equation}
|\delta\Psi^{+}\rangle\ =\ \frac{|\Psi_{1}\rangle}{||\Psi^{+}||}
\end{equation}
Thus it is easy to see that,
\begin{equation}
||\delta\Psi^{+}||^{2}\ \leq \frac{1}{||\Psi^{+}||^{2}}
\exp[-2\frac{\hbar\epsilon^{2}}{a^{2}}(2\pi L^{2})\sum_{m}\frac{|\tilde{\triangle\epsilon}^{(I)}(m)|^{2}}{f(m)}]\sum_{\vec{\triangle l}\in S_{1}(\epsilon a)}c(\triangle l)^{2}[\sum_{j}|F_{j}(\triangle l)|]^{2} 
\end{equation}
\\
Thus if we can show that\\
\\ 
\noindent{\bf (i)} $\exp[-2\frac{\hbar\epsilon^{2}}{a^{2}}(2\pi L^{2})\sum_{m}\frac{|\tilde{\triangle\epsilon}^{(I)}(m)|^{2}}{f(m)}]$ is bounded
\footnote{This has already been shown in the main text in the equations (\ref{eq:jan15-1}) and (\ref{eq:oct26-0}).}  and\\ 
\noindent{\bf (ii)} $|F_{j}(\vec{\triangle l})|_{\vec{\triangle l}\in S_{1}}\leq F_{max}|\ \forall\ j$, %where $F_{max}\rightarrow 0$ in the continuum.\\ 
\\
Then we get the following upper bound on $||\delta\Psi^{+}||$. 
\begin{equation}\label{eq:jan10-5}
\begin{array}{lll}
||\delta\Psi^{+}||^{2}\\
\vspace*{0.1in}
\leq \frac{N}{||\Psi^{+}||^{2}}\exp\left[-2\frac{\hbar\epsilon^{2}}{a^{2}}(2\pi L^{2})\sum_{m}\frac{|\tilde{\triangle\epsilon}^{(I)}(m)|^{2}}{f(m)}\right]|F_{max}|^{2}
\sum_{\vec{\triangle l}\in S_{1}(\epsilon a)}c(\vec{\triangle l})^{2}\\
\vspace*{0.1in}
<\ N\exp\left[-2\frac{\hbar\epsilon^{2}}{a^{2}}(2\pi L^{2})\sum_{m}\frac{|\tilde{\triangle\epsilon}^{(I)}(m)|^{2}}{f(m)}\right]|F_{max}|^{2}
\end{array}
\end{equation}
where $\frac{\sum_{\vec{\triangle l}\in S_{1}(\epsilon a)}c(\vec{\triangle l})^{2}}{||\Psi||^{2}}\ <\ 1$ has been used, and N is defined in (\ref{eq:nov17-2}).\\
If we can also show that $F_{max}\rightarrow 0$ in the continuum, then (\ref{eq:jan10-5}) would imply that $lim_{a\rightarrow 0}||\delta\Psi^{+}||=0$.\\
Rest of this appendix is devoted to establishing $F_{max}$ and showing that $lim_{a\rightarrow 0}F_{max}\ =\ 0$.\\

%this in conjunction with the bound on $||\Psi_{2}||$ shows that $\frac{||\hat{a}_{n}|\Psi\rangle||}{||\Psi||}\rightarrow 0$.\\
%In this appendix we show that $F_{j}(\triangle l)$ in (\ref{eq:nov17-2}) are bounded from above by $F_{max}$ which in-turn vanishes as $a\rightarrow 0$.\\ 
Let us recap (\ref{eq:oct14-4}). 
\begin{equation}
\begin{array}{lll}
|\Psi_{1}\rangle\ =\\
\vspace*{0.1in}
\frac{1}{2i\epsilon L}\sum_{I}\left(e^{ink_{I}}\sum_{\vec{\triangle l}\in S_{1}}\right.\\
\vspace*{0.1in}
\hspace*{1.0in}\left.\left[e^{-\frac{i\hbar\epsilon^{2}}{2}(\triangle l_{I} + \triangle l_{I+1})}c(\vec{\triangle l}-\vec{\triangle\epsilon}^{(I)})\ -\ 
e^{\frac{i\hbar\epsilon^{2}}{2}(\triangle l_{I} + \triangle l_{I+1})}c(\vec{\triangle l}+\vec{\triangle\epsilon}^{(I)})\right]\right)\\
\vspace*{0.1in}
\hspace*{4.7in}|\vec{\triangle l}\rangle
\end{array}
\end{equation}
We extract out the sub-leading terms by re-expressing the above equation as,
\begin{equation}
\begin{array}{lll}
|\Psi_{1}\rangle\ = \\
\vspace*{0.1in}
\frac{1}{2i\epsilon L}\sum_{I}\left(e^{ink_{I}}\sum_{\vec{\triangle l}\in S_{1}}c(\vec{\triangle l})\right.
\exp[-\frac{\hbar\epsilon^{2}}{a^{2}}2\pi L^{2}\sum_{m}\frac{|\tilde{\triangle \epsilon}^{(I)}(m)|^{2}}{f(m)}]\\
\vspace*{0.1in}
\hspace*{0.4in}\left[(1-\frac{i\hbar\epsilon^{2}}{2}(\triangle l_{I} + \triangle l_{I+1})) + 
\left(e^{-\frac{i\hbar\epsilon^{2}}{2}(\triangle l_{I} + \triangle l_{I+1})}-(1-\frac{i\hbar\epsilon^{2}}{2}(\triangle l_{I} + \triangle l_{I+1}))\right)\right]\\
\vspace*{0.1in}
\hspace*{0.4in} \cdot\left[(1-\frac{\hbar\epsilon^{2}}{a^{2}}(2\pi L^{2})\sum_{m}\frac{[\tilde{\triangle l}(m)\tilde{\triangle\epsilon}^{(I)}(-m)+c.c.]}{f(m)})\right. +\\
\vspace*{0.1in} 
\hspace*{0.4in} \left.\left(\exp[-\frac{\hbar\epsilon^{2}}{a^{2}}(2\pi L^{2})\sum_{m}\frac{[\tilde{\triangle l}(m)\tilde{\triangle\epsilon}^{(I)}(-m)+c.c.]}{f(m)}]-
(1-\frac{\hbar\epsilon^{2}}{a^{2}}(2\pi L^{2})\sum_{m}\frac{[\tilde{\triangle l}(m)\tilde{\triangle\epsilon}^{(I)}(-m)+c.c.]}{f(m)})\right)\right]\\
\vspace*{0.1in}
\hspace*{0.4in}-\left[(1+\frac{i\hbar\epsilon^{2}}{2}(\triangle l_{I} + \triangle l_{I+1})) + 
\left(e^{+\frac{i\hbar\epsilon^{2}}{2}(\triangle l_{I} + \triangle l_{I+1})}-(1+\frac{i\hbar\epsilon^{2}}{2}(\triangle l_{I} + \triangle l_{I+1}))\right)\right]\\
\vspace*{0.1in}
\hspace*{0.4in}\cdot\left[(1+\frac{\hbar\epsilon^{2}}{a^{2}}(2\pi L^{2})\sum_{m}\frac{[\tilde{\triangle l}(m)\tilde{\triangle\epsilon}^{(I)}(-m)+c.c.]}{f(m)})\right. +\\
\vspace*{0.1in} 
\hspace*{0.4in}\left.\left.\left(\exp[+\frac{\hbar\epsilon^{2}}{a^{2}}(2\pi L^{2})\sum_{m}\frac{[\tilde{\triangle l}(m)\tilde{\triangle\epsilon}^{(I)}(-m)+c.c.]}{f(m)}]-
(1+\frac{\hbar\epsilon^{2}}{a^{2}}(2\pi L^{2})\sum_{m}\frac{[\tilde{\triangle l}(m)\tilde{\triangle\epsilon}^{(I)}(-m)+c.c.]}{f(m)}\right)\right]\right)\\
\vspace*{0.1in}
\hspace*{4.7in}|\vec{\triangle l}\rangle
\end{array}
\end{equation}
It is easy to show that if we write all the sub-leading terms in the form 
\begin{equation}
\exp[-\frac{\hbar\epsilon^{2}}{a^{2}}(2\pi L^{2})\sum_{m}\frac{|\tilde{\triangle\epsilon}^{(I)}(m)|^{2}}{f(m)}]\sum_{j=1}^{N}\sum_{\triangle l}
c(\vec{\triangle l})F_{j}(\vec{\triangle l})|\vec{\triangle l}\rangle
\end{equation}
then, the $F_{j}$'s are given by,
\begin{equation}
\begin{array}{lll}
F_{1}\ =\\
\vspace*{0.1in}
\frac{1}{2i\epsilon L}\sum_{I}e^{ink_{I}}\left[\left(e^{-\frac{i\hbar\epsilon^{2}}{2}(\triangle l_{I} + \triangle l_{I+1})}-
(1-\frac{i\hbar\epsilon^{2}}{2}(\triangle l_{I} + \triangle l_{I+1}))\right)\right]\\
\vspace*{0.1in}
F_{2}\ =\\
\vspace*{0.1in}
\frac{1}{2i\epsilon L}\sum_{I}e^{ink_{I}}\left(-\frac{\hbar\epsilon^{2}}{a^{2}}(2\pi L^{2})\sum_{m}\frac{[\tilde{\triangle l}(m)\tilde{\triangle\epsilon}^{(I)}(-m)+
c.c.]}{f(m)}\right)\\
\hspace*{1.7in}\cdot\left[e^{-\frac{i\hbar\epsilon^{2}}{2}(\triangle l_{I} + \triangle l_{I+1})}-(1-\frac{i\hbar\epsilon^{2}}{2}(\triangle l_{I} + \triangle l_{I+1}))\right]\\
\vspace*{0.1in}
F_{3}\ =\\
\vspace*{0.1in}
\frac{1}{2i\epsilon L}\sum_{I}e^{ink_{I}}
\left[\exp[-\frac{\hbar\epsilon^{2}}{a^{2}}(2\pi L^{2})\sum_{m}\frac{[\tilde{\triangle l}(m)\tilde{\triangle\epsilon}^{(I)}(-m)+c.c.]}{f(m)}]-\right.\\
\vspace*{0.1in}
\hspace*{2.7in}\left.\left(1-\frac{\hbar\epsilon^{2}}{a^{2}}(2\pi L^{2})\sum_{m}\frac{[\tilde{\triangle l}(m)\tilde{\triangle\epsilon}^{(I)}(-m)+c.c.]}{f(m)}\right)\right]\\
\vspace*{0.1in}
F_{4}\ =\\
\vspace*{0.1in}
\frac{1}{2i\epsilon L}\sum_{I}e^{ink_{I}}\left[\left(e^{-\frac{i\hbar\epsilon^{2}}{2}(\triangle l_{I} + \triangle l_{I+1})}-
(1-\frac{i\hbar\epsilon^{2}}{2}(\triangle l_{I} + \triangle l_{I+1}))\right)\right.\\
\vspace*{0.1in}
\hspace*{0.4in}\left.\cdot\left(\exp[-\frac{\hbar\epsilon^{2}}{a^{2}}(2\pi L^{2})\sum_{m}\frac{[\tilde{\triangle l}(m)\tilde{\triangle\epsilon}^{(I)}(-m)+c.c.]}{f(m)}]-
(1-\frac{\hbar\epsilon^{2}}{a^{2}}(2\pi L^{2})\sum_{m}\frac{[\tilde{\triangle l}(m)\tilde{\triangle\epsilon}^{(I)}(-m)+c.c.]}{f(m)})
\right)\right]\\
F_{5}\ =\\
\vspace*{0.1in}
\frac{1}{2i\epsilon L}\sum_{I}e^{ink_{I}}\left[(-\frac{i\hbar\epsilon^{2}}{2}(\triangle l_{I} + \triangle l_{I+1}))\right.\\
\vspace*{0.1in}
\hspace*{0.4in}\left.\cdot\left(\exp[-\frac{\hbar\epsilon^{2}}{a^{2}}(2\pi L^{2})\sum_{m}\frac{[\tilde{\triangle l}(m)\tilde{\triangle\epsilon}^{(I)}(-m)+c.c.]}{f(m)}]-
(1-\frac{\hbar\epsilon^{2}}{a^{2}}(2\pi L^{2})\sum_{m}\frac{[\tilde{\triangle l}(m)\tilde{\triangle\epsilon}^{(I)}(-m)+c.c.]}{f(m)})\right)\right]\\
\vspace*{0.1in}
%F_{5}\ =\\
%\vspace*{0.1in}
%\frac{1}{2i\epsilon L}\sum_{I}e^{ink_{I}}
%\left[\exp[-\frac{\hbar\epsilon^{2}}{a^{2}}(2\pi L^{2})\sum_{m}\frac{[\tilde{\triangle l}(m)\tilde{\triangle\epsilon}^{(I)}(-m)+c.c.]}{f(m)}]-
%\left(1-\frac{\hbar\epsilon^{2}}{a^{2}}(2\pi L^{2})\sum_{m}\frac{[\tilde{\triangle l}(m)\tilde{\triangle\epsilon}^{(I)}(-m)+c.c.]}{f(m)}\right)\right]\\
%\vspace*{0.1in}
F_{6}\ =\\
\vspace*{0.1in}
\frac{1}{2i\epsilon L}\sum_{I}e^{ink_{I}}\left[(-\frac{i\hbar\epsilon^{2}}{2}(\triangle l_{I} + \triangle l_{I+1}))\right.\\
\vspace*{0.1in}
\hspace*{0.8in}\left.\left(-\frac{\hbar\epsilon^{2}}{a^{2}}(2\pi L^{2})\sum_{m}\frac{[\tilde{\triangle l}(m)\tilde{\triangle\epsilon}^{(I)}(-m)+
c.c.]}{f(m)}\right)\right]
\end{array}
\end{equation}
There are 6 more terms obtained by replacing $\frac{-i\hbar\epsilon^{2}}{2}(\triangle l_{I} + \triangle l_{I+1})$ by $\frac{i\hbar\epsilon^{2}}{2}(\triangle l_{I} + \triangle l_{I+1})$
and $-\frac{\hbar\epsilon^{2}}{a^{2}}(2\pi L^{2})[...]$ by $\frac{\hbar\epsilon^{2}}{a^{2}}(2\pi L^{2})[...]$ in the above terms.\\
We now show that $|F_{j}|\ <\ F_{max}$ $\forall\ j$.
\pagebreak

\subsection*{Bound on $|F_{1}|$}

\begin{equation}
F_{1}(\vec{\triangle l})\ =\ \frac{1}{2i\epsilon L}\sum_{I}e^{ink_{I}}[(e^{-\frac{i\hbar\epsilon^{2}}{2}(\triangle l_{I} + \triangle l_{I+1})}-(1-\frac{i\hbar\epsilon^{2}}{2}(\triangle l_{I} + \triangle l_{I+1})))]\\
\end{equation}
\begin{equation}\label{eq:nov23-4}
\begin{array}{lll}
|F_{1}|\leq \frac{1}{2\epsilon L}\sum_{I}\sum_{n=2}^{\infty}|\frac{[\frac{\hbar\epsilon^{2}}{2}(\triangle l_{I} + \triangle l_{I+1})]^{n}}{n!}|\\
\vspace*{0.2in}
\hspace*{0.3in} <\ \frac{1}{2\epsilon L}\sum_{I}(\frac{\hbar\epsilon^{2}}{2}(\triangle l_{I} + \triangle l_{I+1}))^{2}e^{\frac{\hbar\epsilon^{2}}{2}(\triangle l_{I} + \triangle l_{I+1})}\\
\vspace*{0.2in}
\hspace*{0.3in} <\ \frac{2}{2\epsilon L}\sum_{I}(\frac{\hbar\epsilon^{2}}{2}(\triangle l_{I} + \triangle l_{I+1}))^{2}
\end{array}
\end{equation}
Where we have used the a simple inequality, $\alpha^{2}e^{\alpha}\ <\ 2\alpha^{2}$ if $\alpha << 1$.\\
The exponent in our case is  $\frac{\hbar\epsilon^{2}}{2}(\triangle l_{I} + \triangle l_{I+1})$. That it is arbitrarily small can be understood by
estimating $\epsilon^{2}(\triangle l_{I})$ for all $\vec{\triangle l}\in {\bf S}_{1}$.
\begin{equation}
\begin{array}{lll}
\frac{\epsilon}{a}(\triangle l_{I})\ <\ \frac{C}{a^{\frac{1+\delta}{2}}}\\
\vspace*{0.2in}
\Rightarrow \epsilon(\triangle l_{I})\ <\ \frac{C}{a^{\frac{\delta-1}{2}}}\\
\vspace*{0.2in}
\Rightarrow \epsilon^{2}(\triangle l_{I})\ <\ \frac{C\epsilon}{a^{\frac{\delta-1}{2}}}
\end{array}
\end{equation}
As $\epsilon\ =\ C_{0}a^{\triangle}$ with $\triangle\ > 4$ (and $C_{0}$ being a finite dimensionful constant that we fix once and for all) 
implies that $\epsilon^{2}(\triangle l_{I})\rightarrow\ 0$ 
%if we choose $\epsilon\ >\ C\ a^{\frac{\delta-1}{2}}$ for some finite(dimensionful) constant C. Ofcourse this means $\epsilon$ 
%should go to zero faster then a.\\
Let us go back to (\ref{eq:nov23-4}). Using a simple inequality $\sum_{I}(\triangle l_{I} + \triangle l_{I+1})^{2} < 4\sum_{I}(\triangle l_{I})^{2}$ we get,
\begin{equation}\label{eq:nov23-6}
\begin{array}{lll}
|F_{1}|<\ \frac{8}{2\epsilon L}(\frac{\hbar\epsilon^{2}}{2})^{2}\sum_{I}(\triangle l_{I})^{2}\\
\vspace*{0.2in}
\hspace*{0.4in}=\ \hbar^{2}\frac{4\epsilon a^{2}}{L}\sum_{I}(\frac{\epsilon^{2}}{a^{2}}(\triangle l_{I})^{2})\\
\vspace*{0.2in}
\hspace*{0.4in}<\ \hbar^{2}\frac{4\epsilon a^{2}}{L}\frac{1}{a^{1+\delta}}
\end{array}
\end{equation}
which tends to zero as $\frac{\epsilon}{a^{\delta-1}} = C_{0}a^{\triangle + 1 - \delta}$.

\pagebreak
\subsection*{Bound on $|F_{2}|$}

\begin{equation}
\begin{array}{lll}
F_{2}\ =\ \frac{1}{2i\epsilon L}\sum_{I}e^{ink_{I}}(-\frac{\hbar\epsilon^{2}}{a^{2}}(2\pi L^{2})\sum_{m}\frac{[\tilde{\triangle l}(m)\tilde{\triangle\epsilon}^{(I)}(-m)+c.c.]}{f(m)}\\
\hspace*{0.7in} [(e^{-\frac{i\hbar\epsilon^{2}}{2}(\triangle l_{I} + \triangle l_{I+1})}-(1-\frac{i\hbar\epsilon^{2}}{2}(\triangle l_{I} + \triangle l_{I+1})))]\\
\end{array}
\end{equation}
\\
Following estimate will be essential for putting bounds on $|F_{2}|$
\\
\begin{equation}\label{eq:nov23-5}
\hbar\frac{\epsilon^{2}}{a^{2}}(2\pi L^{2})\sum_{m}[\frac{[\tilde{\triangle l}(m)\tilde{\triangle\epsilon}^{(I)}(-m)+c.c.]}{f(m)}]\leq\ \hbar\frac{\epsilon^{2}}{a^{2}}(2\pi L^{2})\sqrt{\sum_{m}|\tilde{\triangle\epsilon}^{(I)}(m)|^{2}}\sqrt{\sum_{m}\frac{|\tilde{\triangle l}(m)|^{2}}{f(m)}}\\
\end{equation}

It is easy to show that 
\noindent{\bf (i)} $\sum_{m}|\tilde{\triangle\epsilon}^{(I)}(m)|^{2}\ =\ 2a$.\\
And using $f(m)\ >\ \frac{1}{2}\ \forall\ m$, we also have\\
\noindent{\bf (ii)}
\begin{equation}
\begin{array}{lll}
\frac{\epsilon^{2}}{a^{2}}\sum_{m}\frac{|\tilde{\triangle l}(m)|^{2}}{f(m)}\ <\ 2\frac{\epsilon^{2}}{a^{2}}\sum_{m}|\tilde{\triangle l}(m)|^{2}\\
\vspace*{0.2in}
\hspace*{1.0in} = 2\frac{\epsilon^{2}}{a^{2}}a\sum_{J}(\triangle l)_{J}^{2}\\
\vspace*{0.2in}
\hspace*{1.0in} < 2\frac{1}{a^{\delta}}
\end{array}
\end{equation}
Using (\noindent{\bf (i)} and \noindent{\bf (ii)} in (\ref{eq:nov23-5}),
\begin{equation}\label{eq:nov23-7}
\begin{array}{lll}
\hbar\frac{\epsilon^{2}}{a^{2}}(2\pi L^{2})\sum_{m}[\frac{[\tilde{\triangle l}(m)\tilde{\triangle\epsilon}^{(I)}(-m)+c.c.]}{f(m)}]<\ \hbar\frac{\epsilon}{a}(2\pi L^{2})\sqrt{2a}\sqrt{2}\frac{1}{a^{\frac{\delta}{2}}}\\
\vspace*{0.2in}
\hspace*{2.2in}=\ \hbar(2\pi L^{2})\sqrt{2}\frac{\epsilon}{a^{\frac{\delta+1}{2}}}
\end{array}
\end{equation}

Now it is straightforward to put an upper bound on $|F_{2}|$,

\begin{equation}
\begin{array}{lll}
|F_{2}|\ <\ \frac{1}{2\epsilon L}\sum_{I}|\frac{\hbar\epsilon^{2}}{a^{2}}(2\pi L^{2})\sum_{m}\frac{[\tilde{\triangle l}(m)\tilde{\triangle\epsilon}^{(I)}(-m)+c.c.]}{f(m)}|\\
\hspace*{0.7in} 2|\frac{\hbar\epsilon^{2}}{2}(\triangle l_{I} + \triangle l_{I+1})|^{2}\\
\vspace*{0.2in}
\hspace*{0.2in}<\ \hbar(2\pi L^{2})\sqrt{2}\frac{\epsilon}{a^{\frac{\delta+1}{2}}}\frac{1}{2\epsilon L}\sum_{I}2|\frac{\hbar\epsilon^{2}}{2}(\triangle l_{I} + \triangle l_{I+1})|^{2}\\
\vspace*{0.1in}
\hspace*{0.2in}<\  \hbar(2\pi L^{2})\sqrt{2}\frac{\epsilon}{a^{\frac{\delta+1}{2}}}\frac{1}{2\epsilon L}\hbar^{2}\epsilon^{2}a^{2}\frac{1}{a^{1+\delta}}\\
\vspace*{0.1in}
\hspace*{0.2in}=\ \hbar^{3}(\pi L)\sqrt{2}\epsilon^{2}\frac{1}{a^{\frac{-3+\delta}{2}}}
\end{array}
\end{equation}
which clearly converges to zero at the rate $a^{\triangle + 3 -\delta}$.
%will converge to zero if we choose $\epsilon\ =\ C\ a^{\triangle}$ with $\triangle\ >\ \frac{\delta-3}{4}$.
\pagebreak

\subsection*{Bound on $|F_{3}|$}
\begin{equation}
\begin{array}{lll}
F_{3}\ =\\
\vspace*{0.1in}
\frac{1}{2i\epsilon L}\sum_{I}e^{ink_{I}}\\
\vspace*{0.1in}
\hspace*{0.5in}\left[\exp[-\frac{\hbar\epsilon^{2}}{a^{2}}(2\pi L^{2})\sum_{m}\frac{[\tilde{\triangle l}(m)\tilde{\triangle\epsilon}^{(I)}(-m)+c.c.]}{f(m)}]-
\left(1-\frac{\hbar\epsilon^{2}}{a^{2}}(2\pi L^{2})\sum_{m}\frac{[\tilde{\triangle l}(m)\tilde{\triangle\epsilon}^{(I)}(-m)+c.c.]}{f(m)}\right)\right]
\end{array}
\end{equation}
Let 
\begin{equation}
\alpha_{I}(\vec{\triangle l})\ :=\ \frac{\hbar\epsilon^{2}}{a^{2}}(2\pi L^{2})\sum_{m}\frac{[\tilde{\triangle l}(m)\tilde{\triangle\epsilon}^{(I)}(-m)+c.c.]}{f(m)}
\end{equation}
Applying  Cauchy Schwarz, 
\begin{equation}\label{eq:jan11-6}
\begin{array}{lll}
|\alpha_{I}(\vec{\triangle l})|\ <\ 2 \frac{\hbar\epsilon^{2}}{a^{2}}(2\pi L^{2})\sqrt{\sum_{m}|\tilde{\triangle\epsilon}^{(I)}(m)|^{2}}
\sqrt{\sum_{m}\frac{|\tilde{\triangle l}(m)|^{2}}{f(m)}}\\
\vspace*{0.1in}
\hspace*{1.0in}= 2 \frac{\hbar\epsilon}{a}(2\pi L^{2})\sqrt{\sum_{m}|\tilde{\triangle\epsilon}^{(I)}(m)|^{2}}
\sqrt{\frac{\epsilon^{2}}{a^{2}}\sum_{m}\frac{|\tilde{\triangle l}(m)|^{2}}{f(m)}}\
\end{array}
\end{equation}
Notice that
\begin{equation}\label{eq:jan11-6*}
|\tilde{\triangle\epsilon}^{(I)}(m)|^{2}\ =\ 2a\\
\end{equation}
and,
\begin{equation}\label{eq:jan11-6**}
\begin{array}{lll}
\frac{\epsilon^{2}}{a^{2}}\sum_{m}\frac{|\tilde{\triangle l}(m)|^{2}}{f(m)}\ <\ 2\frac{\epsilon^{2}}{a^{2}}\sum_{m}|\tilde{\triangle l}(m)|^{2}\\
\vspace*{0.1in}
\hspace*{1.0in}\ = 2a\sum_{I}\frac{\epsilon^{2}}{a^{2}}(\triangle l_{I})^{2}\\
\vspace*{0.1in}
\hspace*{1.0in}\ <\ 2a\frac{C}{a^{1+\delta}}
\end{array}
\end{equation}
where in the first line we have used $f(m) > \frac{1}{2}\ \forall\ m$, in the second line, the forumla for Fourier transform and in the third line, 
we have used the fact that $\vec{\triangle l}\in {\bf S}_{1}(\epsilon, a)$.\\
Using  (\ref{eq:jan11-6*}), (\ref{eq:jan11-6**}) in (\ref{eq:jan11-6}) we get,
\begin{equation}\label{eq:jan11-9}
\begin{array}{lll}
|\alpha_{I}(\vec{\triangle l})|\ <\ 2 \frac{\hbar\epsilon}{a}(2\pi L^{2})(2a)\frac{\sqrt{C}}{a^{\frac{1+\delta}{2}}}\\
\vspace*{0.1in}
\hspace*{1.0in} = (8\pi L^{2})\frac{\sqrt{C}\hbar\epsilon}{a^{\frac{1+\delta}{2}}}
\end{array}
\end{equation}
Once again we use the same trick we used in (\ref{eq:nov23-4}) to conclude that for sufficiently small $|\alpha_{I}(\vec{\triangle l})|$, 
\begin{displaymath}
|e^{-\alpha_{I}}-(1-\alpha_{I})|\ <\ 2|\alpha_{I}|^{2}
\end{displaymath}
Whence,
\begin{equation}
\begin{array}{lll}
|F_{3}|\ <\ \frac{\pi}{\epsilon a}2\ (8\pi L^{2})^{2}\frac{C\hbar^{2}\epsilon^{2}}{a^{1+\delta}}\\
\vspace*{0.1in}
\hspace*{1.0in} =\ C\hbar^{2}\pi(8\pi L^{2})^{2}\frac{\epsilon}{a^{2+\delta}}
\end{array}
\end{equation}
as $2 > \delta > 1$ and $\epsilon\sim a^{\triangle}$ with $\triangle\ >\ 4$ implies that $|F_{3}|$ is bounded from above and tends to zero as a tends to zero.

\subsection*{Bound on $|F_{4}|$}
\begin{equation}
\begin{array}{lll}
F_{4}\ =\ \frac{1}{2i\epsilon L}\sum_{I}e^{ink_{I}}[(e^{-\frac{i\hbar\epsilon^{2}}{2}(\triangle l_{I} + \triangle l_{I+1})}-(1-\frac{i\hbar\epsilon^{2}}{2}(\triangle l_{I} + \triangle l_{I+1})))\\
\vspace*{0.1in}
\hspace*{0.4in} (\exp[-\frac{\hbar\epsilon^{2}}{a^{2}}(2\pi L^{2})\sum_{m}\frac{[\tilde{\triangle l}(m)\tilde{\triangle\epsilon}^{(I)}(-m)+c.c.]}{f(m)}]-(1-\frac{\hbar\epsilon^{2}}{a^{2}}(2\pi L^{2})\sum_{m}\frac{[\tilde{\triangle l}(m)\tilde{\triangle\epsilon}^{(I)}(-m)+c.c.]}{f(m)})
)]
\end{array}
\end{equation}
Let $\beta_{I}\ =\ \frac{\hbar\epsilon^{2}}{2}(\triangle l_{I} + \triangle l_{I+1})$, $\omega_{I}\ =\ \frac{\hbar\epsilon^{2}}{a^{2}}(2\pi L^{2})\sum_{m}\frac{[\tilde{\triangle l}(m)\tilde{\triangle\epsilon}^{(I)}(-m)+c.c.]}{f(m)}]$,
then using the identity that we have used many times so far,
\begin{equation}
|e^{if}-(1-if)|\ <\ 2f^{2}\ \forall f << 1
\end{equation}
we get,
\begin{equation}
\begin{array}{lll}
|F_{4}|\ <\ \frac{4}{2\epsilon L}\sum_{I}\alpha_{I}^{2}\beta_{I}^{2}
\end{array}
\end{equation}
Whence 
\begin{equation}
|F_{4}|\ <\ \frac{4}{2\epsilon L}(\alpha_{I}^{2})_{max}(\beta_{I}^{2})_{max}A
\end{equation}

Once again using $\sum_{I}(\triangle l_{I} + \triangle l_{I+1})^{2} < 4\sum_{I}(\triangle l_{I})^{2}$ and the fact that 
$\frac{\epsilon^{2}}{a^{2}}(\sum_{I}\triangle l_{I}^{2})_{max} = \frac{C}{a^{1+\delta}}$ we have,

\begin{equation}
\begin{array}{lll}
(\beta_{I}^{2})_{max}\ <\ \frac{\hbar^{2}\epsilon^{4}}{4}4\frac{a^{2}}{\epsilon^{2}}\frac{C}{a^{1+\delta}}\\
\vspace*{0.1in}
\hspace*{1.0in}=\ C\hbar^{2}\frac{\epsilon^{2}}{a^{\delta-1}}
\end{array}
\end{equation}
and from (\ref{eq:jan11-9}),
\begin{equation}
(\alpha_{I}^{2})_{max}\ <\ C\hbar^{2}(8\pi L^{2})^{2}\frac{\epsilon^{2}}{a^{1+\delta}}
\end{equation}
Thus 
\begin{equation}
|F_{4}|\ <\ C^{2}\hbar^{4}4\pi(8\pi L^{2})^{2}\frac{\epsilon^{3}}{a^{2\delta+1}}
\end{equation}
Once again with $\triangle > 4$, $|F_{4}|$ tends to zero as $a\rightarrow\ 0$.

\pagebreak

\subsection*{Bound on $|F_{5}|$}
\begin{equation}
\begin{array}{lll}
F_{5}\ =\\
\vspace*{0.1in}
\frac{1}{2i\epsilon L}\sum_{I}e^{ink_{I}}\left[(-\frac{i\hbar\epsilon^{2}}{2}(\triangle l_{I} + \triangle l_{I+1}))\right.\\
\vspace*{0.1in}
\hspace*{0.8in}\left.\cdot\left(\exp[-\frac{\hbar\epsilon^{2}}{a^{2}}(2\pi L^{2})\sum_{m}\frac{[\tilde{\triangle l}(m)\tilde{\triangle\epsilon}^{(I)}(-m)+c.c.]}{f(m)}]-
(1-\frac{\hbar\epsilon^{2}}{a^{2}}(2\pi L^{2})\sum_{m}\frac{[\tilde{\triangle l}(m)\tilde{\triangle\epsilon}^{(I)}(-m)+c.c.]}{f(m)})\right)\right]\\
\end{array}
\end{equation}
Using same techniques that we have used for the terms above, it is easy to see that 
\begin{equation}
\begin{array}{lll}
|F_{5}|\ <\\
\vspace*{0.1in}
\hspace*{0.5in}2\frac{\pi}{\epsilon a}(\hbar\epsilon^{2})(\triangle l_{I})_{max}(\alpha_{I}^{2})_{max}\\
\vspace*{0.1in}
\hspace*{0.5in}<\ \frac{\pi}{\epsilon a}(\hbar\epsilon^{2})\frac{1}{\epsilon}\frac{1}{a^{\frac{\delta-1}{2}}}C\hbar^{2}(8\pi L^{2})^{2}\frac{\epsilon^{2}}{a^{1+\delta}}\\
\vspace*{0.1in}
\hspace*{0.5in}=\ C\hbar^{2}(8\pi L^{2})^{2}\pi\frac{\epsilon^{2}}{a^{\frac{3}{2}(\delta+1)}}\\
\vspace*{0.1in}
\hspace*{0.5in}=\ C\hbar^{2}(8\pi L^{2})^{2}\pi\ C_{0}\ a^{\left(2\triangle -\ \frac{3}{2}(\delta+1)\right)} 
\end{array}
\end{equation}
as $\triangle > 4$, $|F_{5}|\rightarrow\ 0$.

\subsection*{Bound on $|F_{6}|$}
$|F_{6}|$ is straightforward to estimate in light of the previous computations.
\begin{equation}
\begin{array}{lll}
F_{6}\ =\\
\vspace*{0.1in}
\frac{1}{2i\epsilon L}\sum_{I}e^{ink_{I}}\left[(-\frac{i\hbar\epsilon^{2}}{2}(\triangle l_{I} + \triangle l_{I+1}))\right.\\
\vspace*{0.1in}
\hspace*{0.8in}\left.\left(-\frac{\hbar\epsilon^{2}}{a^{2}}(2\pi L^{2})\sum_{m}\frac{[\tilde{\triangle l}(m)\tilde{\triangle\epsilon}^{(I)}(-m)+
c.c.]}{f(m)}\right)\right]
\end{array}
\end{equation}
We supress all the details. Interested reader can easily re-trace the missing steps.
\begin{equation}
\begin{array}{lll}
|F_{6}| < \frac{\pi}{\epsilon a}\hbar\epsilon^{2}(\triangle l_{I})_{max}\frac{\hbar\epsilon^{2}}{a^{2}}(2\pi L^{2})|\alpha_{I}|_{max}\\
\vspace*{0.2in}
\hspace*{1.0in}<\frac{\pi}{\epsilon a}(\hbar\epsilon^{2})\frac{1}{\epsilon}\frac{1}{a^{\frac{\delta-1}{2}}}\sqrt{C}\hbar\ (8\pi L^{2})\frac{\epsilon}{a^{\frac{1+\delta}{2}}}\\
\vspace*{0.2in}
\hspace*{1.0in}=\ \pi\sqrt{C}\hbar^{2}\ (8\pi L^{2})\frac{\epsilon}{a^{1+\delta}}
\end{array}
\end{equation}
Once again it is easy to see that with $\epsilon=C_{0}a^{\triangle}$,    $|F_{6}|\rightarrow\ 0$.\\

The remaining 6 terms, that are obtained by replacing $\frac{-i\hbar\epsilon^{2}}{2}(\triangle l_{I} + \triangle l_{I+1})$ by $\frac{i\hbar\epsilon^{2}}{2}(\triangle l_{I} + \triangle l_{I+1})$
and $-\frac{\hbar\epsilon^{2}}{a^{2}}(2\pi L^{2})[...]$ by $\frac{\hbar\epsilon^{2}}{a^{2}}(2\pi L^{2})[...]$ in the above terms can be estimated in exactly analogous ways.\\
All $|F_{j}|$'s are dimensionless quantities, and in the limit $a\rightarrow\ 0$ they all tend to zero. Whence we can assume an existence of $|F_{max}|$ which tends to
zero in the continuum. This proves the assertion we stated in the beginning of this appendix.

\pagebreak
\section*{C. Proof of $\lim_{a\rightarrow\ 0}||\Psi_{2}||=0$ }
Consider a disjoint union of the set of all matter charges into two mutually exclusive subsets, 
$S_{\mathbf{1}}(\epsilon, a) := \{\vec{\triangle l}|\frac{\epsilon^{2}}{a^{2}}\sum_{I}a(\triangle l_{I})^{2}\ <\ \frac{C}{a^{\delta}}\}$, 
$S_{\mathbf{2}}(\epsilon, a)\ =\ S(\epsilon, a)-S_{\mathbf{1}}(\epsilon, a)$. 
(\ref{eq:nov-14}) can be written in a schematic form as,
\begin{equation}\label{eq:nov-17}
\begin{array}{lll}
\hat{a}_{n}^{poly}|\Psi^{+}\rangle\ =\ \frac{1}{2i\epsilon L}\sum_{\vec{\triangle l}\in S(\epsilon, a)}\sum_{I}e^{ink_{I}}
\left[e^{-\frac{i\hbar}{2}\alpha(\vec{\triangle\epsilon},\vec{\triangle l})}c(\vec{\triangle l}-\vec{\triangle \epsilon}^{(I)})\right.\\
\vspace*{0.1in}
\hspace*{1.7in}\left.-\ e^{+\frac{i\hbar}{2}\alpha(\vec{\triangle\epsilon},\vec{\triangle l})}c(\vec{\triangle l}+\vec{\triangle \epsilon}^{(I)})\right]|\vec{\triangle l}\rangle\\
\vspace*{0.1in}
=\ \left[\sum_{\vec{\triangle l}\in S_{\mathbf{1}}} + \sum_{\vec{\triangle l}\in S_{\mathbf{2}}}\right]\\
\vspace*{0.1in}
\hspace*{1.0in}\sum_{I}e^{ink_{I}}\left[e^{-\frac{i\hbar}{2}\alpha(\vec{\triangle\epsilon},\vec{\triangle l})}c(\vec{\triangle l}-\vec{\triangle \epsilon}^{(I)})\right.\\
\vspace*{0.1in}
\hspace*{1.3in}\left.-\ e^{+\frac{i\hbar}{2}\alpha(\vec{\triangle\epsilon},\vec{\triangle l})}c(\vec{\triangle l}+\vec{\triangle \epsilon}^{(I)})\right]|\vec{\triangle l}\rangle\\
\vspace*{0.1in}
=\ |\Psi_{1}\rangle + |\Psi_{2}\rangle
\end{array}
\end{equation}

\noindent{\bf Lemma}:Show that,
$||\Psi_{2}||\rightarrow\ 0$ rapidly as $a\rightarrow 0$

\noindent{\bf Proof}:\\
Note the following inequality
\begin{equation}
\begin{array}{lll}
||\Psi_{2}||^{2} = \frac{1}{4\epsilon^{2}{L}^{2}}\sum_{\vec{\triangle l}\in S_{\mathbf{2}}}\sum_{I,J}e^{in(k_{I}-k_{J})}\\
\vspace*{0.1in}
\hspace*{1.3in} [e^{-\frac{i\hbar}{2}\alpha(\vec{\triangle\epsilon}^{(I)},\vec{\triangle l})}c(\vec{\triangle l}-\vec{\triangle \epsilon}^{(I)})-e^{+\frac{i\hbar}{2}\alpha(\vec{\triangle\epsilon}^{(I)},\vec{\triangle l})}c(\vec{\triangle l}+\vec{\triangle \epsilon}^{(I)})]\\
\vspace*{0.1in}
\hspace*{1.7in} [e^{-\frac{i\hbar}{2}\alpha(\vec{\triangle\epsilon}^{(J)},\vec{\triangle l})}c(\vec{\triangle l}-\vec{\triangle \epsilon}^{(J)})-e^{+\frac{i\hbar}{2}\alpha(\vec{\triangle\epsilon}^{(J)},\vec{\triangle l})}c(\vec{\triangle l}+\vec{\triangle \epsilon}^{(I)})]\\
\vspace*{0.1in}
\leq \frac{1}{4\epsilon^{2}{L}^{2}}\sum_{\vec{\triangle l}\in S_{\mathbf{2}}}\big[\sum_{I}[e^{-\frac{i\hbar}{2}\alpha(\vec{\triangle\epsilon}^{(I)},\vec{\triangle l})}c(\vec{\triangle l}-\vec{\triangle \epsilon}^{(I)})-e^{+\frac{i\hbar}{2}\alpha(\vec{\triangle\epsilon}^{(I)},\vec{\triangle l})}c(\vec{\triangle l}+\vec{\triangle \epsilon}^{(I)})]\\
\vspace*{0.1in}
\hspace*{1.7in}[\sum_{J}[e^{-\frac{i\hbar}{2}\alpha(\vec{\triangle\epsilon}^{(J)},\vec{\triangle l})}c(\vec{\triangle l}-\vec{\triangle \epsilon}^{(J)})-e^{+\frac{i\hbar}{2}\alpha(\vec{\triangle\epsilon}^{(J)},\vec{\triangle l})}c(\vec{\triangle l}+\vec{\triangle \epsilon}^{(J)})]\big]\\
\vspace*{0.1in}
\leq \frac{1}{4\epsilon^{2}{L}^{2}}\sum_{\vec{\triangle l}\in S_{\mathbf{2}}}\big[\sum_{I}|e^{-\frac{i\hbar}{2}\alpha(\vec{\triangle\epsilon}^{(I)},\vec{\triangle l})}c(\vec{\triangle l}-\vec{\triangle \epsilon}^{(I)})-e^{+\frac{i\hbar}{2}\alpha(\vec{\triangle\epsilon}^{(I)},\vec{\triangle l})}c(\vec{\triangle l}+\vec{\triangle \epsilon}^{(I)})|\big]^{2}\\
\vspace*{0.1in}
=\frac{1}{4\epsilon^{2}{L}^{2}}\sum_{\vec{\triangle l}\in S_{\mathbf{2}}}\big[\sum_{I}\sqrt{c(\vec{\triangle l}-\vec{\triangle \epsilon}^{(I)})^{2} + c(\vec{\triangle l}+\vec{\triangle \epsilon}^{(I)})^{2}}\big]^{2}\\
\end{array}
\end{equation}
Thus
\begin{equation}\label{eq:nov23-3}
\begin{array}{lll}
||\Psi_{2}||\leq\ \frac{1}{4\epsilon L}\sum_{\vec{\triangle l}\in S_{\mathbf{2}}}\left[\sum_{I}\sqrt{c(\vec{\triangle l}-
\vec{\triangle \epsilon}^{(I)})^{2} + c(\vec{\triangle l}+\vec{\triangle \epsilon}^{(I)})^{2}}\right]\\
\vspace*{0.1in}
\hspace*{0.1in} \leq \frac{1}{4\epsilon L}\sum_{\vec{\triangle l}\in S_{\mathbf{2}}}\sum_{I} 
\left[c(\vec{\triangle l}-\vec{\triangle \epsilon}^{(I)}) +  c(\vec{\triangle l}+\vec{\triangle \epsilon}^{(I)})\right]
\end{array}
\end{equation}

Note however that,
\begin{equation}\label{eq:nov23-1}
\begin{array}{lll}
c(\vec{\triangle l}\pm\vec{\triangle\epsilon}^{(I)})\ =\ \exp[-\hbar\frac{\epsilon^{2}}{a^{2}}(2\pi L^{2})\sum_{m}\frac{|\tilde{\triangle l}(m)|^{2} + 2 \pm (\tilde{\triangle l}(m)\tilde{\triangle\epsilon}^{(I)}(-m) + c.c.)}{f(m)}]\\
\vspace*{0.1in}
\hspace*{0.5in} < \exp[-\hbar\frac{\epsilon^{2}}{a^{2}}(2\pi L^{2})\frac{1}{A}\sum_{m}(|\tilde{\triangle l}(m)|^{2} \pm (\tilde{\triangle l}(m)\tilde{\triangle\epsilon}^{(I)}(-m) + c.c.))\\
\vspace*{0.1in}
\hspace*{0.5in} = \exp[-\hbar\frac{\epsilon^{2}}{a^{2}}(2\pi L^{2})\frac{a}{A}\sum_{J}((\triangle l_{J})^{2}\pm (\triangle l_{J})(\triangle\epsilon^{(I)})_{J})\\
\vspace*{0.1in}
\hspace*{0.5in} < \exp[-\hbar\frac{\epsilon^{2}}{a^{2}}(2\pi L^{2})\frac{4\pi L}{A^{2}}\sum_{J}(\triangle l_{J})^{2}) - \sqrt{\sum_{J}(\triangle l_{J})^{2})}
\end{array}
\end{equation}

where in the first step we used $f(n) < n_{max}=A\ \forall\ n$. In the second step $\sum_{m}\tilde{G}(m)\tilde{G}(-m)\ =\ \sum_{J}a\ G_{J}^{2}$ and in the third step 
$\pm\sum_{J}(\triangle l_{J})(\triangle\epsilon^{(I)})_{J}\geq -\sqrt{2}\sqrt{\sum_{J}(\triangle l)_{J}^{2}}$ via Cauchy-Schwarz. ($\triangle\epsilon^{(I)}$ is a vector of length $\sqrt{2}$).

Also note that, as for small enough $\epsilon$, a and for all $\vec{\triangle l}\in S_{2}$ it is easy to see that,
\begin{equation}\label{eq:nov23-2}
\frac{1}{2}\sum_{J}(\triangle l)_{J}^{2} > \sqrt{\sum_{J}(\triangle l)_{J}^{2}}
\end{equation}
Whence
\begin{equation}
c(\vec{\triangle l}\pm\vec{\triangle\epsilon}^{(I)})\ <\ \exp[-\hbar\frac{\epsilon^{2}}{a^{2}}(2\pi L^{2})\frac{4\pi L}{A^{2}}\sum_{J}\frac{(\triangle l_{J})^{2}}{2}]
\end{equation}
Using this bound in (\ref{eq:nov23-3}) we see that,
\begin{equation}
||\Psi_{2}||\ <\ \frac{1}{4\epsilon L}2A\ \sum_{\vec{\triangle l}\in S_{\mathbf{2}}}\exp[-\hbar\frac{\epsilon^{2}}{a^{2}}(2\pi L^{2})\frac{4\pi L}{A^{2}}\sum_{J}\frac{(\triangle l_{J})^{2}}{2}]
\end{equation}
However as we have shown in appendix-A, the R.H.S vanishes in the limit $\epsilon, a\rightarrow\ 0$.\\

\end{document}